\documentclass[11pt,a4paper]{article}
\pdfoutput=1
\usepackage{jcappub}

\usepackage{bm}
\usepackage{amsfonts}
\usepackage{subfigure}
\usepackage{feynmp}

\makeatletter
\def\endfmffile{
  \fmfcmd{\p@rcent\space the end.^^J
          end.^^J
          endinput;}
  \if@fmfio
    \immediate\closeout\@outfmf
  \fi
  \ifnum\pdfshellescape=\@ne
    \immediate\write18{mpost \thefmffile}
  \fi}

\renewcommand\({\left(}
\renewcommand\){\right)}

\newcommand{\be}{\begin{equation}}
\newcommand{\ee}{\end{equation}}
\newcommand{\bea}{\begin{eqnarray}}
\newcommand{\eea}{\end{eqnarray}}

\begin{document}



\title{Higher-order massive neutrino perturbations in large-scale structure}

\author[a,b]{Florian F\"uhrer}
\author[c]{and  Yvonne Y.~Y.~Wong}

\affiliation[a]{Institut f\"ur Theoretische Physik, Universit\"at Heidelberg,
D-69120 Heidelberg, Germany}
\affiliation[b]{Institut f\"ur Theoretische Teilchenphysik und Kosmologie, RWTH Aachen,
D-52056 Aachen, Germany}
\affiliation[c]{School of Physics, The University of New South Wales,
Sydney NSW 2052, Australia}

\emailAdd{fuehrer@thpys.uni-heidelberg.de, yvonne.y.wong@unsw.edu.au}

\abstract{We develop a higher-order perturbation theory for large-scale structure formation involving a free-streaming hot or warm dark matter species. We focus on the case of mixed cold dark matter and massive neutrinos, although our approach is applicable also to a single warm dark matter species. In order to capture the suppressed growth of neutrino density perturbations on small scales, we account for the full momentum dependence of the phase space distribution using the Vlasov equation, and derive from it a formal closed-form nonlinear equation for the neutrino density. Using a systematic perturbative expansion of this equation we compute high-order corrections to the neutrino density contrast without the explicit need to  track the perturbed neutrino momentum distribution. We calculate the leading-order total matter bispectrum for several neutrino masses. Using our result as a benchmark, we test the accuracy of the fluid approximation and a linear approximation used in perturbative and $N$-body analyses, as well as a new hybrid approach that combines the exact linear evolution with the nonlinear structure of the fluid equations. Aiming at $\lesssim1\%$ accuracy, we find that the total matter bispectrum with a low neutrino mass $m=0.046$~eV can be reproduced by all but the fluid approximation, while for larger neutrino masses $m=0.46\to 0.93$~eV only the hybrid approach has the desired accuracy on a large range of scales. This result serves as a cautionary note that approximate nonlinear models of neutrino clustering that reproduce the gross features of some observables may not suffice for precision calculations, nor are they guaranteed to apply to other observables. All of the approximation schemes fail to reproduce the bispectrum of the neutrino density perturbations at better than 20\% accuracy across all scales, indicating that an exact treatment of nonlinear neutrino perturbations is necessary.
}

\maketitle


\section{Introduction}

The dual discoveries of atmospheric and solar neutrino flavour oscillations at the turn of the 21st century and subsequent confirmations of flavour oscillations by terrestrial accelerator and reactor neutrino experiments have by now established unequivocally that at least one neutrino has a mass $m_i$ ($i=1,2,3$) exceeding 0.057~eV~\cite{bib:pdg}.  Measurements of the tritium $\beta$-decay end-point spectrum, on the other hand, impose an upper bound on the effective electron neutrino mass of $m_{ee}  \equiv \sum_{i}  |U_{ei}|^2 m_i
< 2.2$~eV~\cite{Kraus:2004zw,Lobashev:1999tp},
where $U_{\alpha i}$ ($\alpha = e, \mu, \tau$) is the neutrino mixing matrix.  Together these limits predict a present-day neutrino energy density 
 $\Omega_{\nu}h^2=\sum_i m_i/(94~{\rm eV})$ in  the range $0.0006 < \Omega_\nu h^2 <0.074$,
 \footnote{To estimate the maximum $\Omega_\nu h^2$  we have made use of large mixing, i.e.,  $|U_{e i}| \sim |U_{\mu i}| \sim |U_{\tau i}|  \sim O(0.1)$, so that ${\rm max}\sum_i m_i  \sim 3 m_{ee} \sim 7$~eV.}
making the neutrino an inevitable and potentially sizeable component of the cosmic dark matter.

Cosmology itself also provides an independent constraint on $\Omega_\nu h^2$ and hence the neutrino mass sum $\sum_i m_i$ via the phenomenon of neutrino free-streaming.  See, e.g., \cite{Lesgourgues:2006nd,Hannestad:2006zg,Wong:2011ip} for recent reviews.
In linear perturbation theory free-streaming causes the scalar spacetime metric perturbations on length scales smaller than a certain $m_i$-dependent ``free-streaming scale'' to decay away.  Such a scale-dependent decay manifests itself in the cosmic microwave background (CMB) temperature anisotropies as an enhancement of power in the temperature autocorrelation spectrum at large $\ell$ multipoles; for the large-scale matter distribution, its main effect is to attenuate the growth of density perturbations on small scales.  The non-observation so far of these scale-dependent effects in various CMB and galaxy redshift surveys have allowed us to constrain $\sum_i m_i$ to at most $O(1)$~eV in $\Lambda$CDM-cosmologies and variants thereof, the precise number depending on the details of the cosmological model adopted in the statistical inference and the data combination used.  See, e.g., \cite{Abazajian:2011dt,Lesgourgues:2014zoa} for a summary of pre- and post-Planck bounds.  The next generation of multi-purpose galaxy/cluster/cosmic shear surveys such as the Large Synoptic Survey Telescope%
\footnote{\tt http://www.lsst.org/lsst/}
and the ESA Euclid mission%
\footnote{\tt http://sci.esa.int/euclid/}
are expected to improve these limits by at least an order of magnitude: conservative estimates place the $1\sigma$ sensitivity to $\sum_i m_i$ in the $0.02 \to 0.03$~eV region~\cite{Hamann:2012fe,Audren:2012vy}, sufficient to even ``measure'' the minimum predicted neutrino mass sum with $2 \sigma+$ significance.

Realisation of this last goal, however, hinges crucially on our being able to predict the observable quantities to an adequate level of accuracy.  To this end, the need to go beyond linear perturbation theory in the computation of matter perturbations on scales relevant for galaxy redshift and cosmic shear surveys cannot be overemphasised.
Higher-order perturbation theory for cold dark matter (CDM)-only cosmologies has been explored in great detail, including extensions to resummation and renormalisation group schemes. 
See, e.g.,~\cite{Bernardeau:2001qr} for a review and~\cite{Crocce:2005xy,Crocce:2005xz,McDonald:2006hf,Matsubara:2007wj,Matarrese:2007wc,Baumann:2010tm,Pietroni:2011iz}
for a sample of recent works.  
Free-streaming massive neutrinos with a velocity dispersion, however, demand a different treatment, and attempts to incorporate them in a 
higher-order calculation so far consist in approximations that may not be well justified.  The works of~\cite{Saito:2008bp,Wong:2008ws,Lesgourgues:2009am},
for example, assume the neutrinos to have density and velocity perturbations only to linear order.  The analyses of~\cite{Shoji:2009gg,Shoji:2010hm,Blas2014}
attempt to include nonlinear neutrino perturbations by modelling the free-streaming neutrinos as a fluid with a sound speed; while this approach 
reproduces qualitatively the effect of suppressed perturbation growth, the effective sound speed remains an {\it ad hoc} quantity that, in principle, needs to be fixed order by order in the perturbative series.%
\footnote{The linear analysis of mixed CDM+massive neutrino cosmologies in~\cite{Ringwald:2004np} shows that the steady-state solution for the neutrino density perturbations 
does have a Jeans'-scale-like quantity from which one can extract an effective sound speed for the neutrinos.  There is however no persuasive reason that this same sound speed should apply to all orders.}

In this paper, we develop from first principles a higher-order perturbation theory for free-streaming particles
in structure formation and apply it to the case of massive neutrinos. We focus on perturbations in the neutrino energy density, and formulate the theory in a way that avoids the explicit evaluation of the perturbed neutrino momentum distribution in real time (in contrast to the recent work of~\cite{Dupuy:2013jaa}).
When combined with standard perturbation theory for CDM, this theory can be used to calculate observables such as the $N$-point statistics of the matter density perturbations in mixed cold+hot dark matter cosmologies without further, uncontrolled approximations.  We apply the theory to compute the leading-order total matter bispectrum, and use our result as a benchmark against which to test the validity of the approximation schemes of~\cite{Saito:2008bp,Wong:2008ws,Lesgourgues:2009am,Shoji:2009gg,Shoji:2010hm} discussed above.
As the full nonlinear theory is quite computationally intensive, we also investigate a hybrid approximation scheme that  combines elements of the full theory and the simpler fluid approach.

Lastly, we note that perturbative analyses such as this one can also inform $N$-body simulations of large-scale structure.  This is especially so in view of the
large amount of simulation noise incurred by the particle realisation of neutrinos free-streaming at some 10\% of the speed of light at initialisation, 
e.g.,~\cite{Brandbyge:2008rv,Villaescusa-Navarro:2013pva}.  Although a good number of approximation schemes have been advocated to circumvent the noise 
problem,~e.g.,~\cite{Brandbyge:2008js,Upadhye:2013ndm,AliHaimoud:2012vj,Hannestad:2011td}, these schemes need to be grounded in theory and the extent of their validity investigated.  In this work, we shall identify the perturbative limits of some of these approximation schemes, and compare them with our full nonlinear theory of neutrino perturbations.

The paper is structured as follows.   We review in section~\ref{sec:CDM} the standard perturbation theory for CDM, and extend the theory to the case of two fluids with disparate effective sound speeds.  In section~\ref{sec:boltz}, starting from the collisionless Boltzmann equation, we develop our perturbation theory for the neutrino density contrast, and generalise this framework to mixed CDM+neutrino cosmologies in section~\ref{sec:cdmneu}.  The hybrid approach which combines elements of the exact theory and the fluid approximation is presented  section~\ref{sec:hybrid}.  In section~\ref{sec:diagram} we introduce a diagrammatic representation of the theory and the resulting $N$-point functions, while in section~\ref{sec:applications} we apply the theory to compute, in particular, the leading-order matter bispectrum in a mixed CDM+neutrino cosmology, and discuss the validity of various approximation schemes.  Section~\ref{sec:conclusions} contains our conclusions. We assume the Newtonian limit of cosmological perturbation theory throughout this work.


\section{Fluid equations}\label{sec:CDM}

Consider an ensemble of identical nonrelativistic particles permeating all space in an expanding universe.  In the continuum limit, the time evolution of the density contrast $\delta(\bm{x},\tau)$ and peculiar velocity $\bm{u}(\bm{x},\tau)$ at a comoving spatial coordinate point $\bm{x}$ is governed by  the continuity and Euler equations (e.g.,~\cite{Bernardeau:2001qr}),
\begin{equation}
\begin{aligned}
& \frac{\partial \delta (\bm{x},\tau)}{\partial \tau} + \nabla \cdot \{ (1+ \delta)  \bm{u}(\bm{x},\tau)\} =0, \\
&  \frac{\partial \bm{u} (\bm{x},\tau)}{\partial \tau} + \mathcal{H}(\tau) \bm{u}(\bm{x},\tau)+ \bm{u} \cdot \nabla \bm{u} = -\nabla \Phi(\bm{x}, \tau) -\frac{1}{1+\delta} \nabla \cdot \{(1+\delta) \bm{\sigma}(\bm{x},\tau) \}.\label{eq:eulerx}
\end{aligned}
\end{equation}
Here, $\tau$ denotes the conformal time, $\mathcal{H}(\tau)  \equiv d \ln a/d \tau = a H$ the conformal Hubble expansion rate, $\bm{\sigma}(\bm{x},\tau)$ the spatial stress tensor, and $\Phi(\bm{x},\tau)$ is the Newtonian gravitational potential, related to the density contrast via the Poisson equation,
\begin{align}
 \nabla^2 \Phi(\bm{x},\tau) = \frac{3}{2} \mathcal{H}^2(\tau) \Omega(\tau) \delta(\bm{x},\tau),\label{eq:poissonx}
\end{align}
where $\Omega(\tau) \equiv \bar{\rho}_m(\tau)/\rho_{\rm crit}(\tau)$ is the (mean) matter density parameter at time $\tau$.

It is useful to rewrite the equations of motion in Fourier space, for which we employ the transformation convention 
\begin{equation}
\begin{aligned}
&f\left(\bm{k}\right) =\mathcal{F}\left[f\left(\bm{x}\right)\right]=\int  d^3x \: f\left(\bm{x}\right)  e^{-i \bm{k}  \cdot \bm{x}},\\
&f\left(\bm{x}\right) =\mathcal{F}^{-1}\left[f\left(\bm{k}\right)\right]=\int  \frac{d^3k}{\left( 2 \pi \right)^3} \: f\left(\bm{k}\right) e^{i \bm{k} \cdot \bm{x}},
\end{aligned}
\end{equation}
for some field $f$. Then, introducing a new, super-conformal time variable $s=s_{\rm in}+\int_{\tau_i}^{\tau} d \tau'/a$, 
and assuming that vorticity vanishes (i.e., $\nabla \times \bm{u}=0$), 
equation~(\ref{eq:eulerx}) can now be equivalently expressed as
\begin{equation}
\begin{aligned}
 &\frac{\partial \delta(\bm{k})}{\partial s}+a\theta(\bm{k})=-a\left[\frac{\bm{k}\cdot \bm{k}_2}{k_2^2}\delta\left(\bm{k}_1\right) \theta\left(\bm{k}_2\right)\right]_{\bm{k}},\\ 
& \frac{\partial \theta(\bm{k})}{\partial s}+a\mathcal{H} \theta(\bm{k}) -a k^2 \Phi(\bm{k}) -a \bm{k}\cdot \bm{\sigma}(\bm{k}) \cdot\bm{k}=\\
& \hspace{30mm} - a\left[\frac{\bm{k}_1\cdot\bm{k}_2\bm{k}\cdot\bm{k}_2}{k_1^2k_2^2} \theta(\bm{k}_1) \theta(\bm{k}_2)\right]_{\bm{k}} -ai\bm{k}\cdot\mathcal{F}\left[\frac{1}{1+\delta} \nabla\delta \cdot \bm{\sigma}\right], \label{eq:euler} 
 \end{aligned}
 \end{equation}
where $\theta(\bm{k}) \equiv i  \bm{k}\cdot \bm{u}$ denotes the divergence of the velocity field, accompanied by 
the Poisson equation,
 \begin{align}
-k^2 \Phi(\bm{k}) = \frac{3}{2} \mathcal{H}^2 \Omega(s) \delta(\bm{k}). \label{eq:poissonk}
\end{align}
For convenience we have adopted in equation~(\ref{eq:euler}) a short-hand notation for the convolution integrals,
\begin{align}
\int \left(\prod_{i=1}^n \frac{d^3k_i}{(2\pi)^3}\right) (2\pi)^3\delta_D\left(\bm{k}-\sum_ {i=1}^n \bm{k}_i\right)f_1(\bm{k}_1) \ldots f_n(\bm{k}_n)\equiv\left[f_1(\bm{k}_1)\ldots f_n(\bm{k}_n)\right]_{\bm{k}}.
\label{eq:convolutionshorthand}
\end{align}
 Where no confusion is likely to arise, we shall not write out explicitly the time dependence of the variables.


\subsection{Effective sound speed and the free-streaming scale}
\label{sec:soundspeed}

As they stand now equations~(\ref{eq:euler}) and~(\ref{eq:poissonk}) do not form a closed system of equations; we have not yet specified the behaviour of the spatial stress tensor $\bm{\sigma}$.  To model CDM particles it is common to assume $\bm{\sigma}=0$ in the mildly nonlinear regime where no shell crossing has yet occurred (see, however,~\cite{Baumann:2010tm,Pietroni:2011iz,McDonald:2009hs})). The assumption always breaks down, however, for any other form of free-streaming dark matter that comes with an intrinsic velocity dispersion, e.g., massive neutrinos, or warm dark matter (WDM) particles.  In such cases, approximating the stress term by an effective sound speed $c_{\rm s}^2$ in the manner
\begin{align}
\bm{\sigma}=c_{\rm s}^2(s)\delta \: \bm{1}\label{eq:soundspeed}
\end{align}
at least permits us to study its effect on the evolution of $\delta$ and $\theta$ qualitatively.
Endowing~$c_{\rm s}^2$ with a $k$-dependence might yield even better results, but at the expense of introducing a non-local term in the equations of motion in real space.

Then, replacing the stress tensor in the {\it linearised} version of equation~\eqref{eq:euler} according to equation~\eqref{eq:soundspeed} yields 
\begin{align}
\label{eq:fluid}
 \frac{\partial^2 \delta^{(1)}}{\partial s^2}+a^2 \left(k^2 c_{\rm s}^2- \frac{3}{2}\mathcal{H}^2\Omega(s)\right)\delta^{(1)}=0,
\end{align}
where we have also made use of the Poisson equation~(\ref{eq:poissonk}).  The role of the stress term is then clear: for those $k$ values at which $k^2 c_{\rm s}^2$ is much smaller than the gravitational source term proportional to $(3/2) \mathcal{H}^2 \Omega(s)$, the linear density contrast $\delta^{(1)}$ grows with time as though $c_{\rm s}^2$ were zero (i.e., like CDM).  At the other extreme where the stress term exceeds the gravitational source term, the growth of~$\delta^{(1)}$ is suppressed.  The demarcation between these two limiting behaviours is called the free-streaming scale $\lambda_{\rm FS} = 2 \pi/k_{\rm FS}$, defined here as
\begin{align}
k_{\rm FS}^2\equiv \frac{3}{2}\frac{\mathcal{H}^2\Omega(s)}{c_{\rm s}^2}\approx\frac{3}{2}\frac{a^2 m^2 \mathcal{H}^2\Omega(s)}{\overline{q^2}},
\end{align}
where we have approximated the effective sound speed by the velocity dispersion of the unperturbed momentum distribution $\bar{f}(q)$ \cite{Shoji:2010hm,Boyanovsky:2008he}, with
\begin{align}
\overline{q^2}\equiv \frac{ \int d^3 q\: q^2 \bar{f}(q)}{ \int d^3 q\:  \bar{f}(q)},
\label{eq:q2bar}
\end{align}
and $m$ is the particle mass.
The approximation $\bm{\sigma}=\frac{\overline{q^2}}{(am)^2}\delta\: \bm{1}$ 
thus corresponds to assuming that the perturbed distribution is the same
as the unperturbed one, but allowing for a spatial dependence of the number
density. 
 At linear order, up to a subdominant source terms%
 \footnote{The subdominance can be confirmed by solving the fluid equations without the source term and then comparing with the solution of the Boltzmann hierarchy. See e.g.,~\cite{Shoji:2010hm}.}
 this corresponds to a velocity expansion up to second order~\cite{Boyanovsky:2008he}.

At linear order the fluid description~(\ref{eq:fluid}) faithfully reproduces the clustering behaviour of free-streaming dark matter
on length scales much larger than the free-streaming scale, i.e., $k \ll k_{\rm FS}$.
On small length scales, $k \gg k_{\rm FS}$, however, the approximation~(\ref{eq:soundspeed}) 
necessarily leads to acoustic oscillations in the density contrast, an artefact that 
renders the fluid approach a poor description especially in WDM scenarios in which the free-streaming WDM is the dominant  dark matter constituent.  Nonetheless, 
for mixed CDM+massive neutrino cosmologies where the latter is subdominant, 
a fluid description for the neutrino component still has some merit in the $k \gg k_{\rm FS}$ regime; we defer a discussion to section~\ref{sec:twofluid}.


\subsection{Higher-order fluid perturbations}\label{sec:pertCDM}

Higher-order perturbation theory for the CDM case (i.e., $\bm{\sigma}=0$) is well known (see, e.g.,~\cite{Bernardeau:2001qr}).  Generalising the theory to include a sound speed is also straightforward~\cite{Shoji:2009gg}, which we review here for completeness.

We begin by defining a doublet, 
\begin{align}
\bm{\varphi}(\bm{k}) 
\equiv\begin{pmatrix}
\delta(\bm{k}) \\
-\theta(\bm{k})
\end{pmatrix}.
\end{align}
Then the equations of motion~\eqref{eq:euler} can be rewritten in a more compact form:
\begin{equation}
\frac{d\varphi_a(\bm{k})}{ds}+\Pi_{ab}(k,s) \varphi_b(\bm{k})=a\left[\gamma_{abc}\left(\bm{k}_1,\bm{k}_2\right) \varphi_b \left(\bm{k}_1\right)\varphi_c\left(\bm{k}_2\right)\right]_{\bm{k}} 
+a c_{\rm s}^2(s)i\bm{k}\cdot
\mathcal{F}\left[\frac{\varphi_1}{1+\varphi_1}\nabla\varphi_1\right]\delta_{a2},
\label{eq:CDMcompact}
\end{equation}
where
\begin{align}
\label{eq:pi}
\bm{\Pi}(k,s) \equiv\begin{pmatrix}
0&&-a\\
-\frac{3}{2} a \mathcal{H}^2 \Omega(s) +ak^2 c_{\rm s}^2(s) &&a \mathcal{H}
\end{pmatrix},
\end{align}
and $\gamma_{abc}(\bm{k}_1,\bm{k}_2)$ is the symmetrised integral kernel with
\begin{equation}
\begin{aligned}
&\gamma_{112}\left(\bm{k}_1,\bm{k}_2\right)= \gamma_{121}\left(\bm{k}_2,\bm{k}_1\right)\equiv \frac{1}{2}\frac{\left(\bm{k}_1+\bm{k}_2\right)\cdot \bm{k}_2}{k_2^2}, 
\label{eq:KernelCDM}\\
&\gamma_{222}\left(\bm{k}_1,\bm{k}_2\right)\equiv \frac{1}{2}\frac{\bm{k}_1\cdot\bm{k}_2 \left(\bm{k}_1+\bm{k}_2\right)^2}{k_1^2 k_2^2}
\end{aligned}
\end{equation}
as its only non-vanishing components.

Equation~\eqref{eq:CDMcompact} can now be solved by way of a Green's function.  Defining the Green's function $g_{ab}(k;s,s')$ via
\begin{equation}
\begin{aligned}
&\frac{dg_{ab}}{ds}\left(k;s,s' \right)+\Pi_{ac}\left(k,s\right) g_{cb}\left(k;s,s'\right)=\delta_{ab} \delta\left(s-s'\right),\label{eq:defGreensCDM}\\
&\quad\quad\quad\quad\quad g_{ab}(k;s,s')=0 \quad \mathrm{if}\quad s'>s,
\end{aligned}
\end{equation}
equation~\eqref{eq:CDMcompact} then has the formal solution
\begin{equation}
\begin{aligned}
\varphi_a\left(\bm{k},s\right)=\: &g_{ab}\left(k;s,s_{\rm in}\right) \varphi_b(\bm{k},s_{\rm in})\\
&+\int^{s}_{s_{\rm in}} ds' \: g_{ab}\left(k;s,s'\right)a(s')\Bigg\{\left[\gamma_{bcd}\left(\bm{k}_1,\bm{k}_2\right)\varphi_c\left(\bm{k}_1,s'\right)\varphi_d\left(\bm{k}_2,s'\right)\right]_{\bm{k}} \label{eq:CDMintegral}\\
&\hspace{50mm} + c_{\rm s}^2(s')i\bm{k}\cdot
\mathcal{F}\left[\frac{\varphi_1}{1+\varphi_1}\nabla\varphi_1\right]\delta_{b2}\Bigg\},
\end{aligned} 
\end{equation}
where $s_{\rm in}$ denotes the initial time. Note that in contrast to the CDM case, the presence of a finite sound speed gives rise to a $k$-dependence in $g_{\rm ab}(k;s,s')$. 
Equation~\eqref{eq:CDMintegral} can be solved iteratively, with the understanding that the stress term can be expanded in powers of $\delta$ thus:
\begin{align}
\mathcal{F}\left[\frac{\delta}{1+\delta}\nabla\delta\right]=\left[i\bm{k}_1\delta(\bm{k}_1)\delta(\bm{k}_2)\right]_{\bm{k}}-\left[i\bm{k}_1\delta(\bm{k}_1)\delta(\bm{k}_2)\delta(\bm{k}_3)\right]_{\bm{k}}+\ldots,
\end{align}
thereby yielding an expansion in powers of $\bm{\varphi}({\bm k},s_{\rm in})$.


\subsection{Two-fluid perturbation theory}
\label{sec:twofluid}

Generalisation of the perturbation theory to two fluids coupled only through gravity, e.g., CDM and massive neutrinos, requires only minimal modifications to the equations of motion.   Firstly, the gravitational potential $\Phi$ in the Poisson equation~(\ref{eq:poissonk}) is now sourced by both CDM and neutrino density perturbations,
$\delta_{\rm C}$ and $\delta_\nu$, which requires that we make the replacement
\begin{align}
 \delta(\bm{k},s) \rightarrow f_{\rm C} \delta_{\rm C}(\bm{k},s)+f_\nu \delta_\nu(\bm{k},s),
\end{align}
where $f_{\rm C}$ and~$f_\nu$ denote the fractions of the total matter density $\Omega(s)$ in the form of CDM and massive neutrinos respectively, and $f_{\rm C}+f_\nu = 1$.%
\footnote{In this definition we have implicitly assumed that CDM and baryons form a single fluid, and the parameter~$f_{\rm C}$
subsumes both fractions of nonrelativistic matter in the form of CDM and in the form of  baryons.}
Note that these fractions are constant  in our nonrelativistic treatment of the neutrinos.

Then, assigning the doublets $\varphi_a$ (subscript) 
and $\varphi^A$ (superscript) 
 to the CDM and the neutrino fluid respectively, 
the Green's functions are defined by 
\begin{equation} 
\begin{aligned}
\label{eq:greens}
&\frac{dg_{ab}}{ds}\left(k;s,s' \right)+\Pi_{ac}\left(s\right) g_{cb}\left(k;s,s'\right)-\delta_{a}^2\frac{3}{2}a\mathcal{H}^2\Omega(s) f_\nu {g^1}_{b}(k;s,s')=0,\\
&\frac{d{g^A}_{b}}{ds}\left(k;s,s' \right)+\Pi^{AC}\left(k,s\right) {g^C}_{b}\left(k;s,s'\right)-\delta^A_{2}\frac{3}{2}a\mathcal{H}^2\Omega(s) f_{\mathrm{C}}g_{1b}(k;s,s')=0,\\
&\frac{dg^{AB}}{ds}\left(k;s,s' \right)+\Pi^{AC}\left(k,s\right) g^{CB}\left(k;s,s'\right)-\delta^{A}_2\frac{3}{2}a\mathcal{H}^2\Omega(s) f_{\mathrm{C}}{g_1}^B(k;s,s')=0,\\
&\frac{d{g_a}^B}{ds}\left(k;s,s' \right)+\Pi_{ac}\left(s\right) {g_c}^B\left(k;s,s'\right)-\delta_{a}^2\frac{3}{2}a\mathcal{H}^2\Omega(s) f_\nu g^{1B}(k;s,s')=0,
\end{aligned}
\end{equation}
where the matrices $\Pi_{ab}(s)$ and $\Pi^{AB}(k,s)$ take the form given in equation~(\ref{eq:pi}), but with the replacements
\begin{equation}
\begin{aligned}
&\Pi_{21} (s)=-\frac{3}{2} a \mathcal{H}^2 \Omega(s) f_{\rm C},  \\
&\Pi^{21}(k,s)= -\frac{3}{2} a \mathcal{H}^2 \Omega(s) f_\nu +ak^2 c_{\rm s}^2(s),
\end{aligned}
\end{equation}
and
\begin{equation}
\begin{aligned}
&g_{ab}\left(k;s,s'\right)\underset{ s \rightarrow s'} { \longrightarrow }\delta_{ab}, &\quad
&g^{AB}\left(k;s,s'\right)\underset{ s \rightarrow s'} { \longrightarrow }\delta^{AB},\\
&{g_a}^B\left(k;s,s'\right)\underset{ s \rightarrow s'} { \longrightarrow }0,&\quad
&{g^A}_b\left(k;s,s'\right)\underset{ s \rightarrow s'} { \longrightarrow }0,\\
&g(k;s,s')=0 \quad \mathrm{if}\quad s'>s
\end{aligned}
\end{equation}
constitute the initial conditions.
The full nonlinear equations then have the formal solutions
\begin{equation}
\begin{aligned}
\varphi_a\left(\bm{k},s\right)=\: & g_{ab}\left(k;s,s_{\rm in}\right) \varphi_b(\bm{k},s_{\rm in})+{g_a}^B\left(k;s,s_{\rm in}\right) \varphi^B(\bm{k},s_{\rm in})\\
&+\int^{s}_{s_{\rm in}} ds' \: g_{ab}\left(k;s,s'\right)a(s')\left[\gamma_{bcd}\left(\bm{k}_1,\bm{k}_2\right)\varphi_c\left(\bm{k}_1,s'\right)\varphi_d\left(\bm{k}_2,s'\right)\right]_{\bm{k}}\\
&+\int^{s}_{s_{\rm in}} ds' \: {g_a}^B\left(k;s,s'\right)a(s')\Bigg\{\left[\gamma^{BCD}\left(\bm{k}_1,\bm{k}_2\right)\varphi^C\left(\bm{k}_1,s'\right)\varphi^D\left(\bm{k}_2,s'\right)\right]_{\bm{k}}\\
&\hspace{52mm}+c_{\rm s}^2(s')i\bm{k}\cdot
\mathcal{F}\left[\frac{\delta_{\nu}}{1+\delta_{\nu}}\nabla\delta_{\nu}\right]\delta^{B2}\Bigg\},
\label{eq:2fluid1}
\end{aligned}
\end{equation}
and
\begin{equation}
\begin{aligned}
\varphi^A\left(\bm{k},s\right)=\:& {g^A}_{b}\left(k;s,s_{\rm in}\right) \varphi_b(\bm{k},s_{\rm in})+g^{AB}\left(k;s,s_{\rm in}\right) \varphi^B(\bm{k},s_{\rm in})\\
&+\int^{s}_{s_{\rm in}} ds' \: {g^A}_{b}\left(k;s,s'\right)a(s')\left[\gamma_{bcd}\left(\bm{k}_1,\bm{k}_2\right)\varphi_c\left(\bm{k}_1s'\right)\varphi_d\left(\bm{k}_2,s'\right)\right]_{\bm{k}}\\
&+\int^{s}_{s_{\rm in}} ds' \: g^{AB}\left(k;s,s'\right)a(s')\Bigg\{\left[\gamma^{BCD}\left(\bm{k}_1,\bm{k}_2\right)\varphi^C\left(\bm{k}_1,s'\right)\varphi^D\left(\bm{k}_2,s'\right)\right]_{\bm{k}}\\
&\hspace{52mm}+c_{\rm s}^2(s')i\bm{k}\cdot
\mathcal{F}\left[\frac{\delta_{\nu}}{1+\delta_{\nu}}\nabla\delta_{\nu}\right]\delta^{B2}\Bigg\},
\label{eq:2fluid2}
\end{aligned}
\end{equation}
a perturbative expansion of which up to third order have been presented in~\cite{Shoji:2009gg}.  Later on in sections~\ref{sec:diagram} and~\ref{sec:applications}, we
 shall also be evaluating these expressions up to second perturbative order for the construction of the leading-order matter bispectrum.

As discussed in section~\ref{sec:soundspeed},  for $k \gg k_{\rm FS}$ the fluid approach formally implies acoustic oscillations, and is, at least at linear order, a poor description of the clustering behaviour of a dominant free-streaming dark matter component on these scales.  If the free-streaming dark matter should be subdominant, however, as in the case of massive neutrinos, and the dominant dark matter is cold, then both the fluid description and an exact treatment in terms of the collisionless Boltzmann equation yield at  $k \gg k_{\rm FS}$  the same linear attractor solution, $\delta_\nu^{(1)} \sim (k_{\rm FS}/k)^2 \delta_{\rm C}^{(1)}$~\cite{Basse:2010qp,Ringwald:2004np}.
The goal of the present work, therefore, is to test the validity of the fluid approximation on the transitional length scales $k \sim k_{\rm FS}$, especially at higher perturbative orders, against an exact treatment using the collisionless Boltzmann equation.



\section{Neutrino perturbations from the collisionless Boltzmann equation}
\label{sec:boltz}
The fluid approximation of section~\ref{sec:CDM}, although extremely simple, is inherently unsatisfactory.  
Firstly, there is the question of how one should model the effective sound speed.  Indeed, our present choice of $c_{\rm s}^2 = \overline{v^2}$, where
$\overline{v^2}$ is the velocity dispersion of {\it unperturbed} momentum distribution implies that the effective sound speed may not be the same 
at all perturbative orders.  Secondly, artificial acoustic oscillations in the density contrast will likely be present before the solution reaches the attractor.
Both issues are expected to impact most strongly on the phenomenology at the transitional length scales $k \sim k_{\rm FS}$.

We therefore begin from first principles, and treat the non-vanishing
stress tensor properly by following 
the evolution of the full momentum distribution $f(\bm{x},\bm{q},s)$ as dictated by the (nonrelativistic) collisionless Boltzmann equation (e.g.,~\cite{Bertschinger:1993xt}),
\begin{align}
\frac{\partial f}{\partial s}+\frac{\bm{q}}{m}\cdot\nabla f - a^2 m\left(\nabla \Phi \right) \cdot \frac{\partial f}{\partial \bm{q}}=0\mathrm{.}
\label{eq:nonrelBoltzmann}
\end{align}
Here, the momentum variable $\bm{q}$ is related to the physical momentum $\bm{p}$ by
$\bm{q}=a\:\bm{p}$, and $m$ is the mass of the free-streaming dark matter particle---hitherto loosely termed the ``neutrino mass''.

As usual we split the distribution function into a
homogeneous and isotropic background and a perturbation, $f(\bm{x},\bm{q},s) = \bar{f}(q) + \delta f(\bm{x},\bm{q},s)$,  where, for massive neutrinos, the background component
is given by the ultra-relativistic Fermi--Dirac distribution,
\begin{align}
\bar{f}(q)=\frac{1}{6 \pi \zeta (3) T^3} \frac{1}{1+\exp(q/T)},
\end{align}
with $T=(4/11)^{1/3}\ T_{\mathrm{CMB}}\approx 1.68\times10^{-4}$~eV representing the present-day neutrino temperature,
$\zeta$ the Riemann zeta function, and the distribution has been normalised such that $\int d^3q\: \bar{f}(q)=1$.
The perturbation $\delta f$, normalised here as  
\begin{align}
\delta (\bm{x},s)\equiv \frac{\delta \rho(\bm{x},s)}{\bar{\rho}(s)}=\int d^3q   \: \delta f(\bm{x},\bm{q},s),
\end{align}
where $\bar{\rho}$ is the homogeneous energy density, follows the collisionless Boltzmann equation~(\ref{eq:nonrelBoltzmann}).  In Fourier space, this is
\begin{align}
\frac{\partial \delta f}{\partial s}+i  \frac{\bm{q}}{m} \cdot \bm{k} \delta f - V(s) \delta(\bm{k}) \bm{v} \cdot \frac{\partial\bar{f}}{\partial \bm{q}}=
V(s) \left[ \delta( \bm{k}_1) \bm{v}_1 \cdot \frac{\partial \delta f}{\partial \bm{q}}(\bm{k}_2)\right]_{\bm{k}},
\label{eq:boltzmannfourier}
\end{align}
where $V(s) \equiv -(3/2)im a^2(s) \mathcal{H}^2(s)\Omega(s)$, $\bm{v}_i \equiv \bm{k}_i/k_i^2$, and 
we have gathered all the the linear and nonlinear terms on the left- and right-hand side respectively. Note that in writing equation~(\ref{eq:boltzmannfourier}) we have assumed for brevity only the free-streaming dark matter
contributes to the Newtonian gravitational potentials~$\Phi$; indeed, equation~(\ref{eq:boltzmannfourier}) would be exact if 
the dark matter content of the universe consists of only one single species of WDM. The generalisation to a mixed CDM+massive neutrino cosmology will be discussed in  section~\ref{sec:cdmneu}.


\subsection{Gilbert's Equation}\label{sec:gilbert}

Setting the right-hand side to zero, the {\it linearised} version of equation~(\ref{eq:boltzmannfourier})  is formally solved by~\cite{Bertschinger:1993xt,Brandenberger:1987kf}
\begin{equation}
\delta f^{(1)}\left(\bm{k},\bm{q},s\right)=  \tilde{g} \left(\bm{k},\bm{q};s,s_{\rm in}\right)\delta f^{(1)}\left(\bm{k},\bm{q},s_{\rm in}\right)  +\int_{s_{\rm in}}^s ds'\, V(s')  \tilde{g}\left(\bm{k},\bm{q};s,s'\right) \delta^{(1)}(\bm{k},s')\bm{v} \cdot \frac{\partial \bar{f}}{\partial \bm{q}},
\label{eq:formalsolution}
\end{equation}
where
$\tilde{g}\left(\bm{k},\bm{q};s,s'\right) \equiv \exp[{-i\bm{q}\cdot\bm{k}\left(s-s'\right)/m}]$
is the solution of the collisionless Boltzmann equation in the free-streaming limit (i.e., formally $\delta=0$).

\begin{figure}[t]
	\centering
		\includegraphics{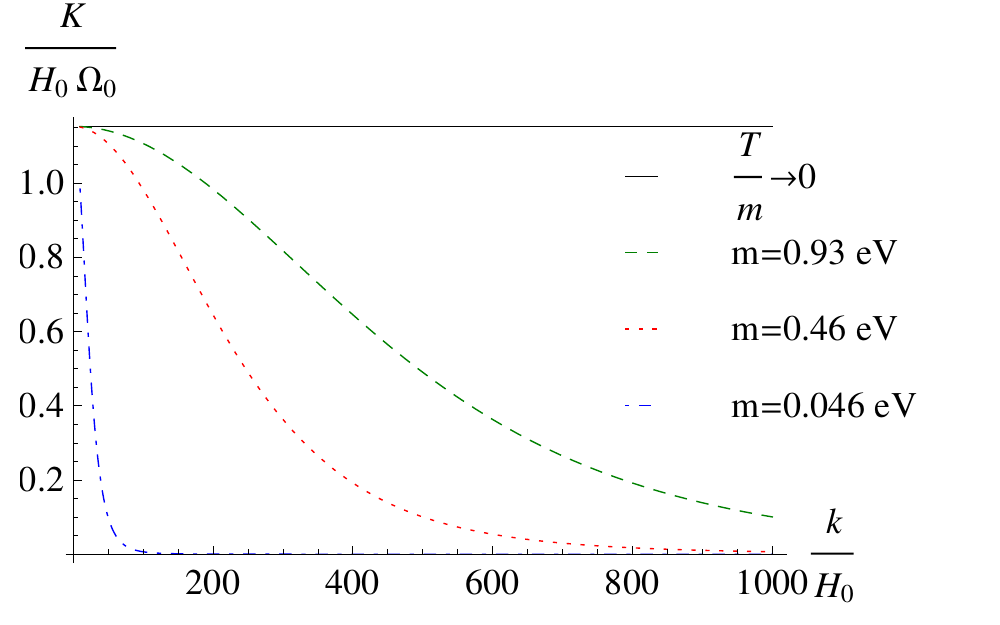}
	\caption{The kernel $K(k; s,s')$ as functions of the wavenumber~$k$ for a sample of neutrino masses~$m$ at $s'=s(a=1/10)$ and $s=s_0=s(a_0)$.
	 The limit $T/m\rightarrow 0$ corresponds to the CDM limit.  Observe that the kernel decreases with growing $k$, and the smaller the neutrino mass the more rapid the decrease.  This is a manifestation of neutrino free-streaming, which translates into a reduction of perturbation growth on small length scales.}
	\label{fig:kernels}
\end{figure}

Integrating equation~\eqref{eq:formalsolution} over momentum $\bm{q}$ then gives an integral equation for the density contrast 
$\delta$~\cite{Brandenberger:1987kf,Bertschinger:1993xt}
\begin{align}
\delta^{(1)}\left(\bm{k},s\right)&=I(\bm{k},s)+\int_{s_{\rm in}}^s ds'\: K\left(k;s, s'\right) \delta^{(1)} \left(\bm{k},s'\right), \label{eq:gilbert}
\end{align}
known as Gilbert's equation.  Here, the source term reads
\begin{align}
I\left(\bm{k},s\right)=\int d^3q \: e^{-i\bm{q}\cdot\bm{k}\left(s-s_{\rm in}\right)/m}\delta f^{(1)}\left(\bm{k},\bm{q},s_{\rm in}\right),
\label{eq:source}
\end{align}
while the integral kernel is given by
\begin{equation}
\begin{aligned}
K\left(k; s,s'\right)&=\frac{3}{2} a^2(s')\mathcal{H}^2(s') \Omega(s') \left(s-s'\right) \int d^3q \:    e^{-i\bm{q}\cdot\bm{k}\left(s-s'\right)/m} \, \bar{f}(q)\\
& =\frac{3}{2} a^2(s')\mathcal{H}^2(s')  \Omega(s') \left(s-s'\right) F\left(\frac{Tk(s-s')}{m} \right),
\label{eq:kkernel}
\end{aligned}
\end{equation}
where~\cite{watts}
\begin{equation}
\begin{aligned}
\label{eq:polygamma}
F(x)&= \frac{4}{3 \zeta(3)} \sum_{\alpha=1}^\infty \frac{(-1)^{\alpha+1}  \alpha}{(\alpha^2+  x^2)^2}\\
&=\frac{i}{12\zeta(3) x} \left\{ \Psi_1\left(\frac{1+ix}{2} \right) -\Psi_1\left(\frac{1-ix}{2} \right)+\Psi_1\left(-\frac{ix}{2} \right) -\Psi_1\left(\frac{ix}{2} \right)
\right\}
\end{aligned}
\end{equation}
follows from integrating the ultra-relativistic Fermi--Dirac distribution, 
and $\Psi_1(y)$ is a poly\-gamma function of order $1$. Figure~\ref{fig:kernels} shows $K(k;s,s')$ for several representative values of the neutrino mass~$m$, as functions of~$k$, 
at $s=s(a_0=1)$ and $s'=s(a=1/10)$.

Schematically, the solution to equation~\eqref{eq:gilbert} can be written as
\begin{align}
\delta^{(1)}(\bm{k},s)=\int_{s_{\rm in}}^s ds' \: G(k;s,s') I(\bm{k},s'),
\label{eq:solutionGilbert}
\end{align} 
where $G(k;s,s')$ is understood to be a solution of
\begin{align}
G\left(k,s,s'\right)-\int_{s_{\rm in}}^s dz\: K\left(k;s, z\right) G\left(k,z,s'\right)&=\delta_D(s-s'), \label{eq:GreensFunctionGilbert}
\end{align}
which is comparable to the definition of a Green's function, and causality implies $G(k; s,s')=0$ for $s'>s$.%
\footnote{Even if no Green's function-like solution $G(s,s')$ exists, the numerical solution of~\eqref{eq:gilbert} can still formally be written in the form~\eqref{eq:solutionGilbert}. See appendix~\ref{sec:nystroem}.} 
Then, substituting equation~\eqref{eq:solutionGilbert} into equation~\eqref{eq:formalsolution} gives the solution to the linearised Boltzmann equation,
\begin{equation}
\begin{aligned}
\label{eq:linsolution}
\delta
f^{(1)}\left(\bm{k},\bm{q},s\right)=\: &\tilde{g} (\bm{k},\bm{q}; s,s_{\rm in})
\delta f^{(1)} \left(\bm{k},\bm{q},s_{\rm in}\right)+ \int_{s_{\rm in}}^s ds'\: V(s') \tilde{g} (\bm{k},\bm{q};s,s')  \\
 & \times  
 \bm{v} \cdot \frac{\partial \bar{f}}{\partial \bm{q}} \int_{s_{\rm in}}^{s'} ds''\: G(k;s',s'') \int d^3q' \: 
 \tilde{g} (\bm{k},\bm{q}'; s'',s_{\rm in})
 \delta f^{(1)}\left(\bm{k},\bm{q}',s_{\rm in}\right),
 \end{aligned}
 \end{equation}
in terms only of the initial perturbed distribution function~$\delta f^{(1)}(\bm{k},\bm{q},s_{\rm in})$.

\subsection{Perturbative expansion}\label{sec:perturbative}

It is straightforward to generalise the formal solution~(\ref{eq:formalsolution}) to the full, nonlinear case.  Restoring the nonlinear right-hand side in equation~\eqref{eq:boltzmannfourier}
and following the same steps that led to equation~(\ref{eq:formalsolution}) yield a similar formal solution save for an additional nonlinear term:
\begin{equation}
\begin{aligned}
\delta f\left(\bm{k},\bm{q},s\right)=\: &\tilde{g}\left(\bm{k},\bm{q};s,s_{\rm in}\right) \delta f \left(\bm{k},\bm{q},s_{\rm in}\right) +\int_{s_{\rm in}}^s ds'\: V(s')   \tilde{g}\left(\bm{k},\bm{q};s,s'\right) \delta(\bm{k},s')\bm{v}  \cdot \frac{\partial \bar{f}}{\partial \bm{q}} \\
&+\int_{s_{\rm in}}^s ds'\: V(s')  \tilde{g}\left(\bm{k},\bm{q};s,s'\right) \left[ \delta(\bm{k}_1,s')\bm{v}_1 \cdot \frac{\partial \delta f}{\partial \bm{q}}\left(\bm{k}_2,\bm{q},s'\right)\right]_{\bm{k}}.
\label{eq:formalsolutionnonlin}
\end{aligned}
\end{equation}
Integrating over momentum $\bm{q}$ likewise  adds a nonlinear term to Gilbert's equation~\eqref{eq:gilbert}:
\begin{equation}
\begin{aligned}
\label{eq:deltaformalsol}
\delta\left(\bm{k},s\right)=\: & I(\bm{k},s)+\int_{s_{\rm in}}^s ds'\: K\left(k;s,s'\right)  \delta\left(\bm{k},s'\right)\\
&+\int_{s_{\rm in}}^s ds'\: V(s') \int d^3q  \: \tilde{g}\left(\bm{k},\bm{q};s,s'\right) \left[\delta(\bm{k}_1,s')\bm{v}_1 \cdot \frac{\partial \delta f}{\partial \bm{q}}\left(\bm{k}_2,\bm{q},s'\right)\right]_{\bm{k}},
\end{aligned}
\end{equation}
which, after integration by parts and using that $\int_\mathcal{V} d^3 q \, \partial  (\tilde{g}  \delta f)/\partial \bm {q}  =\oint_{\mathcal{S}(\mathcal{V})}  d\bm{A}  \ \tilde{g}  \delta f$ vanishes
on the surface $\mathcal{S}(\mathcal{V}) \to \infty$, is equivalently 
\begin{equation}
\begin{aligned}
\label{eq:deltaformalsolIn}
\delta\left(\bm{k},s\right)=& \int_{s_{\rm in}}^s ds' \:G(k;s,s')I(\bm{k},s') 
-\int_{s_{\rm in}}^s ds' \:G(k;s,s') \int_{s_{\rm in}}^{s'} ds_1\: V(s_1)  \\
& \hspace{30mm} \times \Bigg[ \delta(\bm{k}_1,s_1) 
 \int d^3q  \: \bm{v_1} \cdot \frac{\partial \tilde{g}\left(\bm{k},\bm{q};s',s_1\right)}{\partial \bm{q}} \delta f\left(\bm{k}_2,\bm{q},s_1\right)\Bigg]_{\bm{k}},
\end{aligned}
\end{equation}
where
\begin{align}
 \frac{\partial \tilde{g}\left(\bm{k},\bm{q};s',s_1\right)}{\partial \bm{q}} = -i  \frac{\bm{k}}{m}(s'-s_1) \tilde{g}\left(\bm{k},\bm{q};s',s_1\right),
\end{align}
and  $G(k;s,s')$ is the Green's function of equation~(\ref{eq:solutionGilbert}).

Observe that the nonlinearity in equation~\eqref{eq:deltaformalsolIn} resides solely in the coupling of $\delta(\bm{k}_1)$ and  $\delta f (\bm {k}_2,\bm{q})$.  Thus, to solve for the density 
contrast to second order, $\delta^{(1+2)}(\bm{k}) \equiv \delta^{(1)}(\bm{k})+\delta^{(2)}(\bm{k})$,  we need simply to replace $\delta (\bm{k}_1)$ and $\delta f(\bm{k}_2,\bm{q})$ with their linear approximations~(\ref{eq:solutionGilbert}) and~(\ref{eq:linsolution}) in equation~\eqref{eq:deltaformalsolIn}. 
This second order solution $\delta^{(1+2)}(\bm{k})$ can then be incorporated into equation~\eqref{eq:formalsolutionnonlin} in order to establish the distribution function $\delta f^{(1+2)}(\bm{k},\bm{q})$ also to second order.   Putting $\delta f^{(1+2)}(\bm{k},\bm{q})$ back into equation~\eqref{eq:deltaformalsolIn} then gives the desired $\delta^{(1+2+3)}(\bm{k})$.
Formally, the iterative procedure just outlined can be expressed as
\begin{equation}
\begin{aligned}
 \label{eq:deltaformalsolInsecond}
\delta\left(\bm{k},s\right)=&\int_{s_{\rm in}}^s ds' \:G(k;s,s')I(\bm{k},s')\\
&+\int_{s_{\rm in}}^s ds' \:G(k;s,s')\int_{s_{\rm in}}^{s'} ds_1\left[
\tilde{\Gamma}^{(1)}\left(\bm{k}_1,\bm{k}_2;s',s_1\right)
\delta(\bm{k}_1,s_1) \right]_{\bm{k}}\\
&+\int_{s_{\rm in}}^s ds' \:G(k;s,s')\int_{s_{\rm in}}^{s'} ds_1\int_{s_{\rm in}}^{s'} ds_2\left[
\Gamma^{(2)}\left(\bm{k}_1,\bm{k}_2;s',s_1,s_2\right)
\delta(\bm{k}_1,s_1)
 \delta(\bm{k}_2,s_2)\right]_{\bm{k}}\\
&+\int_{s_{\rm in}}^s ds'
\:G(k;s,s') \int_{s_{\rm in}}^{s'} ds_1\int_{s_{\rm in}}^{s'} ds_2\: \Theta(s_1-s_2) V(s_1) V(s_2)  \Bigg[\delta(\bm{k}_1,s_1) \delta(\bm{k}_2,s_2)\\
& \hspace{2mm}\times  \int d^3q  \; \bm{v}_2 \cdot 
\frac{\partial}{\partial \bm{q}} \left( \bm{v}_1 \cdot \frac{\partial \tilde{g}\left(\bm{k}, \bm{q};s',s_1\right)}{\partial \bm{q}}  \tilde{g}\left(\bm{k}_2+\bm{k}_3,\bm{q}; s_1,s_2\right)  \right) 
 \delta f\left(\bm{k}_3,\bm{q},s_2\right)\Bigg]_{\bm{k}},
\end{aligned}
\end{equation}
where we have essentially inserted the formal solution~\eqref{eq:formalsolutionnonlin} for $\delta f (\bm{k},\bm{q},s)$ into~\eqref{eq:deltaformalsolIn},
reduced a set of nested convolution integrals in the last term according to
\begin{align}
\Bigg[ f(\bm{k}_1) g(\bm{K}) \Big[h(\bm{k}_2) j(\bm{k}_3) \Big]_{\bm{K}} 
 \Bigg]_{\bm{k}} = \Bigg[ f(\bm{k}_1) g(\bm{k}_2+\bm{k_3}) h(\bm{k}_2) j(\bm{k}_3)
 \Bigg]_{\bm{k}},
\end{align}
and integrated by parts using a vanishing $\int_\mathcal{V} d^3 q \, \partial 
(\tilde{g}  \tilde{g}  \delta f)/\partial \bm {q}  =\oint_{\mathcal{S}(\mathcal{V})}  d\bm{A}  \ \tilde{g}\tilde{g}\delta f$.

The kernels $\tilde{\Gamma}^{(1)}$ and $\Gamma^{(2)}$ in the second and third term are given respectively by
\begin{equation}
\begin{aligned}
\label{eq:gammatilde1}
\tilde{\Gamma}^{(1)} & \equiv - V(s_1) \int  d^3 q \: \bm{v}_1 \cdot \frac{\partial \tilde{g}\left(\bm{k},\bm{q};s',s_1\right)}{\partial \bm{q}}   \tilde{g}(\bm{k}_2,\bm{q}; s_1,s_{\rm in})
\delta f^{(1)}\left(\bm{k}_2,\bm{q},s_{\rm in} \right)\\
& =  \frac{3}{2}a^2(s_1) \mathcal{H}^2(s_1) \Omega(s_1) \frac{\bm{k}_1}{k_1^2} \cdot \bm{U}^2_1 \int
d^3q\,e^{-i \bm{U}^2_2\cdot\bm{q}/m}\delta
f^{(1)}(\bm{k}_2,\bm{q},s_{\rm in}),
\end{aligned}
\end{equation}
and
\begin{equation}
\begin{aligned}
\label{eq:gamma2}
\Gamma^{(2)}&=   \Theta(s_1-s_2)  V(s_1) V(s_2) \int d^3q \; \bm{v}_2 \cdot 
\frac{\partial}{\partial \bm{q}} \left( \bm{v}_1 \cdot \frac{\partial \tilde{g}\left(\bm{k}, \bm{q};s',s_1\right)}{\partial \bm{q}}  \tilde{g}\left(\bm{k}_2,\bm{q}; s_1,s_2\right)  \right) 
 \bar{f}(q)\\
&= \Theta(s_1-s_2)  \frac{9}{4}a^2(s_1) \mathcal{H}^2(s_1) \Omega(s_1)  a^2(s_2)\mathcal{H}^2(s_2)\Omega(s_2)\frac{\bm{k}_2}{k_2^2}\cdot \bm{U}^2_2 \frac{\bm{k}_1}{k_1^2} \cdot  \bm{U}^2_1\int
d^3q\,e^{-i \bm{U}^2_2 \cdot\bm{q}/m}
\bar{f}(q) \\
&=    \Theta(s_1-s_2) \frac{9}{4}a^2(s_1) \mathcal{H}^2(s_1)  \Omega(s_1) a^2(s_2)\mathcal{H}^2(s_2)\Omega(s_2) \frac{\bm{k}_2}{k_2^2}\cdot \bm{U}^2_2 \frac{\bm{k}_1}{k_1^2} \cdot  \bm{U}^2_1 \, 
F\left(\frac{T U^2_2}{m} \right),
\end{aligned}
\end{equation}
where $\bm{U}^2_1 \equiv (\bm{k}_1+\bm{k}_2)(s'-s_1)$, $\bm{U}^2_2 \equiv (\bm{k}_1+\bm{k}_2)(s'-s_1)+\bm{k}_2(s_1-s_2)$,  $s_2$ is identified with the initial time $s_{\rm in}$ in the case of  $\tilde{\Gamma}^{(1)}$, and~$F(x)$ has been defined in equation~(\ref{eq:polygamma}).
 The initial distribution function $\delta f(\bm{k},\bm{q}, s_{\rm in})=\delta f^{(1)}(\bm{k},\bm{q}, s_{\rm in})$ is always understood to be a linear quantity.
Thus, a perturbative expansion of equation~(\ref{eq:deltaformalsolInsecond}) up to third order can begin by replacing in the last term $\delta(\bm{k}_1)$, $\delta(\bm{k}_2)$ and 
$\delta f(\bm{k}_3,\bm{q})$ with their linear approximations~(\ref{eq:solutionGilbert}) and~(\ref{eq:linsolution}).

Naturally, we need not stop at third order.  Indeed, inserting equation~\eqref{eq:formalsolutionnonlin} again into the last term of equation~(\ref{eq:deltaformalsolInsecond}) 
yields three terms of the form $\delta f(s_{\rm in}) \delta \delta$, $\bar{f} \delta \delta \delta$, and~$\delta f \delta \delta \delta$.  Iterating one more time results in another set of three terms such as
$\delta f(s_{\rm in}) \delta \delta \delta$, $\bar{f} \delta \delta \delta \delta$, and~$\delta f \delta \delta \delta \delta$.  The iteration process can be repeated indefinitely to give
\begin{equation}
\begin{aligned}
\label{eq:gilbertclosedformal}
&\delta(\bm{k},s)= \int_{s_{\rm in}}^s \! ds' \:G(k;s,s') I(\bm{k},s') \\
&\hspace{5mm} +\sum_{n=2}^{\infty} \int_{s_{\rm in}}^s  ds' \:G(k;s,s') \int_{s_{\rm in}}^{s'} \: \prod_{j=1}^{n-1} ds_j \left[\tilde{\Gamma}^{(n-1)} (\bm{k}_1, \ldots, \bm{k}_{n};s',s_1, \ldots,s_{n-1})\prod_{j=1}^{n-1}\delta(\bm{k}_j,s_j)\right]_{\bm{k}}\\
&\hspace{5mm} +\sum_{n=2}^{\infty} \int_{s_{\rm in}}^s  ds'\:G(k;s,s') \int_{s_{\rm in}}^{s'}  \: \prod_{j=1}^{n} ds_j \left[\Gamma^{(n)} (\bm{k}_1, \ldots, \bm{k}_{n};s',s_1, \ldots,s_{n})\prod_{j=1}^{n}\delta(\bm{k}_j,s_j)\right]_{\bm{k}},
\end{aligned}
\end{equation}
where the higher-order kernels 
take the form
\begin{equation}
\begin{aligned}
\tilde{\Gamma}^{(n-1)}=&\left(\prod_{j=1}^{n-1} \Theta(s_{j-1}-s_j) \frac{3 }{2}
a^2(s_j)\mathcal{H}^2(s_j) \Omega(s_j) \frac{\bm{k}_j}{k_j^2}\cdot
\bm{U}^{n}_j\right) \int d^3q \:
e^{-i \bm{U}^{n}_{n}\cdot\bm{q}/m}\delta
f^{(1)}(\bm{k}_{n},\bm{q},s_{\rm in}),
\label{eq:Kernel11}
\end{aligned}
\end{equation}
and
\begin{equation}
\begin{aligned}
\Gamma^{(n)}=&\left(\prod_{j=1}^n  \Theta(s_{j-1}-s_j)  \frac{3 }{2}
a^2(s_j) \mathcal{H}^2(s_j)\Omega(s_j)\frac{\bm{k}_j}{k_j^2}\cdot
\bm{U}^n_j\right) \int d^3q \:
e^{-i\bm{U}^n_n\cdot\bm{q}/m}\bar{f}(q) \\
=&\left(\prod_{j=1}^n   \Theta(s_{j-1}-s_j)  \frac{3 }{2}
a^2(s_j) \mathcal{H}^2(s_j)\Omega(s_j)\frac{\bm{k}_j}{k_j^2}\cdot
\bm{U}^n_j\right) 
\, F\left( \frac{T U^n_n}{m}\right).
\label{eq:KernelInitial}
\end{aligned}
\end{equation}
Here, $\bm{U}^n_j  \equiv \sum_{i=1}^{j} \sum_{l=i}^n \bm{k}_l (s_{i-1} - s_i)$, where $s_0$ is identified with the upper integration limit~$s'$, 
and, in the case of $\tilde{\Gamma}^{(n-1)}$, $s_{n} \equiv s_{\rm in}$. 
Note that we have chosen this particular manner of indexing for $\tilde{\Gamma}^{(n-1)}$ so as to highlight the fact that $\tilde{\Gamma}^{(n-1)}$ is itself already at first order and therefore couples only $n-1$ fields in order to give an $n$th-order~$\delta(\bm{k},s)$.  Observe also that $ \tilde{\Gamma}^{(0)}(\bm{k};s) \equiv I(\bm{k},s)$ and $\Gamma^{(1)} (\bm{k};s,s')\equiv K(k;s,s')$ are but the source function and integral kernel from equations~(\ref{eq:source}) and~(\ref{eq:kkernel}) respectively.
Derivations of the higher-order kernels can be found in appendix~\ref{sec:APkernel}.

The advantage of using equation~\eqref{eq:gilbertclosedformal} for perturbation theory 
is that instead of following the full momentum dependence of the perturbed distribution function $\delta f(\bm{k},\bm{q},s)$, 
which corresponds to infinitely many degrees of freedom that need to be integrated, there is now only one degree of freedom $\delta(\bm{k},s)$, and all
functions are independent of $\bm{q}$.  Nonetheless,  the {\it effects} of the ${\bm q}$-dependence have not been lost: they are now 
stored in two classes of functions~$\tilde{\Gamma}$ and~$\Gamma$, which, as is manifest in equations~(\ref{eq:Kernel11}) and~(\ref{eq:KernelInitial}), 
can be predetermined once the cosmological model and the initial conditions have been chosen.
The price we pay, however, is the non-local time dependence of
equation~\eqref{eq:gilbertclosedformal} in the form of the extra time integrals
compared with the fluid approach.
We note here that the recent analysis of~\cite{Dupuy:2013jaa} also goes beyond the fluid description for massive neutrinos, 
and proposes to follow the neutrino perturbation evolution by decomposing
the distribution function into different flows. However, this
is but a prescription of how to discretise the momentum dependence, and therefore
suffers from the same disadvantages (i.e., $\bm{q}$-integration, etc.) discussed above. 

Lastly, because the initial distribution function $\delta f^{(1)} (\bm{k},\bm{q},s_{\rm in}) = \delta f^{(1)}(\bm{k},q,\mu,s_{\rm in})$, where $\mu \equiv \hat{ \bm{k}} \cdot \hat{\bm{q}}$, 
can be conventionally decomposed in terms of Legendre polynomials~$P_\ell(\mu)$~\cite{Ma:1995ey},
\begin{equation}
\begin{aligned}
\label{eq:decompose}
\delta f^{(1)}(\bm{k},q,\mu,s_{\rm in})&= \sum_{\ell=0}^\infty  (-i)^\ell (2\ell+1) f_\ell (\bm{k},q,s_{\rm in}) P_\ell(\mu), \\
f_{\ell}(\bm{k},q,s_{\rm in})&= \frac{i^\ell}{2} \ \int_{-1}^{1} d\mu \: P_\ell(\mu)\delta f^{(1)}(\bm{k},q,\mu,s_{\rm in}),
\end{aligned}
\end{equation}
it is convenient to expand the first class of kernels $\tilde{\Gamma}^{(n)}$ in the same manner.  This yields, for example,
\begin{equation}
\begin{aligned}
\tilde{\Gamma}^{(0)} 
\equiv \: &  I(\bm{k},s) = \sum_{\ell=0}^\infty  4 \pi (-1)^\ell (2 \ell+1)
\int_0^{\infty}\! dq\: q^2   j_\ell \left(\frac{k q}{m}(s-s_{\rm in}) \right)  f_\ell (\bm{k},q,s_{\rm in}), \\
\tilde{\Gamma}^{(1)}
=\: & \frac{3}{2} a^2(s_1) \mathcal{H}^2(s_1) \Omega(s_1)\frac{\bm{k}_1}{k_1^2} \cdot \bm{U}^2_1\\
& \times 
\sum_{\ell=0}^\infty  4 \pi (-1)^\ell (2 \ell+1)
P_\ell\left(\hat{\bm{k}}_2 \cdot \hat{\bm{U}}^2_2\right) \int_0^{\infty}\! dq\: q^2   j_\ell \left(\frac{U^2_2 q}{m} \right)  f_\ell (\bm{k}_2,q,s_{\rm in}),
\end{aligned}
\end{equation}
where $j_\ell(x)$ is the spherical Bessel function of order $\ell$.  The full derivation and generalisation to $n>1$ can be found in appendix~\ref{sec:ApMultipole}.


\section{Combining CDM and neutrinos}\label{sec:cdmneu}

Having brought the collisionless Boltzmann equation into a more convenient form for perturbative calculations, we are now in the position to generalise
our  theory to the case of mixed CDM and massive neutrinos.
As in the two-fluid treatment of section~\ref{sec:twofluid}, the generalisation consists in replacing $\delta$ with 
$f_{\rm C} \delta_{\mathrm{C}}+f_\nu \delta_{\nu}$ in all occurrences of the gravitational potential~$\Phi$ in the CDM equations of motion and in the collisionless Boltzmann equation~\eqref{eq:boltzmannfourier}. 
Thus, the equations to be solved for this combined system are
\begin{align}
\label{eq:all}
&\frac{\partial \varphi_{a}(\bm{k})}{\partial s}+\Pi_{ab}(k,s) \varphi_b(\bm{k})-\delta_{a2} \frac{3}{2}a \mathcal{H}^2 \Omega(s) f_\nu \delta_{\nu}(\bm{k})
=a\left[\gamma_{abc}(\bm{k}_1,\bm{k}_2) \varphi_b(\bm{k}_1)\varphi_c(\bm{k}_2)\right]_{\bm{k}}, \\
& \delta_{\nu}(\bm{k},s)=I_\nu({\bm k},s) + \int^s_{s_{\rm in}} d s' \: K_\nu (k;s,s') \left\{f_{\rm C} \delta_{\rm C}(\bm{k}_j,s_j)+f_\nu \delta_{\nu}(\bm{k}_j,s_j)\right\} +
S_{\rm \nu}[\bm{\varphi},\delta_{\nu};\bm{k},s],
\label{eq:all2}
\end{align}
where $\bm{\varphi} \equiv \left(\delta_{\rm C}, -\theta_{\rm C} \right)^T$,  $\Pi_{ab}(k,s)$ takes the form given in equation~(\ref{eq:pi}) but with the replacement
$\Pi_{21}= (3/2) a \mathcal{H}^2 \Omega(s) f_{\rm C}$, the linear neutrino source term~$I_\nu(\bm{k}, s)$ and integral kernel~$K_\nu (k; s,s')$ are identically  $I (\bm{k},s)$ and  $K(k;s,s')$
from equations~(\ref{eq:source}) and~(\ref{eq:kkernel}) respectively, and 
\begin{equation}
\begin{aligned}
&S_{\rm \nu}[\bm{\varphi},\delta_{\nu};\bm{k},s]= \\
 &\hspace{5mm}  \sum_{n=2}^{\infty} \: \int_{s_{\rm in}}^{s} \: \prod_{j=1}^{n-1} ds_j  \left[\tilde{\Gamma}^{(n-1)}(\bm{k}_1, \ldots, \bm{k}_{n};s,s_1, \ldots,s_{n-1})\prod_{j=1}^{n-1}\left\{f_{\rm C} \delta_{\rm C}(\bm{k}_j,s_j)+f_\nu \delta_{\nu}(\bm{k}_j,s_j)\right\}
\right]_{\bm{k}} \\
&\hspace{5mm} +\sum_{n=2}^{\infty} \: \int_{s_{\rm in}}^{s} \: \prod_{j=1}^{n} ds_j \left[\Gamma^{(n)} (\bm{k}_1, \ldots, \bm{k}_{n};s,s_1, \ldots,s_{n}) \prod_{j=1}^{n}\left\{f_{\rm C}\delta_{\rm C} (\bm{k}_j,s_j)+f_\nu\delta_{\nu}(\bm{k}_j,s_j)\right\}\right]_{\bm{k}}
\end{aligned}
\end{equation}
is the nonlinear neutrino source term.

As in section~\ref{sec:twofluid}, we can construct a perturbation theory for this two-species system by writing down four Green's functions: two that translate initial perturbations 
within species, and two that map the initial conditions from one species to the other.  The construction is most easily accomplished by first rewriting the CDM evolution equation~\eqref{eq:all} in the form of Gilbert's equation.  Defining the ``free-streaming'' Green's function of CDM as 
\begin{align}
\label{eq:gcfs}
\bm{\tilde g}_{\rm{C}}(s,s')=\begin{pmatrix} 1 &&  a(s')(s-s')\\ 0 &&
\frac{a(s')}{a(s)}\end{pmatrix} ,
\end{align}
so that $\varphi_a(s) = \tilde{g}_{{\rm C},ab} (s, s') \, \varphi_b(s')$  is but the solution of equation~(\ref{eq:all}) in the absence of {\it all} gravitational and non-linear source terms, the formal solution to~(\ref{eq:all}) in terms of $\bm{\tilde g}_{\rm C}$ automatically assumes the Gilbert form:
\begin{equation}
\begin{aligned}
\varphi_a(\bm{k},s)=I_{{\rm C},a}(\bm{k},s)+&\int_{s_{\rm in}}^{s} ds'\:K_{{\rm C},a}(k;s,s') \{f_{\rm C}\delta_{\rm C}(\bm{k},s')+f_{\rm \nu}\delta_{\rm \nu}(\bm{k},s')\} \\
&+\int_{s_{\rm{in}}}^s ds'\: \tilde{g}_{{\rm C},ab}(s,s')S_{{\rm C},b}[\bm{\varphi};\bm{k},s'] \, , \label{eq:gilbertCDM}
\end{aligned}
\end{equation}
where
\begin{equation}
\begin{aligned}
\label{eq:kkernelcc}
I_{{\rm
C},a}(\bm{k},s)&=\tilde{g}_{{\rm C},ab}(s,s_{\rm{in}})\varphi_b(\bm{k},s_{\rm{in}}) , \\
K_{{\rm C},a}(k;s,s')&= \frac{3}{2}a(s')
\mathcal{H}^2(s')\tilde{g}_{{\rm C},a2}(s,s')=\frac{3}{2}a(s')
\mathcal{H}^2(s')\begin{pmatrix}a(s')(s-s')\\ \frac{a(s')}{a(s)}\end{pmatrix}_a,\\
S_{{\rm C}, a}[\bm{\varphi};\bm{k},s]&=a(s)\left[\gamma_{abc}(\bm{k}_1,\bm{k}_2)
\varphi_b(\bm{k}_1)\varphi_c(\bm{k}_2)\right]_{\bm{k}}
\end{aligned}
\end{equation}
play the roles of linear source function, integral kernel, and nonlinear source  respectively.

At linear order  equations~(\ref{eq:all2}) and~(\ref{eq:gilbertCDM})  for the neutrino and CDM density contrasts form a closed set.  We can therefore write their respective formal solutions as
\begin{equation}
\begin{aligned}
\delta^{(1)}_{\rm \nu}(\bm{k}, s)&=\int_{s_{\rm in}}^s ds'\:G_{\rm{\nu \nu}}(k; s,s')I_{\rm \nu}(\bm{k},s')+\int_{s_{\rm in}}^s ds'\:G_{\rm{\nu C}}(k; s,s')I_{{\rm C},1}(\bm{k},s'), \\
\delta^{(1)}_{\rm C}(\bm{k}, s)&=\int_{s_{\rm in}}^s ds'\:G_{\rm{C \nu}}(k; s,s')I_{\rm \nu}(\bm{k},s')+\int_{s_{\rm in}}^s ds'\:G_{\rm{CC}}(k; s,s')I_{{\rm C},1}(\bm{k},s'). 
\label{eq:exactprop}
\end{aligned}
\end{equation}
The CDM velocity divergence~$\theta_{\rm C}({\bm k},s)$, on the other hand, can be constructed from $\delta_{\rm C}({\bm k},s)$ at all orders using the $a=1$ component of equation~(\ref{eq:all}), i.e.,
\begin{equation}
\theta_{\rm C}(\bm{k},s) = -\frac{1}{a(s)} \left\{ \frac{d}{d s}  \delta_{\rm C}(\bm{k},s) - S_{{\rm C},b}[\bm{\varphi};\bm{k},s] \right\}.
\end{equation}
Then, defining the Green's functions
\begin{equation}
\begin{aligned}
{\cal G}_{\nu {\rm C},a}(k;s,s')&\equiv \int_{s_{\rm in}}^s dz \: \Theta(z-s') G_{\rm{\nu
C}}(k;s,z) \tilde{g}_{{\rm C},1a}(z,s'), \\
{\cal G}_{{\rm CC},ab}(k;s,s')& \equiv \begin{pmatrix} 1\\ \frac{1}{a(s)}\frac{d}{ds}
\end{pmatrix}_a \int_{s_{\rm in}}^s dz \: \Theta(z-s') G_{\rm{CC}}(k;s,z) \tilde{g}_{{\rm C},1b}(z,s'), \\
\label{eq:greensfunctionsall}
{\cal G}_{{\rm C} \nu,a}(k;s,s')& \equiv \begin{pmatrix} 1\\\frac{1}{a(s)}\frac{d}{ds}
\end{pmatrix}_a G_{\rm{C \nu}}(k;s,s'), 
\end{aligned}
\end{equation}
the formal solution to the full nonlinear equations~(\ref{eq:all2}) and~(\ref{eq:gilbertCDM}) can now be written as
\begin{equation}
\begin{aligned}
\label{eq:allformal}
\varphi_{a}(\bm{k},s)=& \ {\cal G}_{{\rm CC},ab}(k;s,s_{\rm{in}})\varphi_b(\bm{k},s_{\rm in})+\int_{s_{\rm in}}^s ds'\: {\cal G}_{{\rm C} \nu,a}(k;s,s')I_{\rm \nu}(\bm{k},s')\\
&+\int_{s_{\rm in}}^s ds'\:
{\cal G}_{{\rm CC},ab}(k;s,s')S_{{\rm C},b}[\bm{\varphi};\bm{k},s']+\int_{s_{\rm in}}^s ds'\: {\cal G}_{{\rm C}\nu,a}(k;s,s')S_{\nu}[\bm{\varphi},\delta_{\rm \nu};\bm{k},s'], \\
\delta_\nu(\bm{k},s)=& \  {\cal G}_{\nu  {\rm C},a}(k;s,s_{\rm{in}})\varphi_a(\bm{k},s_{\rm in})+\int_{s_{\rm in}}^s ds'\: G_{\nu \nu}(k;s,s')I_{\nu}(\bm{k},s')\\
&+\int_{s_{\rm in}}^s ds'\: {\cal G}_{\nu {\rm C},a}(k;s,s')S_{{\rm C},a}[\bm{\varphi};\bm{k},s']+\int_{s_{\rm in}}^s ds'\: G_{\rm{\nu \nu}}(k;s,s')S_{\nu}[\bm{\varphi},\delta_{\rm \nu};\bm{k},s'].
\end{aligned}
\end{equation}
Along with equation~\eqref{eq:gilbertclosedformal} this is the main outcome of our first-principles description of CDM+neutrino perturbations, and should be 
be compared with equations~(\ref{eq:2fluid1}) and~(\ref{eq:2fluid2}) derived from the fluid approximation.

Observe that in the limit $f_\nu=0$, equation~(\ref{eq:allformal}) reduces to the standard solution~\eqref{eq:CDMintegral} (with $c_{\rm s}^2=0$), and the Green's functions are identically  ${\cal G}_{{\rm CC},ab} (k;s,s')= g_{ab}(s,s')$.
In the opposite limit $f_{\rm C}=0$, we recover from~(\ref{eq:allformal}) equation~\eqref{eq:gilbertclosedformal} with the identity $G_{\nu\nu}(k;s,s') = G(k;s,s')$.
A perturbative solution of equation~(\ref{eq:allformal}) up to second order will be presented in sections~\ref{sec:diagram} and~\ref{sec:applications}.

\section{Hybrid approach}
\label{sec:hybrid}

The formalism described in sections~\ref{sec:boltz} and~\ref{sec:cdmneu} takes into account all information from the neutrino momentum distribution.
However, its non-locality in time makes it cumbersome to calculate higher order contributions. We therefore propose an approximation scheme that is as simple as the two-fluid perturbation theory described in section~\ref{sec:twofluid}, but can still capture the strong suppression of power at small length scales  without introducing artificial acoustic oscillations.

Our approximation scheme consists of a simple modification to the two-fluid perturbation theory described in section~\ref{sec:twofluid}: while retaining the nonlinear structure of the fluid approximation, we replace the fluid Green's functions $\{g_{ab}, {g^A}_b, g^{AB},{g_a}^B\}$ defined in equation~(\ref{eq:greens}) with the ``exact'' Green's functions $\{G_{\rm CC}, G_{\nu {\rm C}}, G_{\nu \nu},G_{{\rm C} \nu} \}$ of the full theory defined in equation~(\ref{eq:exactprop}).
 To adapt these exact Green's functions to the nonlinear coupling  format of the fluid approximation, 
we first rewrite the neutrino source term~$I({\bm k}, s)$ in equation~(\ref{eq:source}) as
\begin{equation}
\begin{aligned}
I_{\nu}({\bm k},s)&=\int d^3q\: \delta f({\bm k},{\bm q},s_{\rm{in}})\(1-i\frac{{\bm q}\cdot {\bm k}}{m}(s-s_{\rm{in}})+\ldots\)\\
&=\delta_\nu({\bm k},s_{\rm{in}})-a(s_{\rm in}) (s-s_{\rm{in}})\theta_\nu({\bm k},s_{\rm{in}})+\ldots\\
&={\tilde{g}_{\rm C}}^{~\, 1A}(s, s_{\rm in}) \varphi^A({\bm k},s_{\rm{in}})+\ldots,
\end{aligned}
\end{equation}
where we have kept only the first two moments, namely, the density contrast and the velocity divergence, and the free-streaming Green's function~${\tilde{g}_{\rm C}}^{~\, AB}(s,s')$ is formally the same as that given in equation~(\ref{eq:gcfs}).   Following the notations of section~\ref{sec:twofluid}, we use superscript indices to refer to the neutrino component, 
and subscript indices the CDM component.

Then, the exact Green's functions can be reformatted in the same manner as we constructed ${\cal G}_{{\rm CC}, ab}(k;s,s')$ in section~\ref{sec:cdmneu}.  This yields
\begin{equation}
\begin{aligned}
\label{eq:reformat}
g_{ab}(k;s,s')& \equiv \begin{pmatrix} 1\\ \frac{1}{a(s)}\frac{d}{ds}
\end{pmatrix}_a \int_{s_{\rm in}}^s dz \: \Theta(z-s') G_{\rm{CC}}(k;s,z) \tilde{g}_{{\rm C},1b}(z,s'),\\
{g_{a}}^B(k;s,s')& \equiv \begin{pmatrix} 1\\ \frac{1}{a(s)}\frac{d}{ds}
\end{pmatrix}_a \int_{s_{\rm in}}^s dz \: \Theta(z-s') G_{\rm{C\nu}}(k;s,z) \tilde{g}_{{\rm C}}^{~\, 1B}(z,s'),
\\
g^{AB}(k;s,s')& \equiv \begin{pmatrix} 1\\ \frac{1}{a(s)}\frac{d}{ds}
\end{pmatrix}^A \int_{s_{\rm in}}^s dz \: \Theta(z-s') G_{\rm{\nu\nu}}(k;s,z) \tilde{g}_{{\rm C}}^{~\, 1B}(z,s'),\\
{g^{A}}_b(k;s,s')& \equiv \begin{pmatrix} 1\\ \frac{1}{a(s)}\frac{d}{ds}
\end{pmatrix}^{A} \int_{s_{\rm in}}^s dz \: \Theta(z-s') G_{\rm{\nu C}}(k;s,z) \tilde{g}_{{\rm C},1b}(z,s'),
\end{aligned}
\end{equation}
to be used in the formal two-fluid solutions~(\ref{eq:2fluid1}) and ~(\ref{eq:2fluid2}).
We emphasise again that $\{G_{\rm CC}, G_{\nu {\rm C}}, G_{\nu \nu},G_{{\rm C} \nu} \}$ are still to be computed from the the full theory as per equation~(\ref{eq:exactprop});
equation~(\ref{eq:reformat}) merely turns them into a form compatible with the nonlinear structure of equations~(\ref{eq:2fluid1}) and~(\ref{eq:2fluid2}).

At this point we still have the freedom to choose whether or not to retain a non-vanishing sound speed in the nonlinear coupling of the neutrino component.  We test both models in this work, assuming in the non-vanishing case an effective sound speed given by the velocity dispersion of the unperturbed momentum distribution; see equation~(\ref{eq:q2bar}).  As we shall see 
later in section~\ref{sec:bispectrum}, a vanishing sound speed actually turns out to be a better approximation numerically, as far as the tree-level bispectrum is concerned.

Finally, we remark that one needs to be careful when using an {\it ad-hoc} approximation such as this hybrid approach, because unphysical artefacts may spoil the outcome of a calculation.
To rigorously demonstrate that the hybrid approach is a consistent approximation is beyond the scope of this paper.  We shall but briefly comment on two potentially dangerous points, and defer a detailed treatment to a future work. 

Firstly, in standard perturbation theory the extended Galilean symmetry of the fluid equations ensures  that the leading contribution from long wavelength modes cancels among different diagrams at any given order~\cite{Scoccimarro1996,Jain1996}. Importantly, however, neither the linear nor the nonlinear terms in the fluid equations are {\it individually} Galilean invariant; it is the combination that is.
Consequently, modifying the fluid equations only at  linear order---such as we are doing here in the hybrid approach---could potentially violate Galilean invariance  and lead to a non-cancellation of  long-wavelength divergences.

Secondly, the vertex functions in the full theory on small scales are suppressed in comparison with the fluid vertices; compare, for example, $\Gamma^{(1)} = K_\nu(k;s,s')$ for neutrinos in equation~(\ref{eq:kkernel}), and $K_{\rm C}(k;s,s')$ for CDM in equation~(\ref{eq:kkernelcc}).
Since the hybrid approach uses the fluid vertices, it overestimates the importance of small-scale neutrino perturbations, which could potentially induce corrections on the large scales that do not scale as  $\langle\delta^2\rangle\sim k^4$.  However, we argue that because the linear propagators are the correct ones, neutrino perturbations on small scales are still suppressed relative to both CDM perturbations on the same scales and neutrino perturbations on large scales.  Therefore, even without explicitly proving momentum conservation~\cite{Zeldovich1965}, we expect 
the small-scale induced large-scale corrections to respect $\langle\delta^2\rangle\sim k^4$.


\section{Diagrammatic representation and the $N$-point functions}
\label{sec:diagram}

Standard perturbation theory for CDM evolution can be organised in terms of (Feynman) diagrams~\cite{Crocce:2005xy}, an accounting scheme
that is advantageous mainly in the context of renormalisation and resummation approaches to nonlinear structure formation~\cite{Crocce:2005xy,Crocce:2005xz,Matarrese:2007wc}.   The diagrammatic approach is useful too to keep track of the large number of terms encountered in our mixed CDM+massive neutrino scenario, since many of the diagrams have similar topologies.

We review briefly in section~\ref{sec:cdmdiagram} the diagrammatic representation for standard CDM perturbation theory, and extend it to include a nonzero sound speed
in section~\ref{sec:fluiddiagram}.  Diagrams for neutrino perturbations and for mixed CDM+neutrino perturbations will be presented in sections~\ref{sec:neutrinodiagram} and~\ref{sec:combineddiagram} respectively.


\subsection{Standard CDM perturbations}
\label{sec:cdmdiagram}

Following~\cite{Crocce:2005xy}, the building blocks for a diagrammatic representation of standard CDM perturbation theory are
 \begin{eqnarray}
\label{eq:feynmanrulesCDM}
\parbox{24mm}{\begin{fmffile}{diagrams/prop}
\begin{fmfgraph*}(40,20)
\fmfkeep{cdmprop}
\fmfleft{in}
	 \fmfright{out}           
	  \fmf{plain_arrow,label=$\bm{k}$}{in,out}          
	  \fmflabel{$s_2,b$}{in}
	  \fmflabel{$s_1,a$}{out}          
\end{fmfgraph*}
\end{fmffile}} 
& =& g_{ab}\left(k;s_1,s_2\right),\\
\nonumber \\
\nonumber \\
\parbox{18mm}{\begin{fmffile}{diagrams/vert}
\begin{fmfchar*}(20,30)
  \fmfleft{in1,in2}
\fmfright{out}
\fmf{plain_arrow}{in1,v}
\fmf{plain_arrow}{in2,v}
\fmf{plain_arrow}{v,out}
	  \fmflabel{$s'$}{v}
          \fmflabel{$\bm{k}_2,c$}{in1}
	  \fmflabel{$\bm{k}_1,b$}{in2}
	  \fmflabel{$\bm{k},a$}{out}
\end{fmfchar*}
\end{fmffile}} 
&=& a(s')\gamma_{abc}\left(\bm{k}_1,\bm{k}_2\right) (2\pi)^3 \delta \left(\bm{k}-\left(\bm{k}_1+\bm{k}_2\right)\right),
\label{eq:feynmanrulesCDM2} \\
\nonumber \\
\nonumber \\
\parbox{14mm}{\begin{fmffile}{diagrams/ini}
\begin{fmfchar*}(8,10)
\fmfleft{in}
	 \fmfright{out}        
	  \fmf{phantom}{in,out} 
	  \fmflabel{$\bm{k},a$}{out} 
	  \fmfv{decor.shape=circle,decor.filled=empty, decor.size=2thick}{in}
\end{fmfchar*}\end{fmffile}} &=& {\varphi}_a\left(\bm{k}, s_{\rm in}\right),\label{eq:feynmanrulesCDM3}
\end{eqnarray}
which stand for the linear propagator from time $s_2$ to $s_1$, the vertex representing an interaction at time~$s'$, and the initial doublet, respectively.  Constructing a diagram therefore consists in simply pasting these blocks together, and applying at each vertex 
\begin{align}
\int ds' \int \frac{d^3k_i}{(2 \pi)^3}  \frac{d^3 k_j}{(2 \pi)^3},
\end{align}
so as to integrate over all possible pairs of incoming wavevectors $\bm{k}_i$ and $\bm{k}_j$, as well as over all allowed interaction times~$s'$.  
For example, up to second order, the diagrams are 
 \begin{equation}
\begin{aligned}
\parbox{22mm}{\begin{fmffile}{diagrams/1order}
\begin{fmfchar*}(50,20)
\fmfleft{in}
	 \fmfright{out}           
	  \fmf{plain_arrow,label=$\bm{k}$}{in,out} 
	   \fmfv{decor.shape=circle,decor.filled=empty, decor.size=2thick}{in}   
	  \fmflabel{$a$}{out}
	  \fmflabel{$b$}{in}              
\end{fmfchar*}\end{fmffile}}&=g_{ab}\left(k; s,s_{\rm in}\right) \varphi_{b}(\bm{k},s_{\rm in}) = \varphi^{(1)}_a(\bm{k},s),\\
\end{aligned}
\end{equation}
and
\begin{equation}
\begin{aligned}
\\
\hspace{6mm}
\parbox{18mm}{\begin{fmffile}{diagrams/2order}
\begin{fmfchar*}(50,50)
\fmfleft{in}
	   \fmfleft{in1,in2}
\fmfright{out}
\fmf{plain_arrow,label=$\bm{k}_2$}{in1,v}
\fmf{plain_arrow,label=$\bm{k}_1$}{in2,v}
\fmf{plain_arrow,label=$\bm{k}$}{v,out}
	 \fmfv{decor.shape=circle,decor.filled=empty, decor.size=2thick}{in1}
	  \fmfv{decor.shape=circle,decor.filled=empty, decor.size=2thick}{in2}
       \fmflabel{$a$}{out}
       \fmflabel{$f$}{in1}
       \fmflabel{$e$}{in2}
       \fmflabel{$\gamma_{bcd}~s'$}{v}
\end{fmfchar*}\end{fmffile}} \hspace{4mm}
&=\int_{s_{\rm in}}^{s} ds' \: g_{ab}\left(k; s,s'\right)a(s') \\ 
& \hspace{5mm} \times \Big[\gamma_{bcd}\left(\bm{k}_1,\bm{k}_2\right) g_{ce}\left(k_1;s',s_{\rm in}\right)g_{df}\left(k_2; s',s_{\rm in}\right) 
\varphi_{e}\left(\bm{k}_1,s_{\rm in}\right)\varphi_{f}\left(\bm{k}_2,s_{\rm in}\right)\Big]_{\bm{k}} \\
&= \int_{s_{\rm in}}^{s} ds' \: g_{ab}\left(k; s,s'\right)a(s')  \Big[\gamma_{bcd}\left(\bm{k}_1,\bm{k}_2\right) 
\varphi_{c}^{(1)}\left(\bm{k}_1,s'\right)\varphi_{d}^{(1)}\left(\bm{k}_2,s'\right)\Big]_{\bm{k}} \\
&= \varphi_a^{(2)}(\bm{k},s).
\end{aligned}
\end{equation}
Note that in writing the above expressions we have allowed for the possibility of a $k$-dependent linear propagator; in standard CDM perturbation theory, the linear propagator is in fact independent of~$k$.

To quantify the $N$-point functions of the perturbations, we define 
 the power spectrum and the bispectrum  as the connected part of the 2-point and 3-point function respectively:
\begin{equation}
\begin{aligned}
\label{eq:spectra}
\left \langle \varphi_a(\bm{k},s) \varphi_b(\bm{k}',s) \right \rangle_{\rm C} & \equiv (2 \pi)^3 \delta(\bm{k}+ \bm{k}') P_{ab}(k;s), \\
\left \langle \varphi_a(\bm{k}_1,s) \varphi_b(\bm{k}_2,s)   \varphi_c(\bm{k}_3,s)   \right \rangle_{\rm C} & \equiv (2 \pi)^3 \delta(\bm{k}_1+ \bm{k}_2+\bm{k}_3) B_{abc}(k_1,k_2,k_3;s). 
\end{aligned}
\end{equation}
Here, $\langle \cdots \rangle$ denotes an ensemble average, the subscript ``C'' indicates the connected piece, and we have assumed as usual  statistical homogeneity and isotropy.
Then, to construct diagrams for the $N$-point functions simply involves ``glueing'' two or more $\varphi_a^{(n)}$ diagrams together at each open circle. 
If the initial conditions are Gaussian, as is our assumption here,  then the linear initial power spectrum $P_{ab}^{(1)}(k,s_{\rm in})$ alone characterises the statistics, and only two open circles can be amalgamated at any one point.  We denote this amalgamation with a shaded circle, i.e.,
\begin{align}
\parbox{24mm}{\begin{fmffile}{diagrams/initialpowerspectrum}
\begin{fmfchar*}(40,30)
	   \fmfleft{in1}
\fmfright{out1}
\fmf{plain}{v,in1}
\fmf{plain}{v,out1}
\fmfblob{4mm}{v} 
		   \fmflabel{$\bm{k},a$}{in1}
			\fmflabel{$\bm{k}',b$}{out1}            
\end{fmfchar*}\end{fmffile}}
=\left \langle \varphi_a(\bm{k},s_{\rm in}) \varphi_b(\bm{k}',s_{\rm in}) \right \rangle_{\rm C},
\end{align}
which, following from definition~(\ref{eq:spectra}), represents one count of the initial power spectrum $P_{ab}^{(1)}(k,s_{\rm in})$, and ``momentum'' conservation $\bm{k}+\bm{k}'=0$ is implied.

Thus, the recipe for constructing the $n$th order contribution to the connected $N$-point function from Gaussian initial conditions proceeds as follows:
(i) write down $n$ initial power spectra, (ii) use the vertex~(\ref{eq:feynmanrulesCDM2}) to connect any combination of two lines,
(iii) repeat (ii) until $N$ external lines are left, and (iv) keep only connected diagrams that do not contain tadpoles.%
\footnote{A tadpole is a diagram with one external line. The 1-point function is given by these diagrams. In the Newtonian limit tadpole diagrams vanish at all orders in perturbation theory.}
As an illustration, the leading-order diagram of the power spectrum is
\begin{equation}
\begin{aligned}
(2 \pi)^3 P_{ab}^{(1)}(k;s)
& =
\hspace{10mm} \parbox{40mm}{\begin{fmffile}{diagrams/powerspectrum}
\begin{fmfchar*}(80,30)
	   \fmfleft{in1}
\fmfright{out1}
\fmf{plain_arrow,label=$\bm{k}$}{v,in1}
\fmf{plain_arrow,label=$-\bm{k}$}{v,out1}
\fmfblob{4mm}{v} 
		   \fmflabel{$a,s$}{in1}
			\fmflabel{$b,s$}{out1}        
			\fmfv{label=$c~~~~d$,label.angle=90}{v}       
\end{fmfchar*}\end{fmffile}}
\\ 
&=(2 \pi)^3  g_{ac}\left(k; s,s_{\rm in}\right) g_{bd}\left(k; s,s_{\rm in}\right) P_{cd}^{(1)}(k;s_{\rm in}). \\
\end{aligned}
\end{equation}
For the bispectrum we find, to leading order,
\begin{equation}
\begin{aligned}
\\
& (2 \pi)^3 B_{abc}^{(2)}(k,k',|\bm{k}+\bm{k}'|;s)
=2 \times  \left( \hspace{8mm}
\parbox{40mm}{\begin{fmffile}{diagrams/CDMbi}
\begin{fmfchar*}(100,60)
	   \fmfleft{in1}
\fmfright{out1,out2}
\fmftop{v1}
\fmfbottom{v2}
\fmf{plain_arrow,label=$\bm{k}$}{v,in1}
\fmf{plain_arrow,label=$-\bm{k}'$}{v1,v}
\fmf{plain_arrow,label=$\bm{k}+\bm{k}'$}{v2,v}
\fmf{plain_arrow,label=$-\bm{k}-\bm{k}'$}{v2,out1}
\fmf{plain_arrow,label=$\bm{k}'$}{v1,out2}
\fmfblob{4mm}{v1}
\fmfblob{4mm}{v2}
\fmflabel{$a,s$}{in1}
\fmflabel{$c,s$}{out1}
\fmflabel{$b,s$}{out2} 
\fmfv{label=$s'~\gamma_{def}$,label.angle=0}{v}             
\end{fmfchar*}\end{fmffile}} 
\right)+\mathrm{cyclic~perm.}
\\ 
\\
&\hspace{5mm}=  2 \times  (2 \pi)^3 \int_{s_{\rm in}}^{s} ds' \: g_{ad}\left(k; s,s'\right)a(s') 
\gamma_{def}\left(-\bm{k}',\bm{k}+\bm{k}'\right) P_{eb}^{(1)}(k'; s',s) P_{fc}^{(1)} (|\bm{k}+\bm{k}'|;s',s) \\
& \hspace{10mm} + \text{cyclic~permutations},
\end{aligned}
\end{equation} 
where 
\begin{equation}
\begin{aligned}
\langle \varphi_a^{(1)} (\bm{k},s_1) \varphi_b^{(1)}(\bm{k}',s_2) \rangle_{\rm C} &\equiv (2 \pi)^3 \delta(\bm{k}+\bm{k}') P_{ab}^{(1)}(k; s_1,s_2) \\
&= (2 \pi)^3 \delta(\bm{k}+\bm{k}') g_{ac}\left(k; s_1,s_{\rm in}\right) g_{bd}\left(k; s_2,s_{\rm in} \right) P_{cd}^{(1)}\left(k;s_{\rm in}\right)
\end{aligned}
\end{equation}
is an unequal-time power spectrum that can be easily extracted from a linear Boltzmann code.  The symmetrisation factor of ``2''    traces its origin to the decomposition of the initial 4-point function,
\begin{equation}
\begin{aligned}
\langle \varphi_a(\bm{k}_1,s_{\rm in})  \varphi_b(\bm{k}_2,s_{\rm in})   \varphi_c(\bm{k}_3,s_{\rm in})  & \varphi_d(\bm{k}_4,s_{\rm in}) \rangle = \\
(2 \pi)^6 & \big[\delta(\bm{k}_1+\bm{k}_2)  \delta(\bm{k}_3+\bm{k}_4) P_{ab}(k_1;s_{\rm in}) P_{cd}(k_3;s_{\rm in}) \\
&+ \delta(\bm{k}_1+\bm{k}_3)  \delta(\bm{k}_2+\bm{k}_4) P_{ac}(k_1;s_{\rm in}) P_{bd}(k_2;s_{\rm in}) \\
&+\delta(\bm{k}_1+\bm{k}_4)  \delta(\bm{k}_2+\bm{k}_3) P_{ad}(k_1;s_{\rm in}) P_{bc}(k_2;s_{\rm in})\big],
\end{aligned}
\end{equation}
while 
 ``cyclic~permutations'' denote another two terms arising from rotating the  wavevector labels of the external lines.


\subsection{Fluid perturbations}
\label{sec:fluiddiagram}

Extending the standard CDM diagrammatic scheme to include an effective sound speed simply requires that we 
(i) modify the existing 2-vertex in equation~(\ref{eq:feynmanrulesCDM2}) to include an additional term proportional to the sound speed,
\begin{equation}
\begin{aligned}
\label{eq:modified}
\\
\hspace{8mm} \parbox{15mm}{\begin{fmffile}{diagrams/vertcs2}
\begin{fmfchar*}(20,30)
  \fmfleft{in1,in2}
\fmfright{out}
\fmf{plain_arrow}{in1,v}
\fmf{plain_arrow}{in2,v}
\fmf{plain_arrow}{v,out}
	  \fmflabel{$s'$}{v}
          \fmflabel{$\bm{k}_2,c$}{in1}
	  \fmflabel{$\bm{k}_1,b$}{in2}
	  \fmflabel{$\bm{k},a$}{out}
\end{fmfchar*}
\end{fmffile}} 
&=a(s') \left\{\gamma_{abc}\left(\bm{k}_1,\bm{k}_2\right) +\frac{1}{2} c_{\rm s}^2(s') k^2 \delta_{a2}\delta_{b1}\delta_{c1} \right\}(2\pi)^3\delta \left(\bm{k}-\left(\bm{k}_1+\bm{k}_2\right)\right),\\
\\
\end{aligned}
\end{equation}
and (ii) a new vertex that takes $n>2$ incoming lines,
\begin{equation}
\begin{aligned}
\label{eq:newvertex}
\\
\hspace{20mm}
\parbox{25mm}{\begin{fmffile}{diagrams/vertncs}
\begin{fmfchar*}(45,45)
	   \fmfsurround{in1,in2,in3,in4,out1,out2,out3,out5}
\fmf{plain_arrow}{v,in1}
\fmf{plain_arrow}{in2,v}
\fmf{plain_arrow}{in3,v}
\fmf{plain_arrow}{in4,v}
\fmf{plain_arrow}{out1,v}
\fmf{plain_arrow}{out2,v}
\fmf{phantom}{v,out3}
\fmf{plain_arrow}{out5,v}
\fmfv{decoration.size=5mm,decoration.filled=full,label=\mbox{\Large $\stackrel{s'}{\ldots}$},label.angle=-90,label.dist=3mm}{v}
	\fmflabel{$\bm{k},a$}{in1}
	\fmflabel{$\bm{k}_1,b_1$}{in2}
	\fmflabel{$\bm{k}_2,b_2$}{in3}
	\fmflabel{$\bm{k}_3,b_3$}{in4}
	\fmflabel{$\bm{k}_4,b_4$}{out1}
	\fmflabel{$\bm{k}_5,b_5$}{out2}
	\fmflabel{$\bm{k}_n,b_n$}{out5}
\end{fmfchar*}\end{fmffile}}
&=a(s')c_{\rm s}^2(s') \frac{k^2}{n!}\delta_{a2}\delta_{b_1 1}\ldots\delta_{b_n 1}(2\pi)^3\delta \left(\bm{k}-\sum_{j=1}^n\bm{k}_j\right),
\\ 
\\
\end{aligned}
\end{equation}
where the vertices have been symmetrised with respect to interchange of the incoming lines. 
The same rules for building connected $N$-point functions in standard CDM perturbation theory apply also to the fluid case, except that the new vertex~(\ref{eq:newvertex})
now enables the merger of more than two incoming lines.  For example, at one loop the propagator receives a new correction
\begin{equation}
\begin{aligned}
\\
\parbox{35mm}{\begin{fmffile}{diagrams/example4vertex}
\begin{fmfchar*}(60,50)
	   \fmfleft{in1}
\fmfright{out1}
\fmftop{v2}
\fmf{plain_arrow,label=$\bm{k}$}{in1,v}
\fmf{plain_arrow,label=$\bm{k}$}{v,out1}
\fmffreeze
\fmf{plain_arrow,tension=0.5,right,label=$\bm{q}$}{v2,v}
\fmf{plain_arrow,tension=0.5,left,label=$-\bm{q}$}{v2,v}
\fmfblob{4mm}{v2}
\fmflabel{$a,s_1$}{in1}
\fmflabel{$c,s_2$}{out1}
\fmfv{label=$s'$,label.angle=-90}{v}\end{fmfchar*}\end{fmffile}}
\end{aligned}
\end{equation} 
in addition to the usual one-loop correction constructed from two 2-vertices~(\ref{eq:modified}).


\subsection{Neutrino perturbations}
\label{sec:neutrinodiagram}

The neutrino perturbation theory formulated in section~\ref{sec:boltz} can likewise be broken down into diagrammatic building blocks of linear propagators, vertices, and initial fields.
The linear propagator of the theory $G(k; s_1,s_2)$, as defined in equation~(\ref{eq:solutionGilbert}), is represented by
\begin{equation}
\begin{aligned}
\label{eq:neutrinopropagator}
\parbox{20mm}{\begin{fmffile}{diagrams/propneut}
\begin{fmfchar*}(40,20)
\fmfkeep{nuprop}
\fmfleft{in}
	 \fmfright{out}           
	  \fmf{dashes_arrow,label=$\bm{k}$}{in,out}          
	  \fmflabel{$s_2$}{in}
	  \fmflabel{$s_1$}{out}        
	\end{fmfchar*}\end{fmffile}} =G\left(k;s_1,s_2\right),
\end{aligned}
\end{equation}
with the understanding any kind of source term ``$\times$'' eventually attached the propagator (e.g., an initial field, or a vertex) will automatically incur a time integration
from the initial time $s_{\rm in}$ to $s_1$, i.e.,
\begin{equation}
\begin{aligned}
\label{eq:nupropsource}
\parbox{20mm}{\begin{fmffile}{diagrams/propneu2t}
\begin{fmfchar*}(40,20)
\fmfleft{in}
	 \fmfright{out}           
	  \fmf{dashes_arrow,label=$\bm{k}$}{in,out}          
	  \fmflabel{$s_2$}{in}
	  \fmflabel{$s_1$}{out}        
	    \fmfv{decor.shape=cross,decoration.size=4mm}{in}  
\end{fmfchar*}\end{fmffile}} = \int_{s_{\rm in}}^{s_1} ds_2 \: G\left(k;s_1,s_2\right) \;\;
\parbox{7mm}{\begin{fmffile}{diagrams/propneutx}
\begin{fmfchar*}(10,30)
\fmfleft{in}
	 \fmfright{out}           
	  \fmf{phantom}{in,out}               
	    \fmfv{decor.shape=cross,decoration.size=4mm}{in}  
\end{fmfchar*}\end{fmffile}} .
\end{aligned}
\end{equation}
For example, attaching to the propagator the source term $I({\bm k,s'})$ from equation~(\ref{eq:source}),
denoted diagrammatically by an open circle ``$\circ$'' because it contains the initial distribution function~$\delta f(\bm{k},\bm{q},s_{\rm in})$, gives
\begin{equation}
\begin{aligned}
\parbox{18mm}{\begin{fmffile}{diagrams/linneut2}
\begin{fmfchar*}(50,20)
\fmfleft{in}
	 \fmfright{out}           
	  \fmf{dashes_arrow,label=$\bm{k}$}{in,out} 
	 \fmfv{decor.shape=circle,decor.filled=empty, decor.size=2thick}{in}
\end{fmfchar*}\end{fmffile}} &=\int_{s_{\rm in}}^s ds'\:G\left(k; s,s'\right)I(\bm{k},s') = \delta^{(1)} (\bm{k},s),\\
\end{aligned}
\end{equation}
which is also the diagram for the linear order neutrino perturbation~$\delta^{(1)}(\bm{k},s)$.
This time integration marks the first different between neutrino perturbation theory and standard CDM/fluid perturbation theory, the latter of which has no such procedure associated with its propagator $g_{ab}(k;s_1,s_2)$.

Two classes of vertices, given in equations~(\ref{eq:Kernel11}) and~(\ref{eq:KernelInitial}), encode the nonlinear coupling.  These correspond pictorially to
\begin{equation}
\begin{aligned}
 \\
 \label{eq:gamma1}
\hspace{11mm}\parbox{30mm}{\begin{fmffile}{diagrams/gammatilde}
\begin{fmfchar*}(50,50)
	   \fmfsurround{in1,in2,in3,in4,out1,out2,out3,out5}
\fmf{dashes_arrow}{v,in1}
\fmf{dots}{in2,v}
\fmf{dashes_arrow}{in3,v}
\fmf{dashes_arrow}{in4,v}
\fmf{dashes_arrow}{out1,v}
\fmf{dashes_arrow}{out2,v}
\fmf{phantom}{v,out3}
\fmf{dashes_arrow}{out5,v}
  \fmfv{decor.shape=circle,decor.filled=empty, decor.size=2thick}{in2}
\fmfv{decoration.size=5mm,decoration.filled=full,label=\mbox{\Large $\ldots$},label.angle=-90,label.dist=10mm}{v}
	\fmflabel{$\bm{k},s'$}{in1}
	\fmflabel{$\bm{k}_n$}{in2}
	\fmflabel{$\bm{k}_1,s_1$}{in3}
	\fmflabel{$\bm{k}_2,s_2$}{in4}
	\fmflabel{$\bm{k}_3,s_3$}{out1}
	\fmflabel{$\bm{k}_4,s_4$}{out2}
	\fmflabel{$\bm{k}_{n-1},s_{n-1}$}{out5}
\end{fmfchar*}\end{fmffile}}=\tilde{\Gamma}^{(n-1)}_{\rm s}(\bm{k}_1, \ldots, \bm{k}_{n};s',s_1, \ldots,s_{n-1})\  \delta \left(\bm{k}-\sum_{j=1}^n\bm{k}_j\right), \\  
\end{aligned}
\end{equation}
and
\begin{equation}
\begin{aligned}
\\
\label{eq:gamma2diag}
\hspace{10mm}
\parbox{30mm}{\begin{fmffile}{diagrams/gamma}
\begin{fmfchar*}(50,50)
	   \fmfsurround{in1,in2,in3,in4,out1,out2,out3,out5}
\fmf{dashes_arrow}{v,in1}
\fmf{dashes_arrow}{in2,v}
\fmf{dashes_arrow}{in3,v}
\fmf{dashes_arrow}{in4,v}
\fmf{dashes_arrow}{out1,v}
\fmf{dashes_arrow}{out2,v}
\fmf{dashes_arrow}{out5,v}
\fmfv{decoration.size=5mm,decoration.filled=full,label=\mbox{\Large $\ldots$},label.angle=-90,label.dist=10mm}{v}
	\fmflabel{$\bm{k},s'$}{in1}
	\fmflabel{$\bm{k}_1,s_1$}{in2}
	\fmflabel{$\bm{k}_2,s_2$}{in3}
	\fmflabel{$\bm{k}_3,s_3$}{in4}
	\fmflabel{$\bm{k}_4,s_4$}{out1}
	\fmflabel{$\bm{k}_5,s_5$}{out2}
	\fmflabel{$\bm{k}_n,s_n$}{out5}
\end{fmfchar*}\end{fmffile}}=  \Gamma^{(n)}_{\rm s}(\bm{k}_1, \ldots, \bm{k}_{n};s',s_1, \ldots,s_{n}) \ \delta \left(\bm{k}-\sum_{j=1}^n\bm{k}_j\right), \\  \\
\end{aligned}
\end{equation}
where the subscript ``s'' indicates that the vertices are symmetrised versions of the expressions~(\ref{eq:Kernel11}) and~(\ref{eq:KernelInitial}) over  all permutations of their respective $n-1$ and  $n$ incoming wavevectors.
Importantly, both  classes of vertices are non-local in time, meaning that for every incoming line $\bm{k}_j$
one must integrate over $s_j$ from the initial time $s_{\rm in}$ to $s'$.
This is in contrast to the fluid description, in which the $n$-vertex~(\ref{eq:newvertex}) couples all $n$ incoming lines at the same time, so that only one integration over $s'$ from $s_{\rm in}$ to some final time $s$ is required.  Note also that the $\tilde{\Gamma}^{(n-1)}_{\rm s}$ diagram has an additional dotted line, as a reminder that the kernel itself is at first order, the open circle indicating that it is sourced by the initial distribution function~$\delta f({\bm k}_n,\bm{q}, s_{\rm in})$.

Then, combining these vertices with the propagator~(\ref{eq:neutrinopropagator}), and noting that at each vertex we need to perform 
 the usual $\int \prod_j^n d^3 k_j/(2 \pi)^3$ integration for every incoming $\bm{k}_j$ {\it including} the dotted line in the case of $\tilde{\Gamma}^{(n-1)}_{\rm s}$,
we find two second order diagrams:
\begin{equation}
\begin{aligned}
\parbox{18mm}{\begin{fmffile}{diagrams/2orderneut2}
\begin{fmfchar*}(50,50)
\fmfleft{in}
	   \fmfleft{in1,in2}
\fmfright{out}
\fmf{dashes_arrow,label=$\bm{k}_1$}{in1,v}
\fmf{dashes_arrow,label=$\bm{k}$}{v,out}
\fmf{dots,label=$\bm{k}_2$}{in2,v}
  \fmfv{decor.shape=circle,decor.filled=empty, decor.size=2thick}{in1}
    \fmfv{decor.shape=circle,decor.filled=empty, decor.size=2thick}{in2}
\end{fmfchar*}\end{fmffile}}&=\int_{s_{\rm in}}^{s} ds' \: G\left(k;s,s'\right)
\int_{s_{\rm in}}^{s'} ds_1 \int_{s_{\rm in}}^{s_1} dz_1 \: \left[\tilde{\Gamma}^{(1)}_{\rm s}(\bm{k}_1,\bm{k}_2; s',s_1)G\left(k_1;s_1,z_1\right) I(\bm{k}_1,z_1)\right]_{\bm{k}}\\
&=\int_{s_{\rm in}}^{s} ds' \: G\left(k;s,s'\right)
\int_{s_{\rm in}}^{s'} ds_1 \: \left[\tilde{\Gamma}^{(1)}_{\rm s}(\bm{k}_1,\bm{k}_2; s',s_1)\delta^{(1)} (\bm{k}_1,s_1)\right]_{\bm{k}}\\
& = \delta^{(2a)}(\bm{k},s), 
\end{aligned}
\end{equation}
where $\tilde{\Gamma}_{\rm s}^{(1)} = \tilde{\Gamma}^{(1)}$ (because the dotted line is not included in the symmetrisation),
and
\begin{equation}
\begin{aligned}
\parbox{18mm}{\begin{fmffile}{diagrams/2orderneut1}
\begin{fmfchar*}(50,50)
\fmfleft{in}
	   \fmfleft{in1,in2}
\fmfright{out}
\fmf{dashes_arrow,label=$\bm{k}_1$}{in1,v}
\fmf{dashes_arrow,label=$\bm{k}_2$}{in2,v}
\fmf{dashes_arrow,label=$\bm{k}$}{v,out}
  \fmfv{decor.shape=circle,decor.filled=empty, decor.size=2thick}{in1}
    \fmfv{decor.shape=circle,decor.filled=empty, decor.size=2thick}{in2}
\end{fmfchar*}\end{fmffile}}&=\int_{s_{\rm in}}^{s} ds' \: G\left(k;s,s'\right)  \int_{s_{\rm in}}^{s'} ds_1\int_{s_{\rm in}}^{s'} ds_2\int_{s_{\rm in}}^{s_1} dz_1\int_{s_{\rm in}}^{s_2} dz_2 
\\
&\hspace{10mm} \times  \left[\Gamma^{(2)}_{\rm s}(\bm{k}_1,\bm{k}_2; s',s_1,s_2)G\left(k_1;s_1,z_1\right)G\left(k_2;s_2,z_2\right) I(\bm{k}_1,z_1)I(\bm{k}_2,z_2)\right]_{\bm{k}}\\
&=\int_{s_{\rm in}}^{s} ds' \, G\left(k;s,s'\right) \! \int_{s_{\rm in}}^{s'} \! ds_1\! \int_{s_{\rm in}}^{s_1} \! ds_2 
\left[\Gamma^{(2)}_{\rm s}(\bm{k}_1,\bm{k}_2; s',s_1,s_2)\delta^{(1)}(\bm{k}_1,s_1) \delta^{(1)} (\bm{k}_2,s_2)\right]_{\bm{k}}\\
&= \delta^{(2b)}(\bm{k},s),
\end{aligned}
\end{equation}
with the symmetrised kernel
\begin{equation}
\Gamma^{(2)}_{\rm s}(\bm{k}_1,\bm{k}_2; s',s_1,s_2) \equiv \frac{1}{2} \left\{\Gamma^{(2)}(\bm{k}_1,\bm{k}_2; s',s_1,s_2) + \Gamma^{(2)}(\bm{k}_2,\bm{k}_1; s',s_2,s_1) \right\}.
\end{equation}
The full second order density perturbation is thus the sum  $\delta^{(2)}= \delta^{(2a)}+\delta^{(2b)}$.

Defining the power spectrum and the bispectrum as
\begin{equation}
\begin{aligned}
\label{eq:nuspectra}
\left \langle \delta(\bm{k},s) \delta(\bm{k}',s) \right \rangle_{\rm C} & \equiv (2 \pi)^3 \delta(\bm{k}+ \bm{k}') P(k;s), \\
\left \langle \delta(\bm{k}_1,s) \delta(\bm{k}_2,s)   \delta(\bm{k}_3,s)   \right \rangle_{\rm C} & \equiv (2 \pi)^3 \delta(\bm{k}_1+ \bm{k}_2+\bm{k}_3) B(k_1,k_2,k_3;s), 
\end{aligned}
\end{equation}
and assuming again that the initial conditions are Gaussian, the construction of connected $N$-point functions follows the same set of rules discussed in section~\ref{sec:cdmdiagram} for standard CDM perturbation theory.  Then, for the leading-order power spectrum, we find
\begin{equation}
\begin{aligned}
(2 \pi)^3 P^{(1)}(k;s)
& =
 \parbox{40mm}{\begin{fmffile}{diagrams/nupowerspectrum}
\begin{fmfchar*}(80,30)
	   \fmfleft{in1}
\fmfright{out1}
\fmf{dashes_arrow,label=$\bm{k}$}{v,in1}
\fmf{dashes_arrow,label=$-\bm{k}$}{v,out1}
\fmfblob{4mm}{v} 
		\end{fmfchar*}\end{fmffile}}
\\ 
&=  \int^s_{s_{\rm in}} ds' \: G(k;s,s') \int^s_{s_{\rm in}} ds'' \: G(k;s,s'') \langle I(\bm{k}, s') I(-\bm{k},s'')\rangle,
\end{aligned}
\end{equation}
where the $I$-correlator is given in terms of the initial perturbed distribution function as 
\begin{equation}
\begin{aligned}
\langle I(\bm{k}, s') I(-\bm{k},s'')\rangle=&\sum_{\ell,\ell'=0}^{\infty} (4\pi)^2(-1)^{\ell+\ell'}(2\ell+1)(2\ell'+1)\int_{0}^{\infty}dq\:q^2 \int_{0}^{\infty}dq'\:q'^2
\\&\times  j_{\ell}\(\frac{kq}{m}(s'-s_{\rm{in}})\)j_{\ell'}\(\frac{kq'}{m}(s''-s_{\rm{in}})\)\langle f_\ell(k,q,s_{\rm{in}}) f_{\ell'}(k,q',s_{\rm{in}})\rangle
\end{aligned}
\end{equation}
using equation~(\ref{eq:decompose}).  

Similarly, the leading-order bispectrum consists of two parts, $B^{(2)}=B^{(2a)} +B^{(2b)}$.  The first piece is
\begin{equation}
\begin{aligned}
\label{eq:nubi1}
& (2 \pi)^3 B^{(2a)}(k,k',|\bm{k}+\bm{k}'|;s)
= \left( \hspace{1mm}
\parbox{35mm}{\begin{fmffile}{diagrams/nubi1}
\begin{fmfchar*}(100,60)
	   \fmfleft{in1}
\fmfright{out1,out2}
\fmftop{v1}
\fmfbottom{v2}
\fmf{dashes_arrow,label=$\bm{k}$}{v,in1}
\fmf{dots,label=$-\bm{k}'$}{v1,v}
\fmf{dashes_arrow,label=$\bm{k}+\bm{k}'$}{v2,v}
\fmf{dashes_arrow,label=$-\bm{k}-\bm{k}'$}{v2,out1}
\fmf{dashes_arrow,label=$\bm{k}'$}{v1,out2}
\fmfblob{4mm}{v1}
\fmfblob{4mm}{v2}     
\end{fmfchar*}\end{fmffile}}  \hspace{-3mm}+
\hspace{1mm}
\parbox{35mm}{\begin{fmffile}{diagrams/nubi1b}
\begin{fmfchar*}(100,60)
	   \fmfleft{in1}
\fmfright{out1,out2}
\fmftop{v1}
\fmfbottom{v2}
\fmf{dashes_arrow,label=$\bm{k}$}{v,in1}
\fmf{dashes_arrow,label=$-\bm{k}'$}{v1,v}
\fmf{dots,label=$\bm{k}+\bm{k}'$}{v2,v}
\fmf{dashes_arrow,label=$-\bm{k}-\bm{k}'$}{v2,out1}
\fmf{dashes_arrow,label=$\bm{k}'$}{v1,out2}
\fmfblob{4mm}{v1}
\fmfblob{4mm}{v2}     
\end{fmfchar*}\end{fmffile}}
\hspace{-1mm}\right)+ \text{cyc.~perm.} \\ \\
& = \int_{s_{\rm in}}^{s} ds' \: G\left(k;s,s'\right)
\int_{s_{\rm in}}^{s'} ds_1 \:  
\Bigg\{
\langle \tilde{\Gamma}^{(1)}_{\rm s}(\bm{k}+\bm{k}',-\bm{k}'; s',s_1)
 \delta^{(1)}(\bm{k}',s)\rangle P^{(1)}(|\bm{k}+\bm{k}'|;s_1,s) \\
 & \hspace{55mm} 
 +\langle \tilde{\Gamma}^{(1)}_{\rm s}(-\bm{k}',\bm{k}+\bm{k}'; s',s_1)
 \delta^{(1)}(-\bm{k}-\bm{k}',s)\rangle P^{(1)}(k';s_1,s)  \Bigg\} \\
 &\hspace{10mm} + \text{cyclic~permutations}, 
\end{aligned}
\end{equation}
where the 
 unequal-time correlators are given by 
\begin{equation}
\begin{aligned}
& \langle \tilde{\Gamma}^{(1)}_{\rm s}(\bm{k}_1,\bm{k}_2; s',s_1)
\delta^{(1)}(\bm{k}_3,s)\rangle  =  
\frac{3}{2}a^2(s_1)\mathcal{H}^2(s_1)\Omega(s_1) \frac{\bm{k}_1}{k_1^2} \cdot (\bm{k}_1 + \bm{k}_2) (s'-s_1) \\
 &  \hspace{15mm}\times \int d^3 q \: 
e^{-i \bm{q} \cdot \{\bm{k}_1 (s'-s_1)+\bm{k}_2(s'-s_{\rm in})\}/m} \langle\delta f^{(1)}(\bm{k}_2,\bm{q},s_{\rm in})\delta^{(1)}(\bm{k}_3, s)\rangle \delta(\bm{k}_2 + \bm{k}_3)\\
&=\frac{3}{2}a^2(s_1)\mathcal{H}^2(s_1)\Omega(s_1) \frac{\bm{k}_1}{k_1^2} \cdot (\bm{k}_1 + \bm{k}_2) (s'-s_1) \sum_{\ell=0}^{\infty}4\pi(-1)^\ell(2\ell+1)\\
 &  \hspace{15mm}\times \int dq \:q^2 \,
j_\ell \( \frac{U_2^2 q }{m}\) P_\ell\(\hat{\bm k}_2\cdot\hat{\bm
U}_2^2\)\langle f_\ell^{(1)}({\bm k}_2,q,s_{\rm {in}})\delta^{(1)}(\bm{k}_3,
s)\rangle \delta(\bm{k}_2 + \bm{k}_3),
\end{aligned}\end{equation}
with ${\bm U}_2^2=\bm{k}_1 (s'-s_1)+\bm{k}_2(s'-s_{\rm in})$,
and
\begin{equation}
\label{eq:unequalcorrelatorsnu}
\langle \delta^{(1)}(\bm{k},s_1) \delta^{(1)}(\bm{k}',s_2) \rangle_{\rm C} \equiv (2 \pi)^3 \delta(\bm{k}+\bm{k}') P^{(1)}(k;s_1,s_2).
\end{equation}
The second piece  is
\begin{equation}
\begin{aligned}
\label{eq:nubi2}
& (2 \pi)^3 B^{(2b)}(k,k',|\bm{k}+\bm{k}'|;s)
= 2 \times \left(\hspace{2mm}
\parbox{35mm}{\begin{fmffile}{diagrams/nubi2}
\begin{fmfchar*}(100,60)
	   \fmfleft{in1}
\fmfright{out1,out2}
\fmftop{v1}
\fmfbottom{v2}
\fmf{dashes_arrow,label=$\bm{k}$}{v,in1}
\fmf{dashes_arrow,label=$-\bm{k}'$}{v1,v}
\fmf{dashes_arrow,label=$\bm{k}+\bm{k}'$}{v2,v}
\fmf{dashes_arrow,label=$-\bm{k}-\bm{k}'$}{v2,out1}
\fmf{dashes_arrow,label=$\bm{k}'$}{v1,out2}
\fmfblob{4mm}{v1}
\fmfblob{4mm}{v2}     
\end{fmfchar*}\end{fmffile}} \right)
+\mathrm{cyclic~permutations}
\\ 
\\
&= 2 \times (2 \pi)^3  \int_{s_{\rm in}}^{s}  ds' \: G\left(k;s,s'\right) \int_{s_{\rm in}}^{s'} ds_1  \int_{s_{\rm in}}^{s'} ds_2 \: 
 \\& \hspace{30mm} \times 
\Gamma^{(2)}_{\rm s}(\bm{k}+\bm{k}',-\bm{k}'; s',s_1,s_2) P^{(1)}(|\bm{k}+\bm{k}'|;s_1,s)P^{(1)}(k';s_2,s) \\
& \hspace{5mm} + \text{cyclic~permutations},
\end{aligned}
\end{equation}
where the unequal-time correlators are again defined as per equation~(\ref{eq:unequalcorrelatorsnu}).  In both~(\ref{eq:nubi1}) and~(\ref{eq:nubi2})  ``cyclic permutations'' 
denote an additional two terms arising from rotation of the external wavevector labels.


\subsection{Combining CDM and neutrinos}
\label{sec:combineddiagram}

It is straightforward to generalise the formalism discussed in the previous sections to the case of mixed CDM+neutrino perturbations. 
Following from equation \eqref{eq:allformal}, the four linear propagators are represented by
\begin{equation}
\begin{aligned}
\parbox{24mm}{\fmfreuse{cdmprop}} &= {\cal G}_{{\rm CC},ab}(k;s_1,s_2), \\
\parbox{24mm}{\fmfreuse{nuprop}} & = G_{\nu\nu} (k;s_1,s_2), \\
\parbox{24mm}{\begin{fmffile}{diagrams/propneutcdm}
\begin{fmfchar*}(40,20)
\fmfleft{in}
	 \fmfright{out}           
	  \fmf{plain}{in,v}
		\fmf{dashes}{v,out}
		\fmf{phantom_arrow}{in,out}
		\fmfv{label=$\bm{k}$,label.angle=-90}{v}
	  \fmflabel{$s_2,a$}{in}
	  \fmflabel{$s_1$}{out}          
\end{fmfchar*}\end{fmffile}} 
& = {\cal G}_{\nu {\rm C},a}\left(k;s_1,s_2\right),\\   
\parbox{24mm}{\begin{fmffile}{diagrams/propcdmneut}
\begin{fmfchar*}(40,20)
\fmfleft{in}
	 \fmfright{out}           
	  \fmf{dashes}{in,v}
		\fmf{plain}{v,out}
				\fmf{phantom_arrow}{in,out}
		\fmfv{label=$\bm{k}$,label.angle=-90}{v}
	  \fmflabel{$s_2$}{in}
	  \fmflabel{$s_1,a$}{out}          
\end{fmfchar*}\end{fmffile}}
& = {\cal G}_{{\rm C} \nu,a}\left(k;s_1,s_2\right).
\end{aligned}
\end{equation}
Observe that ${\cal G}_{{\rm C} \nu,a}(k;s_1,s_2)$ which connects a CDM perturbation at $s_1$ to a neutrino perturbation at $s_2$ 
 begins as a solid line but ends as a dashed line.  Similarly for the propagator ${\cal G}_{\nu {\rm C},a}(k;s_1,s_2)$ which has the opposite function.  Again,
 attaching a source term to $G_{\nu\nu}(k;s_1,s_2)$ automatically incurs an integration over $s_2$
from the initial time $s_{\rm in}$ to $s_1$ as per equation~(\ref{eq:nupropsource}).  The same procedure applies also to ${\cal G}_{{\rm C}\nu,a}(k;s_1,s_2)$.

Three classes of vertices govern the nonlinear aspect of the theory: the CDM vertex $\gamma_{abc}$, and the two neutrino vertices $\tilde{\Gamma}^{(n-1)}$ and $\Gamma^{(n)}$.
The CDM vertex functions in exactly the same way as in standard CDM-only perturbation theory, i.e., it couples two incoming solid lines to produce one solid outgoing line, 
and is represented by the diagram~(\ref{eq:feynmanrulesCDM2}).   The neutrino vertices, on the other hand, while schematically resembling diagrams~(\ref{eq:gamma1}) and~(\ref{eq:gamma2diag}), can now take any combination of dashed and solid incoming lines to output a single dashed line, weighted by one factor of $f_{\rm C}$ for every incoming solid line and one factor of $f_\nu$ for every incoming dashed lines.

Initial conditions are again represented by open circles ``$\circ$'', and the amalgamation of two such circles form a 2-point function at some initial time $s_{\rm in}$.  In mixed CDM+neutrino cosmologies, there are three types of initial 2-point functions,
\begin{equation}
\begin{aligned}
\label{eq:initialcorrelators}
\parbox{24mm}{\begin{fmffile}{diagrams/initialpowerspectrumcc}
\begin{fmfchar*}(40,20)
	   \fmfleft{in1}
\fmfright{out1}
\fmf{plain}{v,in1}
\fmf{plain}{v,out1}
 \fmflabel{$\bm{k},a$}{in1}
  \fmflabel{$\bm{k}',b$}{out1}
\fmfblob{4mm}{v} 
\end{fmfchar*}\end{fmffile}}
&= \langle \varphi_a(\bm{k},s_{\rm in}) \varphi_b(\bm{k}',s_{\rm in}) \rangle_{\rm C}, \\
\parbox{24mm}{\begin{fmffile}{diagrams/initialpowerspectrumnunu}
\begin{fmfchar*}(40,20)
	   \fmfleft{in1}
\fmfright{out1}
\fmf{dashes}{v,in1}
\fmf{dashes}{v,out1}
\fmflabel{$\bm{k}$}{in1}
  \fmflabel{$\bm{k}'$}{out1}
\fmfblob{4mm}{v}             
\end{fmfchar*}\end{fmffile}}
& = \langle  I_\nu(\bm{k},s_{\rm in}) I_\nu(\bm{k}',s_{\rm in}) \rangle_{\rm C},
\\
\parbox{24mm}{\begin{fmffile}{diagrams/initialpowerspectrumcnu}
\begin{fmfchar*}(40,20)
	   \fmfleft{in1}
\fmfright{out1}
\fmf{plain}{v,out1}
\fmf{dashes}{v,in1}
\fmflabel{$\bm{k}$}{in1}
  \fmflabel{$\bm{k}',a$}{out1}
\fmfblob{4mm}{v} 
\end{fmfchar*}\end{fmffile}} 
& =  \langle I_\nu(\bm{k},s_{\rm in}) \varphi_a(\bm{k}',s_{\rm in}) \rangle_{\rm C} ,
\end{aligned}
\end{equation}
and momentum conservation ${\bm k}+{\bm k}'=0$ is implied.  Again, because of the assumption of Gaussian initial perturbations, only 2-point functions are nonzero at the initial time.

\begin{figure}[t]
	\centering
			\includegraphics[width=0.99\textwidth]{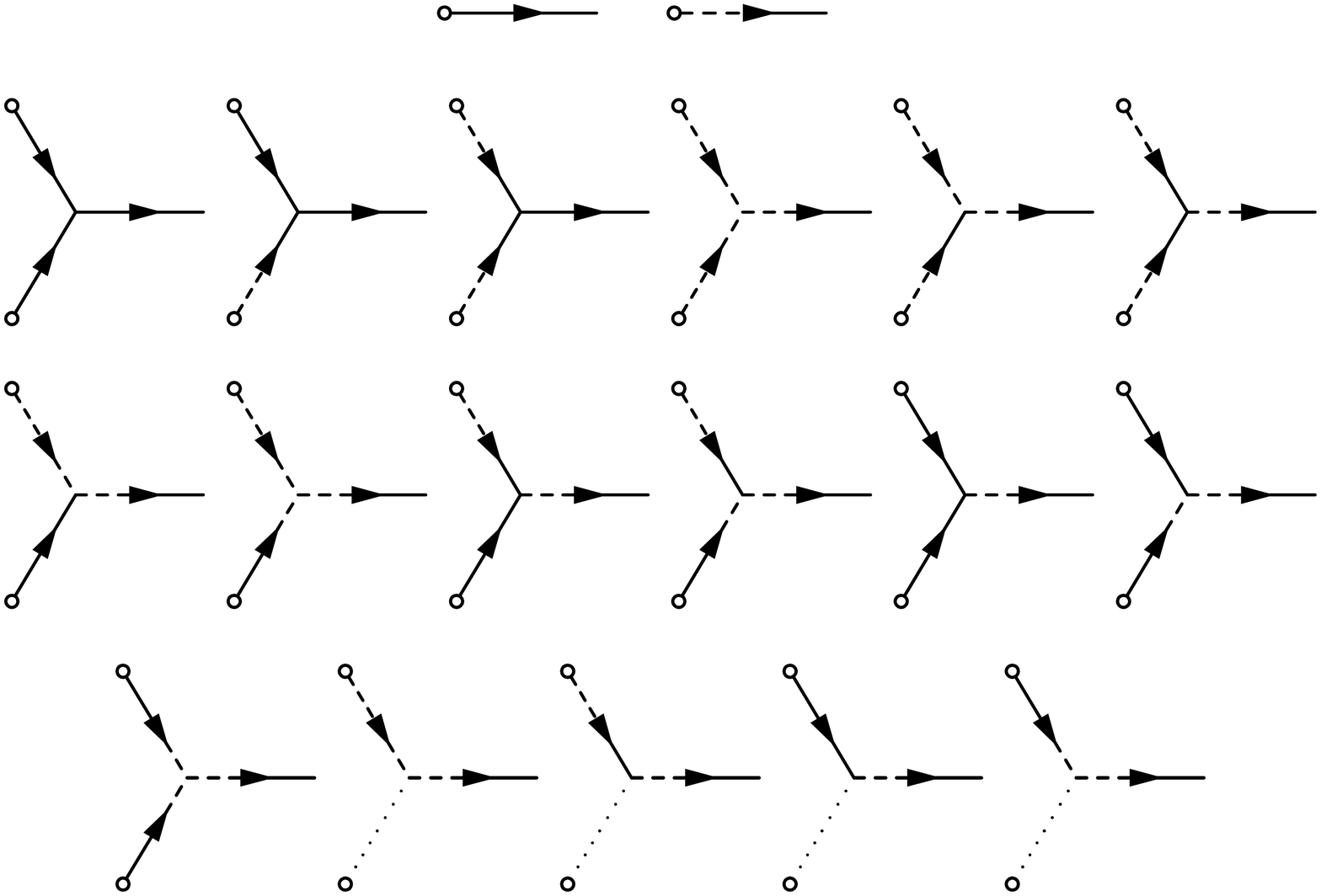}
	\caption{The two first-order and all 17 second-order diagrams contributing to $\varphi_{\rm{C}}$. The diagrams for the leading-order 3-point function $\langle \delta^{(2)}_{\rm C} (\bm{k}_1,s) \delta^{(1)}_{\rm C} (\bm{k}_2,s) \delta^{(1)}_{\rm C} (\bm{k}_2,s) \rangle_{\rm C}$ can be obtained by glueing together the second-order diagrams with one linear diagram at each open circle.}
	\label{fig:diagrams}
\end{figure}

Figure~ \ref{fig:diagrams} shows the two diagrams that contribute to $\varphi_a^{(1)}$ and all 17 diagrams for $\varphi^{(2)}_a$.  Glueing together the diagrams at the open circles as per equation~(\ref{eq:initialcorrelators}), we can form the 2- and 3-point functions $\langle \delta^{(1)}_{\rm C} (\bm{k},s)
\delta^{(1)}_{\rm C} (\bm{k}',s) \rangle_{\rm C}$ and  $\langle \delta^{(2)}_{\rm C} (\bm{k}_1,s) \delta^{(1)}_{\rm C} (\bm{k}_2,s) \delta^{(1)}_{\rm C} (\bm{k}_2,s) \rangle_{\rm C}$,
which contribute respectively to the leading-order term of the total matter spectrum,
\begin{equation}
\begin{aligned}
\left\langle\delta(\bm{k},s)\delta(\bm{k}',s)\right\rangle_{\rm C}  \equiv \: &  (2 \pi)^3 \delta(\bm{k} + \bm{k}') P_\delta (k,s) \\
= \: & f_{\rm C}^2\left\langle\delta_{\rm C}(\bm{k},s)\delta_{\rm C}(\bm{k}',s)\right\rangle_{\rm C}+2f_{\rm C}f_{\rm \nu}\left\langle\delta_{\rm C}(\bm{k},s)\delta_{\rm \nu}(\bm{k}',s)\right\rangle_{\rm C}\\
&+f_{\rm \nu}^2\left\langle\delta_{\rm \nu}(\bm{k},s)\delta_{\rm \nu}(\bm{k}',s)\right\rangle_{\rm C},
\end{aligned}
\end{equation}
and bispectrum
\begin{equation}
\begin{aligned}
\label{eq:combinedbi}
\left\langle\delta(\bm{k}_1,s)\delta(\bm{k}_2,s)\delta(\bm{k}_3,s)\right\rangle_{\rm C}   \equiv \: & (2 \pi)^3 \delta (\bm{k}_1 + \bm{k}_2 + \bm{k}_3)  B_\delta (k_1, k_2, k_3;s) \\
 =\: & f_{\rm C}^3\left\langle\delta_{\rm C}(\bm{k}_1,s)\delta_{\rm C}(\bm{k}_2,s)\delta_{\rm C}(\bm{k}_3,s)\right\rangle_{\rm C}\\
 &+f_{\rm \nu}^3\left\langle\delta_{\rm \nu}(\bm{k}_1,s)\delta_{\rm \nu}(\bm{k}_2,s)\delta_{\rm \nu}(\bm{k}_3,s)\right\rangle_{\rm C}\\
&+2f_{\rm C}^2f_{\rm \nu}\left\langle\delta_{\rm C}(\bm{k}_1,s)\delta_{\rm C}(\bm{k}_2,s)\delta_{\rm \nu}(\bm{k}_3,s)\right\rangle_{\rm C}\\
&+2f_{\rm C}f_{\rm \nu}^2\left\langle\delta_{\rm C}(\bm{k}_1,s)\delta_{\rm \nu}(\bm{k}_2,s)\delta_{\rm \nu}(\bm{k}_3,s)\right\rangle_{\rm C}.
\end{aligned}
\end{equation}
Note that each contributing diagram is weighted by a factor determined by the nature of its external legs: each CDM leg receives a factor~$f_{\rm C}$, while each neutrino leg picks up a factor~$f_\nu$.

Combining perturbations in both the CDM and the neutrino sectors, we find the leading-order total matter bispectrum to be
\begin{equation}
\begin{aligned}
\label{eq:bispectrumcdmneu}
& B_\delta^{(2)} (k,k',|\bm{k}+\bm{k}'|;s) = \\
&2 \int_{s_{\rm in}}^s d s' \: {\cal G}_{{\rm C},a} (k;s,s')
 \gamma_{abc} (-\bm{k}',\bm{k}+\bm{k}') P^{(1)}_{b \delta}(k';s',s) P^{(1)}_{c \delta} (|\bm{k}+\bm{k}'|;s',s) \\
& +2 \int_{s_{\rm in}}^s {\cal G}_\nu (k;s,s') \int_{s_{\rm in}}^{s'} \! d s_1 \int_{s_{\rm in}}^{s'}  \! d s_2 \: \Gamma_{\rm s}^{(2)} (-\bm{k}',\bm{k}+\bm{k}';s',s_1,s_2) P^{(1)}_\delta(k';s_1,s) P^{(1)} _\delta(|\bm{k}+\bm{k}'|;s_2,s) \\
&+\frac{1}{(2 \pi)^3 } \int_{s_{\rm in}}^{s} ds' \: {\cal G}_\nu \left(k;s,s'\right)
\int_{s_{\rm in}}^{s'} ds_1 \:  \Bigg\{
\langle \tilde{\Gamma}^{(1)}_{\rm s}(-\bm{k}',\bm{k}+\bm{k}'; s',s_1)
 \delta^{(1)}(\bm{k}+\bm{k}',s)\rangle P^{(1)}_\delta(k';s_1,s) \\
 & \hspace{50mm} 
 +\langle \tilde{\Gamma}^{(1)}_{\rm s}(\bm{k}+\bm{k}',-\bm{k}; s',s_1)
 \delta^{(1)}(\bm{k}',s)\rangle P^{(1)}_\delta(|\bm{k}+\bm{k}'|;s_1,s)  \Bigg\} \\
 &+ {\rm cyclic \ permutations},
\end{aligned}
\end{equation}
where we have combined the linear propagators to form
\begin{equation}
\begin{aligned}
{\cal G}_{{\rm C},a}(k;s',s) & \equiv f_{\rm C} {\cal G}_{{\rm CC},1a} (k;s,s')+ f_\nu {\cal G}_{\nu {\rm C},a} (k;s,s'), \\
{\cal G}_{\nu} (k;s,s')& \equiv f_{\rm C} {\cal G}_{{\rm C}\nu,1}(k;s,s') + f_\nu  G_{\nu \nu}(k;s,s'), 
\end{aligned}
\end{equation}
and
\begin{equation}
\begin{aligned}
\label{eq:unequaltime}
\left\langle\varphi_b^{(1)}(\bm{k},s')\delta^{(1)} (\bm{k}',s)\right\rangle_{\rm C} &  \equiv  (2 \pi)^3 \delta(\bm{k} + \bm{k}') P^{(1)}_{b\delta} (k;s',s) , \\
\left\langle\delta^{(1)}(\bm{k},s')\delta^{(1)} (\bm{k}',s)\right\rangle_{\rm C} &  \equiv  (2 \pi)^3 \delta(\bm{k} + \bm{k}') P^{(1)}_\delta (k;s',s) 
\end{aligned}
\end{equation}
define the unequal-time correlators.  ``Cyclic permutations'' again denote two additional terms arising from rotation of the external wavevectors $\bm{k}$, $\bm{k}'$ and $\bm{k}+\bm{k}'$.

Lastly, as we shall be comparing in section~\ref{sec:applications}  the bispectrum~(\ref{eq:bispectrumcdmneu}) with that computed from the two-fluid approximation of section~\ref{sec:twofluid}, we give here also the expression for the latter:
\begin{equation}
\begin{aligned}
\label{eq:fluidbi}
&B_\delta^{(2)} (k,k',|\bm{k}+\bm{k}'|;s) = \\
&\hspace{10mm} 2 \int_{s_{\rm in}}^s d s' \: g_{{\rm C},a} (k;s,s')
 \gamma_{abc} (-\bm{k}',\bm{k}+\bm{k}') P^{(1)}_{b \delta}(k';s',s) P^{(1)}_{c \delta} (|\bm{k}+\bm{k}'|;s',s)
\\& \hspace{10mm} +2 \int_{s_{\rm in}}^s d s' \: {g_{\nu}}^{A} (k;s,s')
 \gamma^{ABC} (-\bm{k}',\bm{k}+\bm{k}') {P^B}_{\delta}^{(1)} (k';s',s) {P^C}_{\delta}^{(1)} (|\bm{k}+\bm{k}'|;s',s)\\
 & \hspace{10mm} +  \int_{s_{\rm in}}^s d s' \: {g_{\nu}}^{A} (k;s,s') \: c_{\rm s}^2(s') \: k^2 \: \delta^{A2} \delta^{B1} \delta^{C1} \:
  {P^B}_{\delta}^{(1)} (k';s',s) {P^C}_{\delta}^{(1)} (|\bm{k}+\bm{k}'|;s',s) \\
 & \hspace{10mm} + {\rm cyclic \ permutations},
\end{aligned}
\end{equation}
where, following the convention of section~\ref{sec:twofluid}, superscript indices $(A,B,C)$ refer to neutrino quantities, while subscript indices $(a,b,c)$ refer to their CDM counterparts.  We have again combined the linear fluid propagators to form
\begin{equation}
\begin{aligned}
g_{{\rm C},a}(k;s',s) & \equiv f_{\rm C} g_{1a} (k;s,s')+ f_\nu {g^{1}}_a (k;s,s'), \\
{g_{\nu}}^A (k;s,s')& \equiv f_{\rm C} {g_{1}}^{A}(k;s,s') + f_\nu  g^{1A}(k;s,s'), 
\end{aligned}
\end{equation}
and the unequal-time correlators  ${P^B}_{\delta}^{(1)} (k;s',s)$ and  $P^{(1)}_{b \delta}(k';s',s)$ are given respectively by
\begin{equation}
\begin{aligned}
\left\langle{\varphi^B}^{(1)}(\bm{k},s')\delta^{(1)} (\bm{k}',s)\right\rangle_{\rm C} &  \equiv  (2 \pi)^3 \delta(\bm{k} + \bm{k}') {P^B}_{\delta}^{(1)} (k;s',s) ,
\end{aligned}
\end{equation}
and in equation~(\ref{eq:unequaltime}).  Note that the expression~(\ref{eq:fluidbi}) applies also to the hybrid full theory+fluid approach discussed in section~\ref{sec:hybrid}; we need only to replace the linear fluid propagators with the hybrid propagators defined in equation~(\ref{eq:reformat}).


\section{Application to large-scale structure observables}
\label{sec:applications}

We apply the perturbation theory developed in the previous sections to compute the leading-order total matter power spectrum and bispectrum in mixed CDM+massive neutrino cosmologies in the presence of a cosmological constant $\Lambda$ and assuming a flat spatial geometry.  We take as fixed parameters the present-day $\Lambda$ energy density $\Omega_{\Lambda}=0.728$ and total matter density $\Omega(a=1)=0.272$, the latter number includes the present-day baryon density fixed at $\Omega_{\rm b}=0.0456$.
The primordial perturbations are assumed to be adiabatic, and described by a scale-invariant curvature power spectrum $P_{\cal R}= A_{\rm s} k^{-3}$ (i.e., the scalar spectral index is $n_{\rm s}=1$).  Because we are concerned only with the leading-order terms of the $N$-point functions, the amplitude $A_{\rm s}$ determines only the overall normalisation; without loss of generality
we set it to $A_{\rm s}=1\ h^{-3} \, {\rm Mpc}^{3}$.

We choose an initial time $s_{\rm in}$ corresponding to the scale factor $a=1/10$, and compute the initial conditions using COSMICS~\cite{Bertschinger1995}.
This initial time suffices for our purpose of testing different approximation schemes.  We note however that to reach an accuracy high enough for comparison with observations or $N$-body simulations, the calculation must be initialised  at an earlier time, say $a=1/50$, in order to prevent nonlinear  transients from spoiling the outcome.  At $a = 1/50$ neutrinos
with masses $m\gtrsim 0.0085$~eV are already nonrelativistic; our Newtonian treatment therefore applies at these early times.  
Smaller neutrino masses in principle call for a full relativistic treatment.  However, nonlinear effects should in any case be very small for such light neutrinos; extending our Newtonian treatment  to $a=1/50$ is unlikely to cause problems.

Because we assume in our treatment that at late times CDM and baryons form one single fluid  (which we have loosely termed throughout this work the ``CDM fluid''),
 the initial perturbations output by COSMICS need to be weighted according to
\begin{align}
\delta_{\rm C}(\bm{k},s_{\rm in})=\frac{f_{\rm{cdm}}\delta_{\rm{cdm}}(\bm{k},s_{\rm in}) +f_{\rm{b}}\delta_{\rm{b}}(\bm{k},s_{\rm in}) }{f_{\rm{cdm}}+f_{\mathrm{b}}},
\end{align}
where $f_{\rm b}$ and $f_{\rm cdm}$ denote, respectively, the fractions of the total matter density in the form of  baryons and ``real'' CDM, and $f_{\rm cdm} + f_{\rm b} = f_{\rm C}$.
The same weighting applies also to the initial velocity divergence $\theta_{\rm C}(\bm{k},s_{\rm in})$.
In the neutrino sector, we assume one massive species, whose initial momentum distribution is supplied by COSMICS  in terms of Legendre moments $f_\ell (\bm{k},q,s_{\rm in})$ up to a 
multipole of~$\ell =13$, although for small wavenumbers $k$ the first two or three moments suffice for our purpose~\cite{Shoji:2010hm}.

With these initial conditions we solve the linear  Gilbert's equation numerically using the Nystr\"om method.  See appendix~\ref{sec:nystroem} for details. The time integrals appearing in the higher-order perturbations are performed using either the same quadrature rule as adopted in the Nystr\"om method, or by way of a non-equidistant trapezoidal rule which uses the same nodes as the quadrature rule.


\subsection{Linear power spectrum}
\label{sec:linearpowerspectrum}

\begin{figure}[t]
	\centering	
				\includegraphics[width=0.49\textwidth]{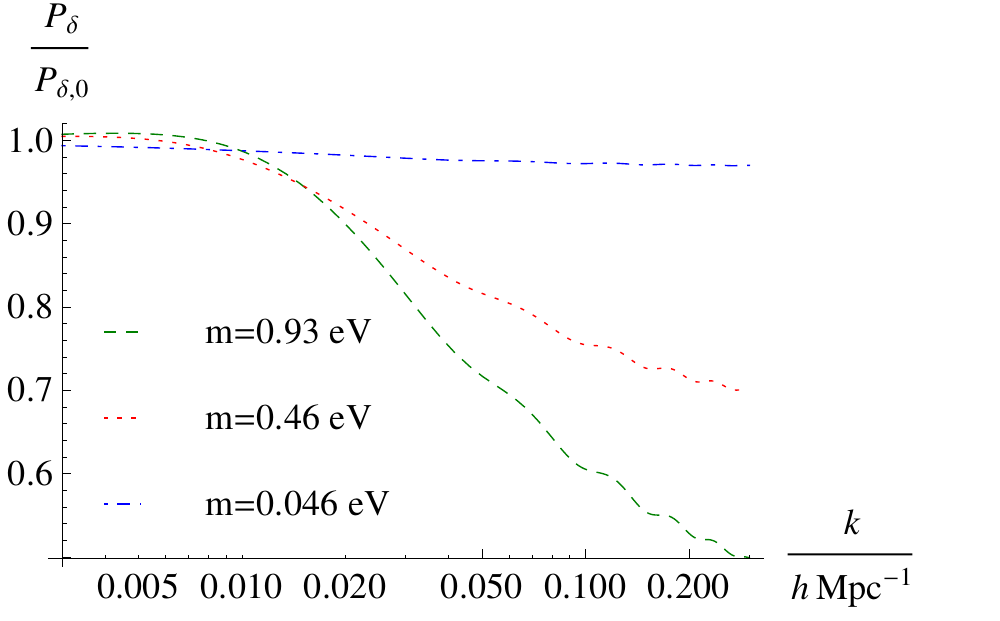}
				\includegraphics[width=0.49\textwidth]{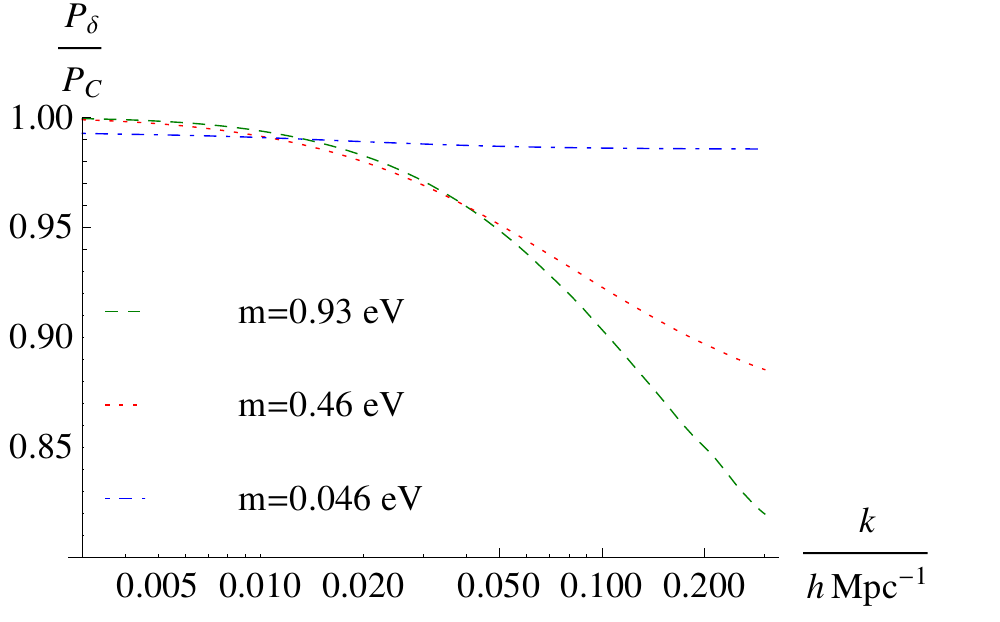}

	\caption{{\it Left}: Total linear matter power spectrum $P_\delta(k)$ in a CDM+neutrino cosmology at $a=1$ for three different neutrino masses,
normalised to the matter power spectrum in a cosmology with massless neutrinos,  $P_{\delta,0}(k)$.  The total present-day matter density has been fixed  at $\Omega=0.272$ in all cases. 
{\it Right}: The same total linear matter power spectra as in the left panel, but now normalised to~$P_{\rm C}(k)$, the power spectrum of CDM density perturbations 
in a fictitious CDM-only cosmology  initialised at $a=1/10$ with the CDM perturbations of the original CDM+massive neutrino cosmology.
Comparing the left and right panels, we see that although most of the power suppression is a consequence of the neutrinos transitioning from a relativistic to nonrelativistic species at early times, free-streaming suppression of the neutrino perturbations at late times can nonetheless be significant.}
	\label{fig:PowerSpectrum}
\end{figure}

It is well known that the presence of massive neutrinos attenuates the growth of structures on small scales and suppresses the matter power spectrum at large wavenumbers.
The left panel of figure~\ref{fig:PowerSpectrum} shows this suppression in the linear matter power spectrum $P_\delta(k)$ at $a=1$ for several choices of the neutrino mass ($m = 0.046, 0.46, 0.93$~eV).  All power spectra have been normalised to~$P_{\delta,0}(k)$, the total matter power spectrum in the case of a vanishing neutrino mass, and 
the total present-day matter density is always held fixed at $\Omega=0.272$.  Our choice of sample neutrino masses spans a range from just below the minimum value suggested by neutrino oscillations experiments, to $\sim 1$~eV motivated by recent suggestions that a $\sim 0.5$~eV-mass sterile neutrino could potentially resolve the conflict between Planck CMB temperature measurements and observations of the cluster abundance and cosmic shear~\cite{Dvorkin:2014lea,Hamann:2013iba,Battye:2013xqa}.

It is instructive to note that the suppressed power at large $k$~values  seen in the left panel of figure~\ref{fig:PowerSpectrum}
is in fact due to two distinct effects: (i) free-streaming suppression of the neutrino perturbations as discussed in section~\ref{sec:soundspeed}, which occurs {\it after} the neutrinos have become nonrelativistic at late times,  and (ii) a general suppression of perturbations of all types on small scales caused by background effects arising from a reduced matter density at early times {\it before} and/or {\it while} the neutrinos transition to a nonrelativistic state~(see, e.g.,~\cite{Lesgourgues:2006nd}).
In order to isolate effect~(i), we plot also in the right panel of figure~\ref{fig:PowerSpectrum} the ratio of the total matter power spectrum $P_\delta(k)$ to the power spectrum of the CDM density perturbations in a fictitious CDM-only cosmology initialised at $a=1/10$ with the CDM perturbations of the original CDM+massive neutrino cosmology,~$P_{\rm C}(k)$. 
Comparing the left and right panels, we see immediately that although most of the power suppression is a consequence of effect~(ii), free-streaming suppression of nonrelativistic neutrino perturbations at late times can nonetheless be significant.

\begin{figure}[t]
	\centering
				\includegraphics[width=0.49\textwidth]{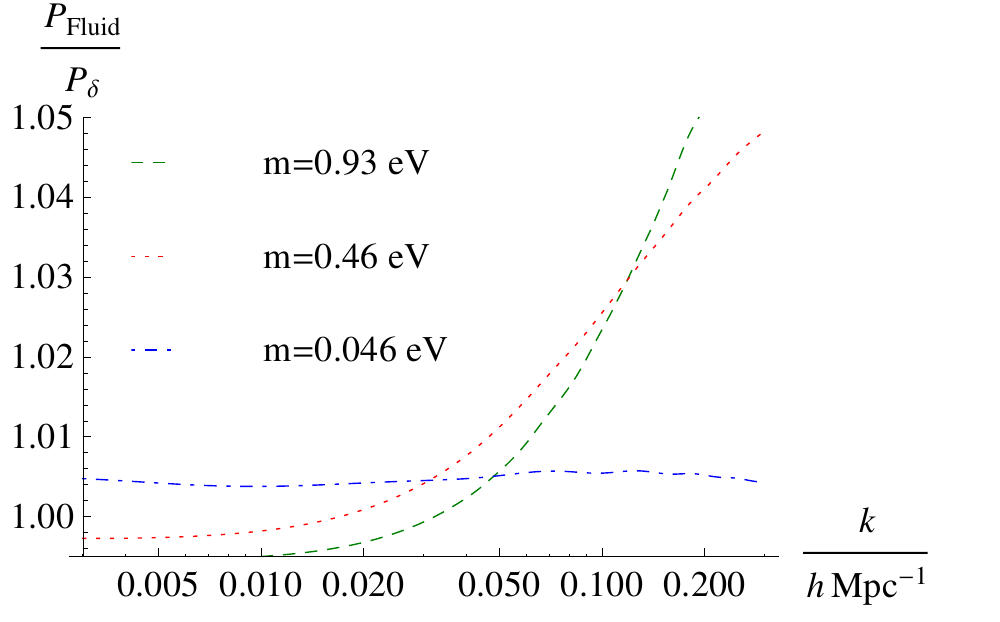}
				\includegraphics[width=0.49\textwidth]{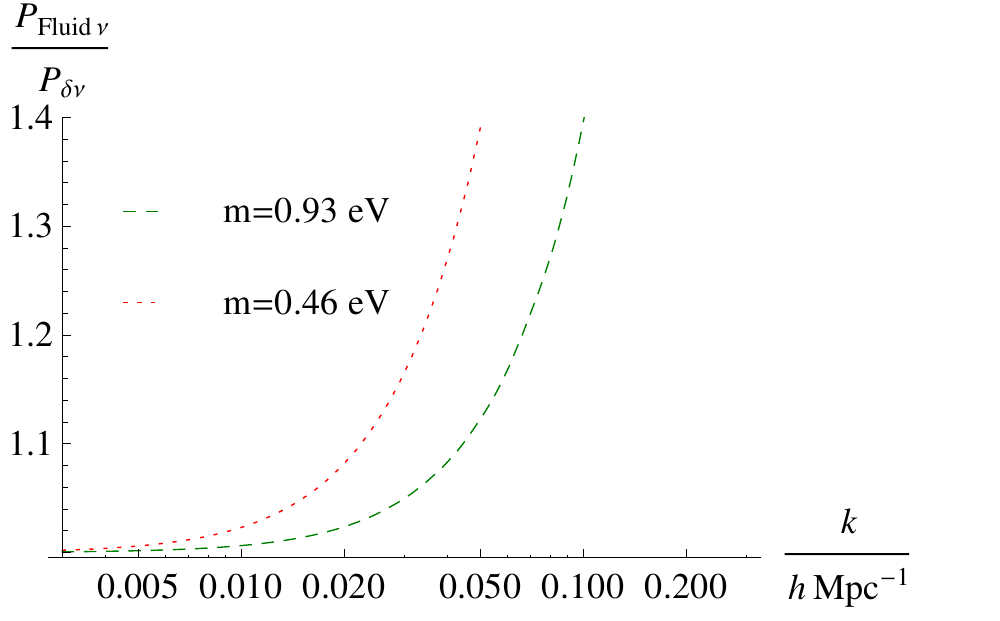}
				\includegraphics[width=0.49\textwidth]{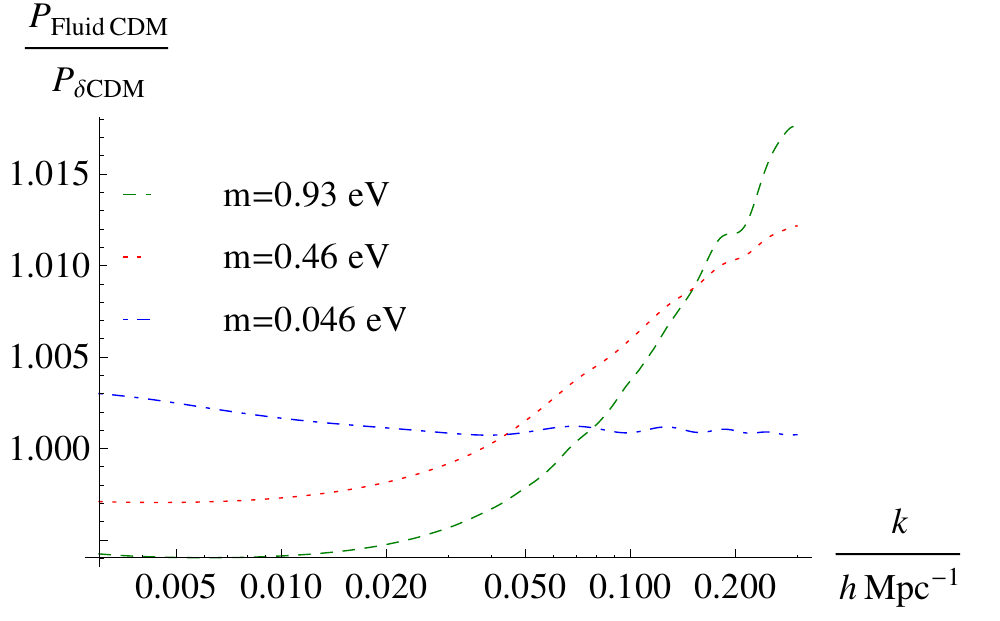}
	\caption{{\it Top left}: Total linear matter power spectrum at $a=1$ for various neutrino masses computed using the two-fluid approximation, normalised to that computed from the full theory.
	{\it Top right}:  The corresponding neutrino power spectrum~$P_\nu(k)  \equiv
	\left\langle\delta_{\nu}\delta_{\nu}\right\rangle/(2 \pi)^3$ computed using the
	fluid approximation, normalised to that from the full theory.  We do not show
	the fluid approximation for the smallest neutrino mass $m=0.046$~eV on
	the right panel, because the error incurred always exceeds
	80\% even on large scales.
	{\it Bottom}: The corresponding CDM power spectrum~$P_{\rm{CDM}}(k)  \equiv
	\left\langle\delta_{\rm{CDM}}\delta_{\rm{CDM}}\right\rangle/(2 \pi)^3$, calculated using the fluid approximation for the neutrinos normalised to the CDM power spectrum from the full theory.}
	\label{fig:PowerSpectrumApp}
\end{figure}

The top left panel of figure~\ref{fig:PowerSpectrumApp} compares the total linear matter power spectrum computed, from $a=1/10$ to $a=1$, using the exact theory,~$P_\delta(k)$, and using the two-fluid approximation of section~\ref{sec:twofluid},~$P_\delta^{\rm Fluid}(k)$.   We note that the authors of~\cite{Shoji:2010hm} also performed a similar test, but neglected the gravitational potential due to the neutrino perturbations.  For small wavenumbers~$k$,  we see that the two treatments yield a difference of less than 1\% in the total matter power spectrum 
 for all neutrino masses considered, thereby confirming the validity of the fluid approximation on length scales greater than the neutrino free-streaming scale.

The fluid approximation begins to break down around the free-streaming scale,
causing the error to grow.  Already at the nonlinear scale of~$k\sim 0.1\ h\:  {\rm Mpc}^{-1}$, we see an error of $\sim 2\%$ for $m = 0.46, 0.93$~eV.
  We emphasise that this number pertains to the {\it total} matter power spectrum: the error on the neutrino perturbations, as manifested in the power spectrum of the neutrino perturbations on the top right panel of figure~\ref{fig:PowerSpectrumApp}, is in fact much larger---about $40$\% at~$k\sim 0.1\ h\:  {\rm Mpc}^{-1}$.
  It is only because neutrinos contribute so subdominant a fraction of the total matter density that the error incurred in CDM power spectrum by the fluid approximation is still less than 1.5\% (bottom panel of figure~\ref{fig:PowerSpectrumApp}), and consequently the total matter power spectrum is still acceptably accurate.  Indeed, in the case of $m=0.046$~eV where the  free-streaming scale evolves from $k_{\rm FS}(s_{\rm in})\approx 2.8\cdot 10^{-4}\ h \: {\rm Mpc}^{-1}$ initially at $a=1/10$ to $k_{\rm FS}(s_0)\approx 9\cdot 10^{-4}\ h\: {\rm Mpc}^{-1}$ today, the fluid approximation incurs an error exceeding 40\% at $k \gtrsim k_{\rm FS}$ in the neutrino power spectrum and thus in principle breaks down on all observable scales; the prediction for the total matter power spectrum, however, still falls within 1\% of the exact theory.   
	
Lastly, while it is of course true that the full {\it linear} theory of neutrino perturbations is widely known---both in the form of Gilbert's equation~(\ref{eq:gilbert}) and the relativistic Boltzmann hierarchy (e.g.,~\cite{Ma:1995ey})---and there is in practice no need to resort to the fluid approximation to compute linear quantities, 
this exercise still highlights the need to be cautious when designing nonlinear models of neutrino clustering.  In particular, the two-fluid approximation is essentially a perturbative version of the Smooth-Particle Hydrodynamic (SPH) model of neutrinos investigated in the simulations of~\cite{Hannestad:2011td}.  Our results show that even though such a model can reproduce the gross features of the total matter power spectrum, ultimately it may not be sufficient for precision ($<1$\%) modelling.


\subsection{Tree-level bispectrum}
\label{sec:bispectrum}

For Gaussian initial conditions, the tree-level bispectrum is the simplest leading-order $N$-point function that arises purely through nonlinear evolution of the 
perturbations at late times.
We use it to study the nonlinearities of neutrino perturbations computed from the exact theory as well as from various approximation schemes.

\begin{figure}[t]
	\centering
				\includegraphics[width=0.49\textwidth]{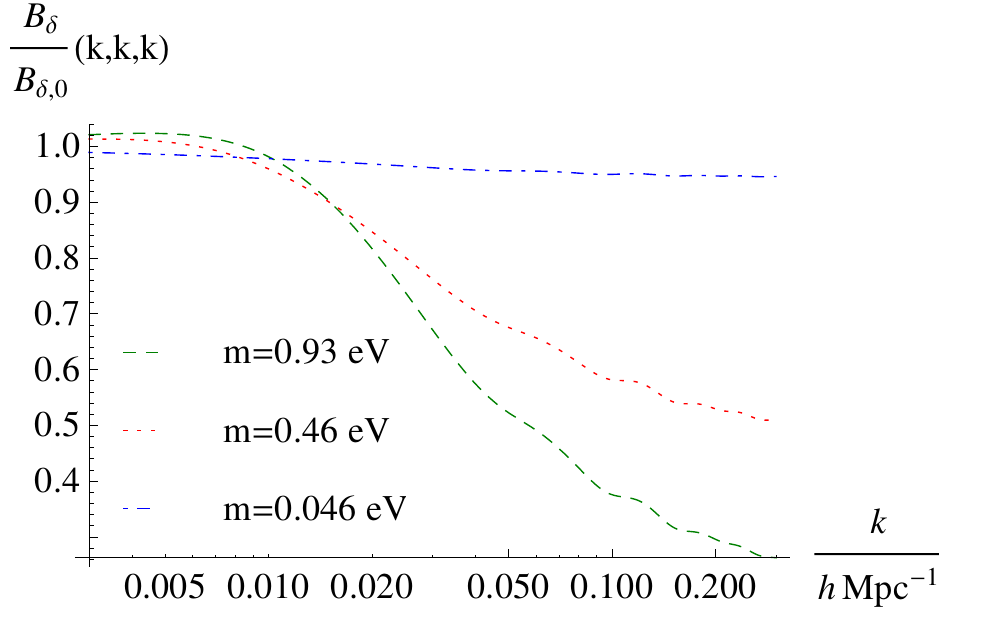}
	\includegraphics[width=0.49\textwidth]{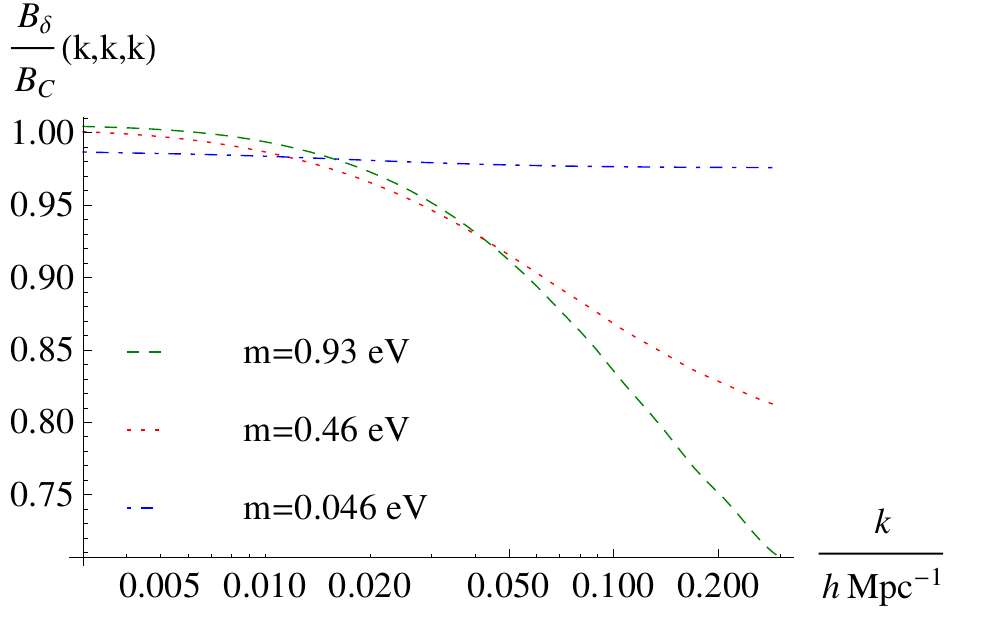}
	\caption{{\it Left}: Leading-order equilateral matter bispectrum $B_\delta(k, k,k)$ in a CDM+neutrino cosmology at $a=1$ for three different neutrino masses,  normalised to 
the equilateral bispectrum in a cosmology with massless neutrinos, $B_{\delta,0}(k, k ,k )$.  The total present-day matter density has been fixed  at $\Omega=0.272$ in all cases. The {\it Right}: The same matter bispectra as in the left panel, but now normalised to the equilateral bispectrum of CDM density perturbations in a fictitious CDM-only cosmology initialised at $a=1/10$ with the CDM perturbations of the original CDM+massive neutrino cosmology, $B_{\rm C}(k, k, k)$.}
	\label{fig:BiSpectrum}
\end{figure}

The left panel of figure~\ref{fig:BiSpectrum} shows the present-day  total matter bispectrum $B_\delta (k,k',|\bm{k}+\bm{k}'|)$  in the equilateral configuration, i.e., $k = k' = |\bm{k}+\bm{k}'|$, for 
several choices of neutrino masses, normalised to the equilateral matter bispectrum in a massless neutrino cosmology,~$B_{\delta, 0}(k,k,k)$.  Again, the total present-day matter density 
has been held fixed in all cases.
As with the total matter power spectrum $P_\delta(k)$, replacing a fraction of CDM with massive neutrinos causes a suppression in the matter bispectrum at large wavenumbers~$k$.  The asymptotic change in the leading-order term in the equilateral configuration appears to be well described by 
\begin{equation}
\frac{\Delta B_\delta}{B_\delta} \sim 13\:  \frac{\Omega_\nu}{\Omega}, 
\end{equation}
to be compared with the analogous asymptotic suppression in the linear matter power spectrum, $\Delta P_\delta/P_\delta \sim 8\: \Omega_\nu/\Omega$.

As in the case of~$P_\delta(k)$, most of the suppression in $B_\delta (k,k,k)$ can in fact be traced back to the transition of neutrinos from a relativistic to a nonrelativistic at early times.  We therefore also plot in the right panel of figure~\ref{fig:BiSpectrum} the same total matter bispectra but now normalised to $B_{\rm C}(k, k, k)$, the equilateral bispectrum of the CDM perturbations in a fictitious CDM-only cosmology initialised at $a=1/10$ with the CDM perturbations of the original CDM+massive neutrino cosmology.

\begin{figure}[t]
	\centering
	\includegraphics[width=0.49\textwidth]{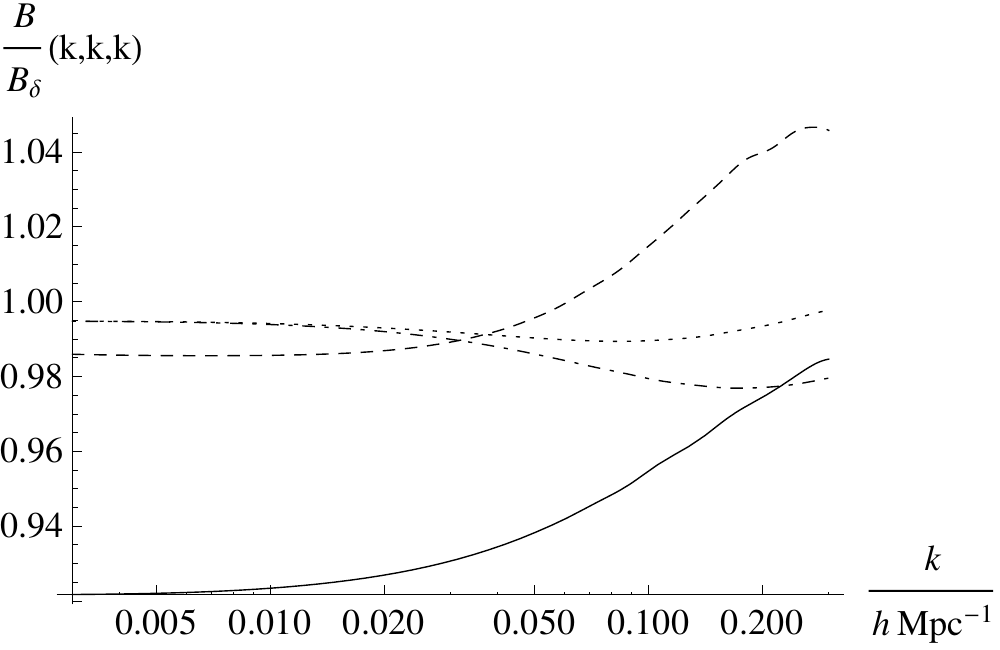}
	\includegraphics[width=0.49\textwidth]{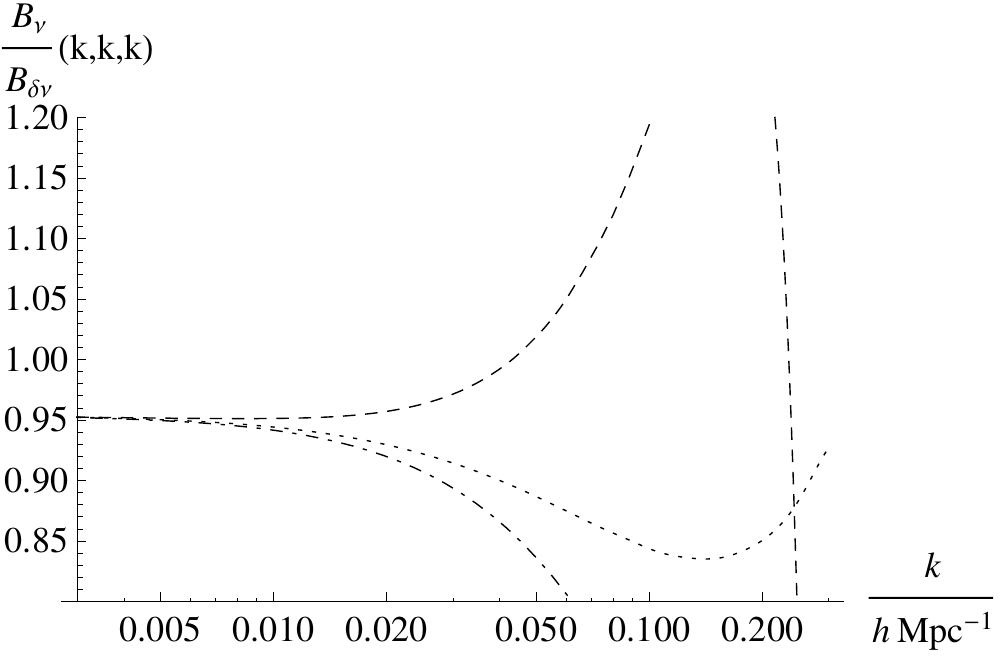}
	\includegraphics[width=0.49\textwidth]{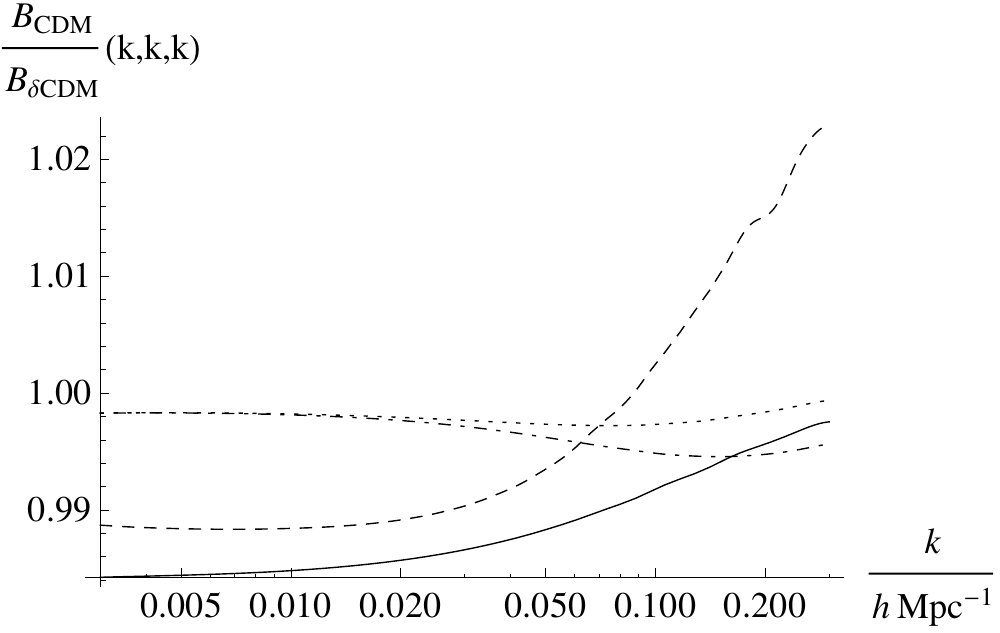}
				\includegraphics[width=0.64\textwidth]{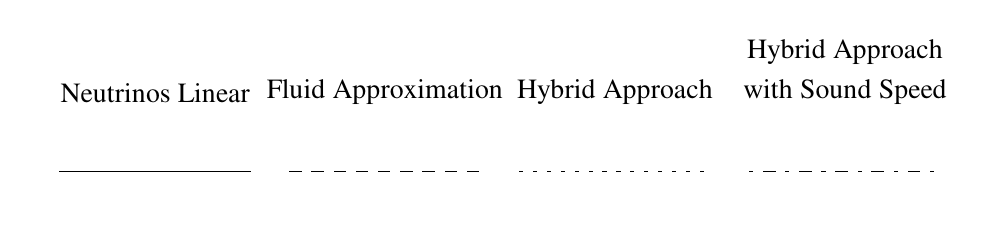}
	\caption{\textit{Top left}: Leading-order equilateral matter bispectrum at $a=1$ for $m=0.93$~eV, computed using various approximations, normalised to that computed from the full theory. \textit{Top right}: The corresponding equilateral neutrino bispectra~$B_\nu(k,k,k)  \equiv \left\langle\delta_{\nu}\delta_{\nu}\delta_{\nu}\right\rangle/(2 \pi)^3$, normalised to the exact result~$B_{\delta \nu}(k,k,k)$. 
	We do not show the neutrino bispectrum computed from the  linear approximation,
		because the error incurred always exceeds $50\%$ on large scales.  
\textit{Bottom}:	The corresponding equilateral CDM bispectra $B_{\rm CDM}(k,k,k)\equiv\langle\delta_{\rm CDM}\delta_{\rm CDM}\delta_{\rm CDM}\rangle/(2 \pi)^3$, again normalised to 
the exact result~$B_{\delta {\rm CDM}}(k,k,k)$.}
	\label{fig:BisSpectrum1}
\end{figure}

\begin{figure}[t]
	\centering
	\includegraphics[width=0.49\textwidth]{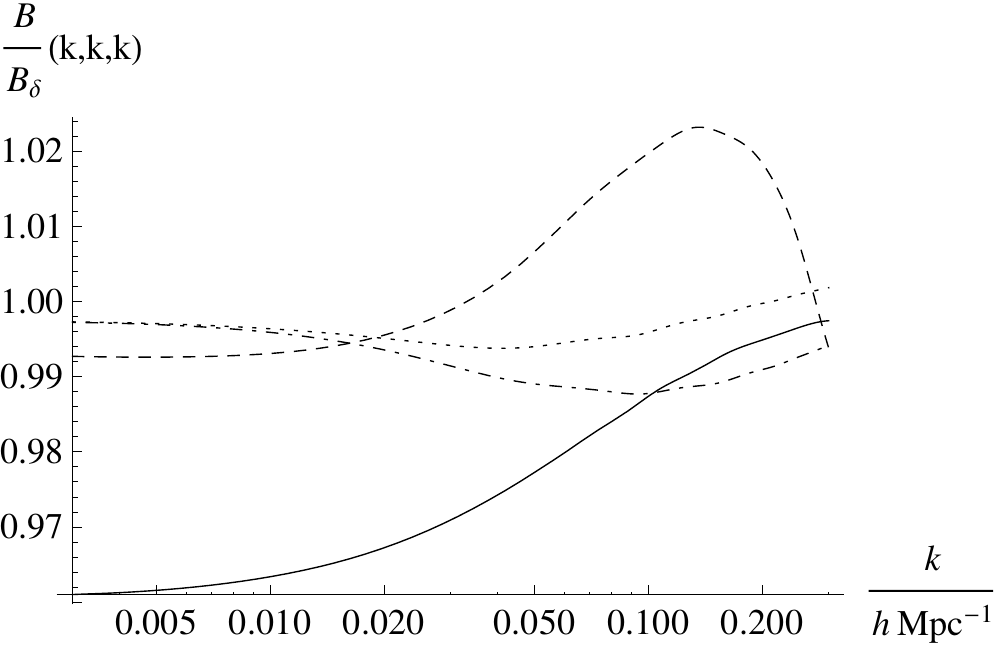}
	\includegraphics[width=0.49\textwidth]{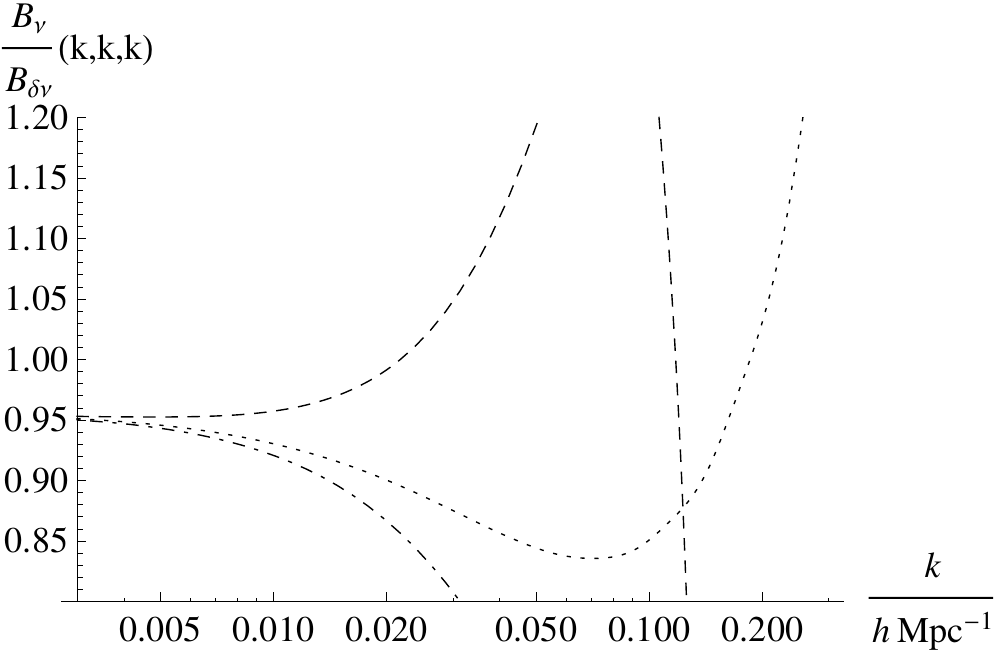}
	\includegraphics[width=0.49\textwidth]{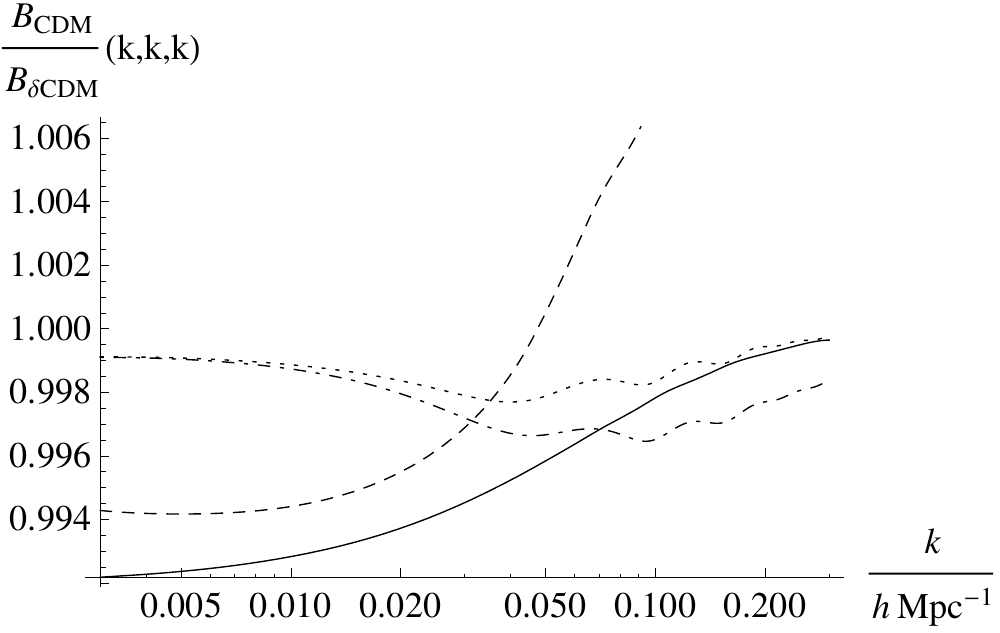}
				\includegraphics[width=0.64\textwidth]{Figures/legend.pdf}
	\caption{Same as figure~\ref{fig:BisSpectrum1}, but for $m=0.46$~eV.}
	\label{fig:BisSpectrum2}
\end{figure}

\begin{figure}[t]
	\centering
		\includegraphics[width=0.49\textwidth]{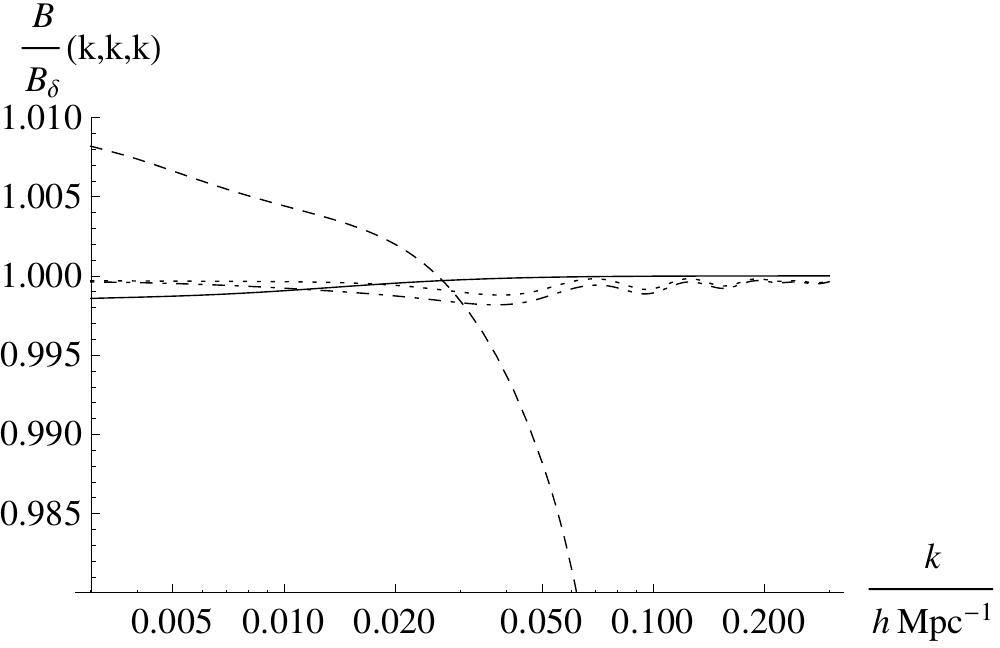}
	\includegraphics[width=0.49\textwidth]{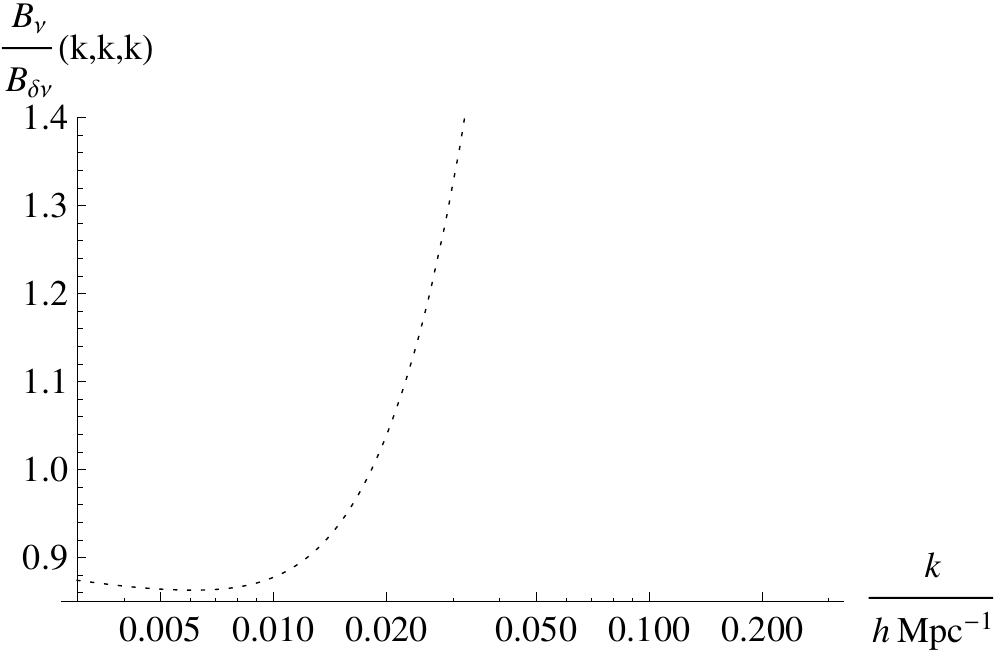}
	\includegraphics[width=0.49\textwidth]{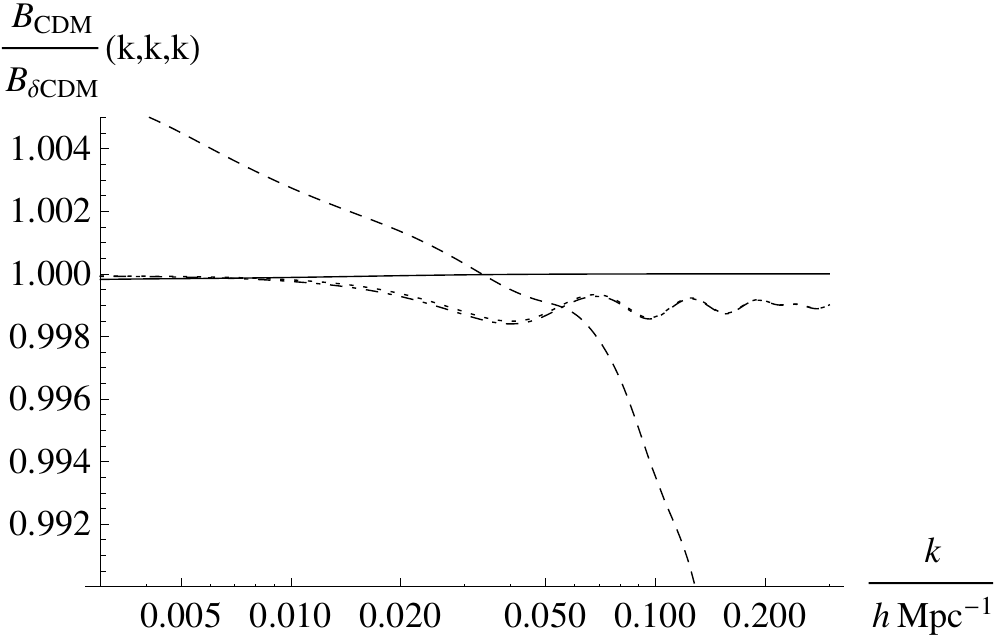}
				\includegraphics[width=0.64\textwidth]{Figures/legend.pdf}
	\caption{Same as figure~\ref{fig:BisSpectrum1}, but for $m=0.046$~eV.  Note that we do not show the neutrino bispectrum computed from the
	 fluid approximation or the hybrid approach with a non-vanishing sound speed, because all scales of interest fall below the free-streaming scale and incur enormous errors.}
	\label{fig:BisSpectrum3}
\end{figure}

The top left panels of figures~\ref{fig:BisSpectrum1} to~\ref{fig:BisSpectrum3}  contrast the leading-order total matter bispectra  computed using various approximations against the exact result
for various neutrino mass values.
  We consider the following approximation schemes:
\begin{itemize}
\item[(i)] {\bf Linear evolution for the neutrino perturbations}. This approximation amounts to neglecting the nonlinear neutrino source term~$S_\nu[\varphi,\delta_\nu; k,s]$  in Gilbert's equations~(\ref{eq:allformal}) for both CDM and neutrinos, so that the equations of motion 
are always linear in the neutrino perturbations. The corresponding leading-order bispectrum is given by equation~(\ref{eq:bispectrumcdmneu}), but formally we set $\Gamma_{\rm s}^{(2)}=\tilde{\Gamma}_{\rm s}^{(1)}=0$.
This linear scheme differs somewhat from that adopted in~\cite{Saito:2008bp,Wong:2008ws,Lesgourgues:2009am} (and their $N$-body analog~\cite{Brandbyge:2008js,Upadhye:2013ndm}), where for the neutrinos all but the linear order  perturbations are set to zero, in that, here, nonlinear coupling of the CDM perturbations can still source higher-order neutrino perturbations. 
It is best compared with the approximation scheme used in the collisionless $N$-body simulations of~\cite{AliHaimoud:2012vj}, where the CDM component is given a particle realisation, while the neutrino perturbations are tracked using the linear equations of motion in the Eulerian frame but with the gravitational potential modified by the nonlinear evolution of the CDM component.

\item[(ii)] {\bf Two-fluid approximation}.  This is the approximation scheme discussed in section~\ref{sec:twofluid}, and the neutrino effective sound speed is chosen to coincide with the 
 velocity dispersion of the unperturbed neutrino momentum distribution~$\overline{q^2}$. 
 The corresponding tree-level matter bispectrum in given by equation~(\ref{eq:fluidbi}). This approximation scheme has previously been used to compute the one-loop matter power spectrum~\cite{Shoji:2009gg}, and can be viewed as a perturbative version of the SPH model of neutrinos investigated in the simulations of~\cite{Hannestad:2011td}.

\item[(iii)]  {\bf Hybrid approach with zero sound speed}.  This approach is discussed in section~\ref{sec:hybrid}, and the leading-order matter bispectrum is formally given by equation~(\ref{eq:fluidbi}), with $c_{\rm s}^2$ set to zero.

\item[(iv)] {\bf Hybrid approach  with a nonzero sound speed}.  Same as above, but with $c_{\rm s}^2$ set to coincide with the 
 velocity dispersion of the unperturbed  neutrino momentum distribution. 
\end{itemize}

Unsurprisingly, the linear approximation~(i) generally provides a poor description of the total matter bispectrum on large scales,  especially for large neutrino masses---the errors exceed 7\% and 4\% for $m=0.93$~eV and 0.46~eV respectively---when the fractional contribution of massive neutrinos to the total matter density becomes large. These large errors in $B(k,k,k)$ come about because for second-order neutrino perturbations, the neutrino nonlinear source terms are in fact larger than the CDM nonlinear source; neglecting them induces an error in the neutrino bispectrum exceeding 50\%.  
On small scales the approximation works well again, as free-streaming suppression causes the already subdominant neutrino perturbations to be even more subdominant on these scales.

For a neutrino mass as small as $0.046$~eV, or equivalently $\Omega_\nu \sim 10^{-3}$, the fractional contribution of massive neutrinos to the total matter density is of order $0.3$\%; in this case the linear approximation appears to reproduce the exact total matter bispectrum on all scales.   It is also useful to note that the error induced in the CDM bispectrum never exceeds 1.6\%, as shown in the bottom panels of figures~\ref{fig:BisSpectrum1} to~\ref{fig:BisSpectrum3}.

Our result suggests that while $N$-body simulations of mixed CDM+neutrino cosmologies that adopt the linear approximation scheme can be useful for calculating nonlinear corrections to the total matter power spectrum on weakly nonlinear scales, they do not work for higher-order $N$-point functions that are intrinsically nonlinear on all scales unless the neutrino mass is exceedingly small.

\medskip

The two-fluid approximation~(ii) reproduces the total matter bispectrum to about $1\to 2$\% accuracy on large scales, and about $2 \to 4$\% at $k\sim 0.1\ h\:  {\rm Mpc}^{-1}$, depending on the neutrino mass, while the error in the corresponding CDM bispectrum is roughly a factor of two smaller.  Again, as with the total matter power spectrum, 
the error incurred in the neutrino fluid by the fluid approximation is in fact huge (see the top right panels of figures~\ref{fig:BisSpectrum1} to~\ref{fig:BisSpectrum3}
for the corresponding bispectra of the neutrino density perturbations); this large error in the neutrino fluid is only masked by the fact that massive neutrinos contribute but a small fraction of the total matter

 Note that the error of the fluid approximation on the total matter bispectrum  appears to ``turn around'' and begin to decrease at  $k\sim 0.1 \to 0.2\ h\:  {\rm Mpc}^{-1}$ for $m = 0.46, 0.93$~eV. The equivalent behaviour can also be seen in the $m=0.046$~eV case but at a larger scale, where the dashed line in the top left plot of figure~\ref{fig:BisSpectrum3} takes a sudden plunge at $k\sim 0.02\ h\:  {\rm Mpc}^{-1}$.  This turnaround, discernible also in the neutrino perturbation bispectra, signals a complete breakdown of the fluid approximation on and below the turnaround scale, and follows from the nonlinear stress term in equations~(\ref{eq:2fluid1}) and~(\ref{eq:2fluid2}) 
overcompensating the usual $\gamma \varphi \varphi$ coupling term.  This overcompensation changes the sign of the nonlinear neutrino source term, which then reduces the net nonlinear source  and causes the two-fluid system to exhibit an acoustic-oscillations-like behaviour {\it in addition to} that already seen at the linear level.

The implications of this result for SPH models of neutrinos in simulations are immediately clear.  We have already seen in section~\ref{sec:linearpowerspectrum} that 
the inherently oscillatory nature of the fluid/sound speed approximation already makes it a less-than-ideal description of the linear evolution of  neutrino perturbations at wavenumbers greater than $k_{\rm FS}$.
Nonlinear evolution enhances this shortcoming, and renders the fluid description poor even for a neutrino mass as small as 0.0046~eV (because all observable $k$-modes in this case 
are greater than $k_{\rm FS}$).
This casts doubts on the usefulness of the SPH model of neutrinos  in simulations.

\medskip

The two hybrid approaches~(iii) and~(iv) are by far the best-performing approximations we have tested in this work, where for the whole $k$~range of interest their respective errors on the total matter bispectra are less than 1\% and 2\%, with the zero-sound-speed version~(iii) as the better performer.  The CDM bispectra are likewise accurate to better than 1\%.  These schemes also fare considerably better than the other approximations in describing the bispectra of the neutrino perturbations up to the free-streaming wavenumber, and although the approach does eventually break down, the breakdown occurs on smaller scales compared with the other schemes.

\medskip

Finally, we remark that although we have assumed a single massive neutrino species in our analysis, 
the main conclusions hold also for three massive neutrino species.  Supposing three degenerate neutrino species of individual mass $m$ instead of one, 
the fraction of free-streaming dark matter goes up by a factor of three.  Consequently, the free-streaming suppression on small scales will be three times larger,
and we expect the errors  incurred in the total matter bispectrum by each approximation scheme to scale up by a factor of three accordingly.
If on the contrary we keep the neutrino density fixed but distribute it equally amongst  three massive neutrino species, 
the free-streaming scale will become three times larger.  While this does not change the free-streaming suppression on small scales and most likely also not the error estimates on those scales, we do expect the the fluid approximation to fail already on scales approximately three times larger.   On the other hand, the linear approximation will hold up to scales three times larger.

\subsection{Nonlinear neutrino density}
\label{sec:nudensity}

Although the majority of future cosmological observations will not be directly sensitive to the neutrino perturbations, we note that new observational techniques have been proposed that make use of the neutrino flow field relative to their CDM counterpart  as a means to measure the neutrino masses~\cite{Zhu:2013tma,Zhu:2014qma}.  In order for these techniques to return physically meaningful constraints, it is essential that we have an accurate way to compute the neutrino perturbations on the nonlinear scales.

For the particular proposal of~\cite{Zhu:2013tma}, the relevant observable quantity is the CDM--neutrino density cross-correlation spectrum, $P_{{\rm C} \nu}(k) \equiv \langle \delta_{\rm C} \delta_\nu \rangle/(2 \pi)^3$.  We have not explicitly calculated the nonlinear corrections to this quantity because the additional time integrals required in the computation of the neutrino
loop corrections are rather difficult to do in comparison with loops in standard perturbation theory.
Nonetheless, because $P_{{\rm C}\nu}$ is directly proportional to the neutrino density perturbations~$\delta_\nu$, we can already glean from the right panel of figure~\ref{fig:BisSpectrum1} and the discussion in section~\ref{sec:bispectrum} that one is likely to grossly misestimate $P_{{\rm C}\nu}$ on scales around and/or below the free-streaming scale using any one of the four approximate methods explored in section~\ref{sec:bispectrum}.  This highlights the need for an exact treatment of nonlinear neutrino perturbations, be it perturbative such as the theory developed in this work, or via $N$-body realisations of the collisionless Boltzmann equation~(\ref{eq:nonrelBoltzmann}).


\section{Conclusions}
\label{sec:conclusions}

We have developed in this work a higher-order perturbation theory for large-scale
structure formation involving a free-streaming hot or warm dark matter species.  The theory 
avoids the need to track the full momentum dependence of the phase space distribution function through 
reformulating the collisionless Boltzmann equation as a nonlinear
generalisation of Gilbert's equation, and is equally applicable to both cases in which the free-streaming dark matter constitutes the dominant or the subdominant 
nonrelativistic energy density.  We have applied our theory to calculate the leading-order total matter bispectrum in CDM+massive neutrino cosmologies with various neutrino masses,
 and because our theory
does not assume $f_\nu/f_{\rm C}\ll 1$, we have been able compute the leading-order bispectrum of the neutrino density perturbations as well.

Using the leading-order bispectrum as a benchmark, we  examined the validity of the fluid/SPH approximation and a linear approximation scheme
previously used in various perturbative analyses and $N$-body simulations of mixed CDM+massive neutrino cosmologies.
Along with these existing approximate schemes, we also tested a hybrid approach proposed in this work, which combines the exact linear evolution of the
free-streaming particles together with the nonlinear coupling structure of the fluid equations. 

Demanding an accuracy of 1\% or better for the total matter bispectrum, we found that only the hybrid approach is able to reproduce the exact result for the whole range of neutrino masses tested  ($m = 0.0046 \to 0.93$~eV).  The fluid approximation performs badly for the entire neutrino mass range, while the linear approximation fails on large scales when the neutrino mass becomes large.  Since these last two approximation schemes were previously adopted in $N$-body simulations of mixed CDM+massive neutrino cosmologies and our investigations here represent their perturbative limits, our results also serve as a 
cautionary note: approximate nonlinear models of neutrino clustering that reproduce the gross features of some observables may not ultimately be  sufficient for precision calculations, nor does their (approximate) validity necessarily extend to other observables.

In contrast, none of the approximation schemes is able to reproduce the bispectrum of the neutrino density perturbations to an accuracy better than $20\%$ across all scales.  
This is potentially problematic for proposed new observational techniques that aim to measure neutrino masses via the relative flow field of neutrinos and CDM, and strongly suggests
the need for an exact treatment of nonlinear neutrino perturbations such as the perturbative theory developed in this work, or via $N$-body solutions of the collisionless Boltzmann equation.
In regard to the former, we expect that adopting a more efficient algorithm for the evaluation of the time integrals would aid in the computation of loop corrections that become important on small scales.  Such a development would also allow us to calculate nonlinear corrections to density correlators in warm dark matter-only cosmologies.

Another possible direction is to develop new approximation schemes that could
reproduce the suppression of power on small scales, the main feature of
free-streaming particles. The hybrid approach proposed in this paper already
takes into account the suppression in the linear evolution. The full theory developed in
this work can be used as a starting point to improve the hybrid approach or to
develop new approximation schemes.

\acknowledgments
A part of this work contributed to the master thesis of FF~at RWTH Aachen University.  FF acknowledges support from the IMPRS-PTFS and the DFG through the TRR33 project ``The Dark Universe''.


\appendix

\section{Kernels for higher-order neutrino perturbations}
\subsection{Generalisation to higher orders}
\label{sec:APkernel}
In order to to derive a general expression for the higher-order kernels, it is instructive to first evaluate equation~(\ref{eq:deltaformalsolInsecond}) by brute force to the next order.  This exercise gives 
\begin{equation}
\begin{aligned}
\tilde{\Gamma}^{(2)} &\equiv  \Theta(s_1-s_2) 
V(s_1) V(s_2) \int  d^3 q \: 
\bm{v}_2 \cdot \frac{\partial}{\partial \bm{q}} 
\left(\bm{v}_1 \cdot \frac{\partial \tilde{g}\left(\bm{k},\bm{q};s',s_1\right)}{\partial \bm{q}}   \tilde{g}(\bm{k}_2+\bm{k}_3,\bm{q}; s_1,s_2)\right) \\
& \hspace{30mm} \times  \tilde{g}(\bm{k}_3,\bm{q}; s_2,s_{\rm in}) \delta f\left(\bm{k}_3,\bm{q},s_{\rm in} \right) \\
&=   \Theta(s_1-s_2) 
\frac{9}{4}a^2(s_1) \mathcal{H}^2(s_1) \Omega(s_1)a^2(s_2) \mathcal{H}^2(s_2) \Omega(s_2)  \frac{\bm{k}_2}{k_2^2} \cdot \bm{U}^3_2   \frac{\bm{k}_1}{k_1^2} \cdot \bm{U}^3_1  \\
& \hspace{30mm} \times \int
d^3q\,e^{-i \bm{U}^3_3\cdot\bm{q}/m}\delta
f(\bm{k}_3,\bm{q},s_{\rm in}), \\
\Gamma^{(3)}& = - \Theta(s_1-s_2)  \Theta(s_2-s_3)V(s_1) V(s_2)V(s_3)  \int d^3q \: \bar{f}(q) \\
& \hspace{10mm} \times \bm{v}_3 \cdot  \frac{\partial}{\partial \bm{q}} \left( \bm{v}_2 \cdot 
\frac{\partial}{\partial \bm{q}} \left( \bm{v}_1 \cdot \frac{\partial \tilde{g}\left(\bm{k}, \bm{q};s',s_1\right)}{\partial \bm{q}}  \tilde{g}\left(\bm{k}_2+\bm{k}_3,\bm{q}; s_1,s_2\right)  \right)
\tilde{g}\left(\bm{k}_3,\bm{q}; s_2,s_3\right)  \right)\\
& =  \Theta(s_1-s_2)  \Theta(s_2-s_3) \frac{27}{8}a^2(s_1) \mathcal{H}^2(s_1) \Omega(s_1)  a^2(s_2)\mathcal{H}^2(s_2) \Omega(s_2) a^2(s_3)\mathcal{H}^2(s_3) \Omega(s_3) 
\\
& \hspace{10mm}\times  \frac{\bm{k}_3}{k_3^2}\cdot \bm{U}^3_3\frac{\bm{k}_2}{k_2^2}\cdot \bm{U}^3_2 \frac{\bm{k}_1}{k_1^2} \cdot  \bm{U}^3_1  \int
d^3q\,e^{-i \bm{U}^3_3 \cdot\bm{q}/m}
\bar{f}(q),
\end{aligned}
\end{equation}
where 
\begin{equation}
\begin{aligned}
\bm{U}^3_1 &\equiv (\bm{k}_1+\bm{k}_2 + \bm{k}_3) (s'-s_1), \\
\bm{U}^3_2 & \equiv  (\bm{k}_1+\bm{k}_2 + \bm{k}_3) (s'-s_1) +  (\bm{k}_2 + \bm{k}_3) (s_1-s_2),\\ 
\bm{U}^3_3 & \equiv  (\bm{k}_1+\bm{k}_2 + \bm{k}_3) (s'-s_1) +  (\bm{k}_2 + \bm{k}_3) (s_1-s_2) + \bm{k}_3(s_2-s_3),
\end{aligned}
\end{equation}
with the understanding that $s_j$  is always integrated over the interval $[s_{\rm in}, s']$, and $s_3$ is identified with $s_{\rm in}$ in the case of $\tilde{\Gamma}^{(2)}$.

A simple pattern emerges when we compare these expressions with equations~(\ref{eq:gammatilde1}) and~(\ref{eq:gamma2}) for $\tilde{\Gamma}^{(1)}$ and $\Gamma^{(2)}$, reproduced here for clarity:
\begin{equation}
\begin{aligned}
\tilde{\Gamma}^{(1)} & \equiv -V(s_1) \int  d^3 q \: \bm{v}_1 \cdot \frac{\partial \tilde{g}\left(\bm{k},\bm{q};s',s_1\right)}{\partial \bm{q}}   \tilde{g}(\bm{k}_2,\bm{q}; s_1,s_{\rm in})
\delta f\left(\bm{k}_2,\bm{q},s_{\rm in} \right) \\
& = \frac{3}{2}a^2(s_1) \mathcal{H}^2(s_1)\Omega(s_1)  \frac{\bm{k}_1}{k_1^2} \cdot \bm{U}^2_1 \int
d^3q\,e^{-i \bm{U}^2_2\cdot\bm{q}/m}\delta
f(\bm{k}_2,\bm{q},s_{\rm in}), \\
\Gamma^{(2)} &=   \Theta(s_1-s_2)  V(s_1) V(s_2) \int d^3q \; \bm{v}_2 \cdot 
\frac{\partial}{\partial \bm{q}} \left( \bm{v}_1 \cdot \frac{\partial \tilde{g}\left(\bm{k}, \bm{q};s',s_1\right)}{\partial \bm{q}}  \tilde{g}\left(\bm{k}_2,\bm{q}; s_1,s_2\right)  \right) 
 \bar{f}(q)\\
&= \Theta(s_1-s_2)  \frac{9}{4}a^2(s_1) \mathcal{H}^2(s_1) \Omega(s_1) a^2(s_2)\mathcal{H}^2(s_2) \Omega(s_2) \frac{\bm{k}_2}{k_2^2}\cdot \bm{U}^2_2 \frac{\bm{k}_1}{k_1^2} \cdot  \bm{U}^2_1 \! \int \!
d^3q\,e^{-i \bm{U}^2_2 \cdot\bm{q}/m}
\bar{f}(q),
\end{aligned}
\end{equation}
with
\begin{equation}
\begin{aligned}
\bm{U}^2_1 &\equiv (\bm{k}_1+\bm{k}_2)(s'-s_1),\\
\bm{U}^2_2 &\equiv (\bm{k}_1+\bm{k}_2)(s'-s_1)+\bm{k}_2(s_1-s_2), 
\end{aligned}
\end{equation}
and the identification $s_2 \equiv s_{\rm in}$  in the case of  $\tilde{\Gamma}^{(1)}$.  At the $n$th order, both classes of kernels  
take $n$ factors of $(3/2) a^2(s_j) \mathcal{H}^2(s_j)$, each evaluated at $n$ different times, $s_1,s_2, \ldots, s_n$, labelled such that $s_{j-1}>s_j$.
There are likewise $n$ factors of $(\bm{k}_j/k_j^2) \cdot \bm{U}^n_j$, where the vectors $\bm{U}^n_j$ arise from the $n$ nested $\bm{q}$-derivatives of the free-streaming 
solution~$\tilde{g}$.  For example, $\bm{U}^3_1$ corresponds to differentiating one~$\tilde{g}$ connecting $s_1$ to $s'$, 
while $\bm{U}^3_3$ represents differentiation of three disparate~$\tilde{g}$'s connecting $s_j$  to $s_{j-1}$, where $j=1,2,3$ ($s_0$ is identified with $s'$).
The exponential that forms part of the $\bm{q}$-integrand collects all of the $n$ (or $n+1$ for $\tilde{\Gamma}^{(n)}$) free-streaming solutions appearing at the corresponding order.  From here it is easy to see that the general forms of $\tilde{\Gamma}^{(n-1)}$ and $\Gamma^{(n)}$ are indeed given by equations~(\ref{eq:Kernel11}) and~(\ref{eq:KernelInitial}).


\subsection{Multipole expansion}
\label{sec:ApMultipole}
To decompose $\tilde{\Gamma}^{(n-1)}$ in terms of multipole moments of initial distribution function, we first note that a plane wave can be expanded as
\begin{align}
e^{-i \bm{x} \cdot \bm{y}}= \sum_{\ell=0}^{\infty}(-i)^\ell (2\ell +1)j_\ell \left(xy\right)P_\ell(\hat{\bm{x}} \cdot \hat{\bm{y}}),
\end{align}
where $\hat{\bm{x}} \equiv \bm{x}/x$ denotes a unit vector, $j_\ell(x)$ is the spherical Bessel function order order~$\ell$, and $P_\ell(\mu)$ is the $\ell$th Legendre polynomial.
Inserting this into the expression~(\ref{eq:KernelInitial}) for $\tilde{\Gamma}^{(n-1)}$ yields
\begin{equation}
\begin{aligned}
\tilde{\Gamma}^{(n-1)}=&\left(\prod_{j=1}^{n-1}
 \Theta(s_{j-1} - s_j)
 \frac{3 }{2}
a^2(s_j)\mathcal{H}^2(s_j) \Omega(s_j) \frac{\bm{k}_j}{k_j^2}\cdot
\bm{U}^{n}_j\right) \sum_{\ell=0}^\infty \sum_{\ell'=0}^\infty 2
(-i)^{\ell'+\ell} (2 \ell+1)(2 \ell'+1) \\
& \hspace{10mm} \times \frac{1}{2} \int d^3q \: j_\ell \left( \frac{U^{n}_{n} q}{m} \right) f_{\ell'}(\bm{k}_{n},q,s_{\rm in}) P_\ell \left( \hat{\bm{U}}_{n}^{n}
\cdot \hat{\bm{q}} \right)
P_{\ell'} \left(\hat{\bm{k}}_{n} \cdot \hat{\bm{q}} \right),
\end{aligned}
\end{equation}
where we have also made use of the Legendre decomposition~(\ref{eq:decompose}) for the initial distribution function~$\delta f(\bm{k}_{n},\bm{q},s_{\rm in})$.
Then, applying  the addition theorem
\begin{align}
P_\ell(\hat{\bm{x}} \cdot \hat{\bm{y}})=\frac{4\pi}{2 \ell+1}\sum_{m=-\ell}^\ell
Y_{\ell m}(\hat{\bm{x}})Y^*_{\ell m}(\hat{\bm{y}})
\end{align}
to the two Legendre polynomials, followed by the orthonormality condition for the spherical harmonics~$Y_{\ell m}( \hat{\bm{x}})$,
\begin{align}
\int d \Omega_{\bm{q}} \: Y_{\ell m}(\hat{\bm{q}})Y^*_{\ell' m'}(\hat{\bm{q}}) = \delta_{\ell \ell'}  \delta_{m m'},
\end{align}
we find the general expression
\begin{equation}
\begin{aligned}
\tilde{\Gamma}^{(n-1)}=&\left(\prod_{j=1}^{n-1} \Theta(s_{j-1} - s_j) \frac{3 }{2}
a^2(s_j)\mathcal{H}^2(s_j) \Omega(s_j) \frac{\bm{k}_j}{k_j^2}\cdot
\bm{U}^{n}_j\right)  \\
& \hspace{2mm}\times \sum_{\ell=0}^\infty  4 \pi (-1)^\ell (2 \ell+1) P_\ell
\left( \hat{\bm{k}}_{n} \cdot \hat{\bm{U}}_{n}^{n}\right) \int dq \: q^2 j_\ell \left( \frac{U^{n}_{n} q}{m} \right) f_{\ell}(\bm{k}_{n},q,s_{\rm in}).
\end{aligned}
\end{equation}
The initial multipole moments~$f_\ell(k_{n},q,s_{\rm in})$ can be extracted from a (linear) Boltzmann code such as COSMICS~\cite{Bertschinger1995}.


\section{Numerical solution of linear integral equations}
\label{sec:nystroem}

We use the Nystr\"om method to solve Gilbert's equation~(\ref{eq:gilbert}) numerically.
For more details, see~\cite{Delves1992}.
Gilbert's equation can be written as a Fredholm equation,
\begin{align} 
\delta \left( s \right) -\int ^{s_{\rm fi}}_{s_{\rm in}} ds'\: \tilde{K}\left(s,s'\right)  \delta \left( s' \right)=I\left( s \right)\mathrm{,} \label{eq:Fredholm}
\end{align}
where $I(s)$ is the same source term defined in equation~(\ref{eq:source}), and 
\begin{align}
\tilde{K}(s,s')=K\left(s,s'\right) \Theta(s-s'),
\end{align}
with $K(s,s')$ given by equation~(\ref{eq:kkernel}).
Note that $\tilde{K}(s,s')$ is continuous, since $\underset{{s\rightarrow
s'}}{\lim}K(s,s')=\underset{{s'\rightarrow s}}{\lim} K(s,s')=0$.
As with differential equations, the Fredholm equation can be solved numerically by discretising the time variable $s$ on $N$~nodes $s_n$, $n=1,\ldots, N$. Then, discretising equation~\eqref{eq:Fredholm} accordingly, we find the matrix equation
\begin{align}
\label{eq:matrixkkk}
\delta \left( s_n \right) -\sum ^{N}_{m=1} w_{nm} \tilde{K}\left(s_n,s_m\right) \delta \left( s_m \right)=I\left( s_n \right),
\end{align}
where $w_{nm}$ denotes the integration weights. In matrix notation equation~(\ref{eq:matrixkkk}) reads
\begin{align}
\left(\bm{1}-\hat{\bm{K}}\right)\cdot \bm{\delta}=\bm{I},
\end{align}
where we have defined
\begin{align}
\hat{K}_{mn}=w_{mn}\tilde{K}(s_n,s_m).
\end{align}
The solution then can be written as
\begin{align}
\bm{\delta}=\left(\bm{1}-\hat{\bm{K}}\right)^{-1}\cdot\bm{I}\equiv\bm{\hat{G}}\cdot\bm{I}\label{eq:GilberSolutionNum},
\end{align}
so that the problem of determining $\delta(s_n)$ reduces to that of a matrix inversion, for which several numerical linear algebra algorithms exist.  Since equation \eqref{eq:GilberSolutionNum} can also be written as
\begin{align}
\delta(s_n)=\sum_{m=1}^N \hat{G}_{nm} I(s_m)=\sum_{m=1}^N G_{nm} w_{nm} I(s_m)\approx \int_{s_{\rm in}}^{s_{\rm fi}} ds'\:G(s_n,s')I(s'),
\end{align}
where
\begin {align}
G(s_n,s_m)\approx G_{nm}=\begin{cases}
   \frac{\hat{G}_{nm}}{w_{nm}}&\rm{if}\quad s_n\geq s_m\\
    0&\rm{if}\quad s_n<s_m\ 
  \end{cases} \, ,
\end{align}
we see from a comparison with equation~(\ref{eq:solutionGilbert}) that this numerical procedure also automatically yields the Green's function~$G(s,s')$.

The integration weights~$w_{nm}$ together with the time nodes $s_n$ are determined by the integration rule.  We list here several possibilities.
\begin{enumerate}
\item  {\it Riemann sum.}~~This is the simplest choice, with equidistant nodes and equal weights, i.e.,
\begin{equation}
\begin{aligned}
s_n&=s_{\rm i}+\frac{s_{\rm fi}-s_{\rm in}}{N}(n-1),\\
w_{nm}&=\frac{s_{\rm fi}-s_{\rm in}}{N}.
\end{aligned}
\end{equation}

\item {\it Trapezoidal rule.}~~A somewhat better choice than the Riemann sum, the time nodes are likewise equidistant, but the weights differ at the extremities: 
\begin{equation}
\begin{aligned}
s_n&=s_{\rm in}+\frac{s_{\rm fi}-s_{\rm in}}{N-1}(n-1), \\
w_{nm}&= \left\{
	\begin{array}{ll}
	\frac{1}{2}\frac{s_{\rm fi}-s_{\rm in}}{N-1}  & \quad \text{if}\ m=1,N \\
	\frac{s_{\rm fi}-s_{\rm in}}{N-1} & \quad \text{otherwise}	
	\end{array} \right. .
\end{aligned}
\end{equation}

\item {\it Gaussian quadrature rule.}~~The continuity of $\tilde{K}(s,s')$ renders the evaluation of the integral also amenable to this integration rule.
Here, the time nodes are given by
\begin{align}
s_n=\frac{s_{\rm in}+s_{\rm fi}}{2}+\frac{s_{\rm in}-s_{\rm fi}}{2} x_k\mathrm{,}
\end{align}
where $x_n$ ($n=1,\ldots,N$) denotes the $N$~zeros of an orthogonal polynomial of degree~$N$ in the interval $[-1,1]$. The weights are given by
\begin{align}
\label{eq:weights}
w_{nm}=w_n=\int^1_{-1} dx \: L_n\left(x \right)\mathrm{,}
\end{align}
where $L_n$ is a polynomial of degree $N$ defined by 
\begin{align}
L_n\left(x\right)=\prod^{N}_{\stackrel{k=1}{k\neq n}} \frac{x-x_k}{x_n-x_k}\mathrm{,}
\end{align}
from which one sees immediately that $L_n(x_n)=1$, and $L_n(x_k)=0$ for all other nodes.  In the case of a Gauss--Legendre quadrature, which we use here, the integral~(\ref{eq:weights}) for weights~$w_n$ reduces to
\begin{align}
w_n=\frac{2}{(1-x_n^2)(P_N'(x_n))^2}\, ,
\end{align}  
where $P_N(x)$ is the $N$th Legendre polynomial, and $P_N'$ its derivative with respect to $x$.
\end{enumerate}

In this work we use a quadrature rule with $200$~time nodes.  Comparing the
results with exactly solvable cases this setting gives an error of order $10^{-4}$, while varying the number of nodes by $\pm 100$ induces a relative difference of order $10^{-5} \to 10^{-6}$.
Adopting instead the trapezoidal rule with the same number of time nodes leads to a similar performance.


\bibliographystyle{utcaps}

\bibliography{newrefs}

\providecommand{\href}[2]{#2}\begingroup\raggedright\begin{thebibliography}{10}

\bibitem{bib:pdg}
{\bfseries Particle Data Group} Collaboration, J.~Beringer {\em et al.},
  ``{Review of Particle Physics (RPP)},''
\href{http://dx.doi.org/10.1103/PhysRevD.86.010001}{{\em Phys.Rev.} {\bfseries
  D86} (2012)  010001}.

\bibitem{Kraus:2004zw}
C.~Kraus, B.~Bornschein, L.~Bornschein, J.~Bonn, B.~Flatt, {\em et al.},
  ``{Final results from phase II of the Mainz neutrino mass search in tritium
  beta decay},'' \href{http://dx.doi.org/10.1140/epjc/s2005-02139-7}{{\em
  Eur.Phys.J.} {\bfseries C40} (2005)  447--468},
\href{http://arxiv.org/abs/hep-ex/0412056}{{\ttfamily arXiv:hep-ex/0412056
  [hep-ex]}}.

\bibitem{Lobashev:1999tp}
V.~Lobashev, V.~Aseev, A.~Belesev, A.~Berlev, E.~Geraskin, {\em et al.},
  ``{Direct search for mass of neutrino and anomaly in the tritium beta
  spectrum},''
\href{http://dx.doi.org/10.1016/S0370-2693(99)00781-9}{{\em Phys.Lett.}
  {\bfseries B460} (1999)  227--235}.

\bibitem{Lesgourgues:2006nd}
J.~Lesgourgues and S.~Pastor, ``{Massive neutrinos and cosmology},''
  \href{http://dx.doi.org/10.1016/j.physrep.2006.04.001}{{\em Phys.Rept.}
  {\bfseries 429} (2006)  307--379},
\href{http://arxiv.org/abs/astro-ph/0603494}{{\ttfamily arXiv:astro-ph/0603494
  [astro-ph]}}.

\bibitem{Hannestad:2006zg}
S.~Hannestad, ``{Primordial neutrinos},''
  \href{http://dx.doi.org/10.1146/annurev.nucl.56.080805.140548}{{\em
  Ann.Rev.Nucl.Part.Sci.} {\bfseries 56} (2006)  137--161},
\href{http://arxiv.org/abs/hep-ph/0602058}{{\ttfamily arXiv:hep-ph/0602058
  [hep-ph]}}.

\bibitem{Wong:2011ip}
Y.~Y. Wong, ``{Neutrino mass in cosmology: status and prospects},''
  \href{http://dx.doi.org/10.1146/annurev-nucl-102010-130252}{{\em
  Ann.Rev.Nucl.Part.Sci.} {\bfseries 61} (2011)  69--98},
\href{http://arxiv.org/abs/1111.1436}{{\ttfamily arXiv:1111.1436
  [astro-ph.CO]}}.

\bibitem{Abazajian:2011dt}
K.~Abazajian, E.~Calabrese, A.~Cooray, F.~De~Bernardis, S.~Dodelson, {\em et
  al.}, ``{Cosmological and Astrophysical Neutrino Mass Measurements},''
  \href{http://dx.doi.org/10.1016/j.astropartphys.2011.07.002}{{\em
  Astropart.Phys.} {\bfseries 35} (2011)  177--184},
\href{http://arxiv.org/abs/1103.5083}{{\ttfamily arXiv:1103.5083
  [astro-ph.CO]}}.

\bibitem{Lesgourgues:2014zoa}
J.~Lesgourgues and S.~Pastor, ``{Neutrino cosmology and Planck},''
\href{http://arxiv.org/abs/1404.1740}{{\ttfamily arXiv:1404.1740 [hep-ph]}}.

\bibitem{Hamann:2012fe}
J.~Hamann, S.~Hannestad, and Y.~Y. Wong, ``{Measuring neutrino masses with a
  future galaxy survey},''
  \href{http://dx.doi.org/10.1088/1475-7516/2012/11/052}{{\em JCAP} {\bfseries
  1211} (2012)  052},
\href{http://arxiv.org/abs/1209.1043}{{\ttfamily arXiv:1209.1043
  [astro-ph.CO]}}.

\bibitem{Audren:2012vy}
B.~Audren, J.~Lesgourgues, S.~Bird, M.~G. Haehnelt, and M.~Viel, ``{Neutrino
  masses and cosmological parameters from a Euclid-like survey: Markov Chain
  Monte Carlo forecasts including theoretical errors},''
  \href{http://dx.doi.org/10.1088/1475-7516/2013/01/026}{{\em JCAP} {\bfseries
  1301} (2013)  026},
\href{http://arxiv.org/abs/1210.2194}{{\ttfamily arXiv:1210.2194
  [astro-ph.CO]}}.

\bibitem{Bernardeau:2001qr}
F.~Bernardeau, S.~Colombi, E.~Gaztanaga, and R.~Scoccimarro, ``{Large scale
  structure of the universe and cosmological perturbation theory},''
  \href{http://dx.doi.org/10.1016/S0370-1573(02)00135-7}{{\em Phys.Rept.}
  {\bfseries 367} (2002)  1--248},
\href{http://arxiv.org/abs/astro-ph/0112551}{{\ttfamily arXiv:astro-ph/0112551
  [astro-ph]}}.

\bibitem{Crocce:2005xy}
M.~Crocce and R.~Scoccimarro, ``{Renormalized cosmological perturbation
  theory},'' \href{http://dx.doi.org/10.1103/PhysRevD.73.063519}{{\em
  Phys.Rev.} {\bfseries D73} (2006)  063519},
\href{http://arxiv.org/abs/astro-ph/0509418}{{\ttfamily arXiv:astro-ph/0509418
  [astro-ph]}}.

\bibitem{Crocce:2005xz}
M.~Crocce and R.~Scoccimarro, ``{Memory of initial conditions in gravitational
  clustering},'' \href{http://dx.doi.org/10.1103/PhysRevD.73.063520}{{\em
  Phys.Rev.} {\bfseries D73} (2006)  063520},
\href{http://arxiv.org/abs/astro-ph/0509419}{{\ttfamily arXiv:astro-ph/0509419
  [astro-ph]}}.

\bibitem{McDonald:2006hf}
P.~McDonald, ``{Dark matter clustering: a simple renormalization group
  approach},'' \href{http://dx.doi.org/10.1103/PhysRevD.75.043514}{{\em
  Phys.Rev.} {\bfseries D75} (2007)  043514},
\href{http://arxiv.org/abs/astro-ph/0606028}{{\ttfamily arXiv:astro-ph/0606028
  [astro-ph]}}.

\bibitem{Matsubara:2007wj}
T.~Matsubara, ``{Resumming Cosmological Perturbations via the Lagrangian
  Picture: One-loop Results in Real Space and in Redshift Space},''
  \href{http://dx.doi.org/10.1103/PhysRevD.77.063530}{{\em Phys.Rev.}
  {\bfseries D77} (2008)  063530},
\href{http://arxiv.org/abs/0711.2521}{{\ttfamily arXiv:0711.2521 [astro-ph]}}.

\bibitem{Matarrese:2007wc}
S.~Matarrese and M.~Pietroni, ``{Resumming Cosmic Perturbations},''
  \href{http://dx.doi.org/10.1088/1475-7516/2007/06/026}{{\em JCAP} {\bfseries
  0706} (2007)  026},
\href{http://arxiv.org/abs/astro-ph/0703563}{{\ttfamily arXiv:astro-ph/0703563
  [astro-ph]}}.

\bibitem{Baumann:2010tm}
D.~Baumann, A.~Nicolis, L.~Senatore, and M.~Zaldarriaga, ``{Cosmological
  Non-Linearities as an Effective Fluid},''
  \href{http://dx.doi.org/10.1088/1475-7516/2012/07/051}{{\em JCAP} {\bfseries
  1207} (2012)  051},
\href{http://arxiv.org/abs/1004.2488}{{\ttfamily arXiv:1004.2488
  [astro-ph.CO]}}.

\bibitem{Pietroni:2011iz}
M.~Pietroni, G.~Mangano, N.~Saviano, and M.~Viel, ``{Coarse-Grained
  Cosmological Perturbation Theory},''
  \href{http://dx.doi.org/10.1088/1475-7516/2012/01/019}{{\em JCAP} {\bfseries
  1201} (2012)  019},
\href{http://arxiv.org/abs/1108.5203}{{\ttfamily arXiv:1108.5203
  [astro-ph.CO]}}.

\bibitem{Saito:2008bp}
S.~Saito, M.~Takada, and A.~Taruya, ``{Impact of massive neutrinos on nonlinear
  matter power spectrum},''
  \href{http://dx.doi.org/10.1103/PhysRevLett.100.191301}{{\em Phys.Rev.Lett.}
  {\bfseries 100} (2008)  191301},
\href{http://arxiv.org/abs/0801.0607}{{\ttfamily arXiv:0801.0607 [astro-ph]}}.

\bibitem{Wong:2008ws}
Y.~Y. Wong, ``{Higher order corrections to the large scale matter power
  spectrum in the presence of massive neutrinos},''
  \href{http://dx.doi.org/10.1088/1475-7516/2008/10/035}{{\em JCAP} {\bfseries
  0810} (2008)  035},
\href{http://arxiv.org/abs/0809.0693}{{\ttfamily arXiv:0809.0693 [astro-ph]}}.

\bibitem{Lesgourgues:2009am}
J.~Lesgourgues, S.~Matarrese, M.~Pietroni, and A.~Riotto, ``{Non-linear Power
  Spectrum including Massive Neutrinos: the Time-RG Flow Approach},''
  \href{http://dx.doi.org/10.1088/1475-7516/2009/06/017}{{\em JCAP} {\bfseries
  0906} (2009)  017},
\href{http://arxiv.org/abs/0901.4550}{{\ttfamily arXiv:0901.4550
  [astro-ph.CO]}}.

\bibitem{Shoji:2009gg}
M.~Shoji and E.~Komatsu, ``{Third-order Perturbation Theory With Non-linear
  Pressure},'' \href{http://dx.doi.org/10.1088/0004-637X/700/1/705}{{\em
  Astrophys.J.} {\bfseries 700} (2009)  705--719},
\href{http://arxiv.org/abs/0903.2669}{{\ttfamily arXiv:0903.2669
  [astro-ph.CO]}}.

\bibitem{Shoji:2010hm}
M.~Shoji and E.~Komatsu, ``{Massive Neutrinos in Cosmology: Analytic Solutions
  and Fluid Approximation},''
  \href{http://dx.doi.org/10.1103/PhysRevD.81.123516,
  10.1103/PhysRevD.82.089901}{{\em Phys.Rev.} {\bfseries D81} (2010)  123516},
\href{http://arxiv.org/abs/1003.0942}{{\ttfamily arXiv:1003.0942
  [astro-ph.CO]}}.

\bibitem{Blas2014}
D.~Blas, M.~Garny, T.~Konstandin, and J.~Lesgourgues, ``{Structure formation
  with massive neutrinos: going beyond linear theory},''
\href{http://arxiv.org/abs/1408.2995}{{\ttfamily arXiv:1408.2995
  [astro-ph.CO]}}.

\bibitem{Ringwald:2004np}
A.~Ringwald and Y.~Y. Wong, ``{Gravitational clustering of relic neutrinos and
  implications for their detection},''
  \href{http://dx.doi.org/10.1088/1475-7516/2004/12/005}{{\em JCAP} {\bfseries
  0412} (2004)  005},
\href{http://arxiv.org/abs/hep-ph/0408241}{{\ttfamily arXiv:hep-ph/0408241
  [hep-ph]}}.

\bibitem{Dupuy:2013jaa}
H.~Dupuy and F.~Bernardeau, ``{Describing massive neutrinos in cosmology as a
  collection of independent flows},''
\href{http://arxiv.org/abs/1311.5487}{{\ttfamily arXiv:1311.5487
  [astro-ph.CO]}}.

\bibitem{Brandbyge:2008rv}
J.~Brandbyge, S.~Hannestad, T.~Haugbolle, and B.~Thomsen, ``{The Effect of
  Thermal Neutrino Motion on the Non-linear Cosmological Matter Power
  Spectrum},'' \href{http://dx.doi.org/10.1088/1475-7516/2008/08/020}{{\em
  JCAP} {\bfseries 0808} (2008)  020},
\href{http://arxiv.org/abs/0802.3700}{{\ttfamily arXiv:0802.3700 [astro-ph]}}.

\bibitem{Villaescusa-Navarro:2013pva}
F.~Villaescusa-Navarro, F.~Marulli, M.~Viel, E.~Branchini, E.~Castorina, {\em
  et al.}, ``{Cosmology with massive neutrinos I: towards a realistic modeling
  of the relation between matter, haloes and galaxies},''
  \href{http://dx.doi.org/10.1088/1475-7516/2014/03/011}{{\em JCAP} {\bfseries
  1403} (2014)  011},
\href{http://arxiv.org/abs/1311.0866}{{\ttfamily arXiv:1311.0866
  [astro-ph.CO]}}.

\bibitem{Brandbyge:2008js}
J.~Brandbyge and S.~Hannestad, ``{Grid Based Linear Neutrino Perturbations in
  Cosmological N-body Simulations},''
  \href{http://dx.doi.org/10.1088/1475-7516/2009/05/002}{{\em JCAP} {\bfseries
  0905} (2009)  002},
\href{http://arxiv.org/abs/0812.3149}{{\ttfamily arXiv:0812.3149 [astro-ph]}}.

\bibitem{Upadhye:2013ndm}
A.~Upadhye, R.~Biswas, A.~Pope, K.~Heitmann, S.~Habib, {\em et al.},
  ``{Large-Scale Structure Formation with Massive Neutrinos and Dynamical Dark
  Energy},'' \href{http://dx.doi.org/10.1103/PhysRevD.89.103515}{{\em
  Phys.Rev.} {\bfseries D89} (2014)  103515},
\href{http://arxiv.org/abs/1309.5872}{{\ttfamily arXiv:1309.5872
  [astro-ph.CO]}}.

\bibitem{AliHaimoud:2012vj}
Y.~Ali-Haimoud and S.~Bird, ``{An efficient implementation of massive neutrinos
  in non-linear structure formation simulations},''
  \href{http://dx.doi.org/10.1093/mnras/sts286}{{\em Mon.Not.Roy.Astron.Soc.}
  {\bfseries 428} (2012)  3375--3389},
\href{http://arxiv.org/abs/1209.0461}{{\ttfamily arXiv:1209.0461
  [astro-ph.CO]}}.

\bibitem{Hannestad:2011td}
S.~Hannestad, T.~Haugbolle, and C.~Schultz, ``{Neutrinos in Non-linear
  Structure Formation - a Simple SPH Approach},''
  \href{http://dx.doi.org/10.1088/1475-7516/2012/02/045}{{\em JCAP} {\bfseries
  1202} (2012)  045},
\href{http://arxiv.org/abs/1110.1257}{{\ttfamily arXiv:1110.1257
  [astro-ph.CO]}}.

\bibitem{McDonald:2009hs}
P.~McDonald, ``{How to generate a significant effective temperature for cold
  dark matter, from first principles},''
  \href{http://dx.doi.org/10.1088/1475-7516/2011/04/032}{{\em JCAP} {\bfseries
  1104} (2011)  032},
\href{http://arxiv.org/abs/0910.1002}{{\ttfamily arXiv:0910.1002
  [astro-ph.CO]}}.

\bibitem{Boyanovsky:2008he}
D.~Boyanovsky, H.~de~Vega, and N.~Sanchez, ``{The dark matter transfer
  function: free streaming, particle statistics and memory of gravitational
  clustering},'' \href{http://dx.doi.org/10.1103/PhysRevD.78.063546}{{\em
  Phys.Rev.} {\bfseries D78} (2008)  063546},
\href{http://arxiv.org/abs/0807.0622}{{\ttfamily arXiv:0807.0622 [astro-ph]}}.

\bibitem{Basse:2010qp}
T.~Basse, O.~E. Bjaelde, and Y.~Y. Wong, ``{Spherical collapse of dark energy
  with an arbitrary sound speed},''
  \href{http://dx.doi.org/10.1088/1475-7516/2011/10/038}{{\em JCAP} {\bfseries
  1110} (2011)  038},
\href{http://arxiv.org/abs/1009.0010}{{\ttfamily arXiv:1009.0010
  [astro-ph.CO]}}.

\bibitem{Bertschinger:1993xt}
E.~Bertschinger, ``{Cosmological dynamics: Course 1},''
\href{http://arxiv.org/abs/astro-ph/9503125}{{\ttfamily arXiv:astro-ph/9503125
  [astro-ph]}}.

\bibitem{Brandenberger:1987kf}
R.~H. Brandenberger, N.~Kaiser, and N.~Turok, ``{Dissipationless Clustering of
  Neutrinos Around a Cosmic String Loop},''
\href{http://dx.doi.org/10.1103/PhysRevD.36.2242}{{\em Phys.Rev.} {\bfseries
  D36} (1987)  2242}.

\bibitem{watts}
E.~Bertschinger and P.~N. Watts, ``{Galaxy formation with cosmic strings and
  massive neutrinos},'' \href{http://dx.doi.org/10.1086/166265}{{\em
  Astrophys.J.} {\bfseries 328} (1988)  23--33}.

\bibitem{Ma:1995ey}
C.-P. Ma and E.~Bertschinger, ``{Cosmological perturbation theory in the
  synchronous and conformal Newtonian gauges},''
  \href{http://dx.doi.org/10.1086/176550}{{\em Astrophys.J.} {\bfseries 455}
  (1995)  7--25},
\href{http://arxiv.org/abs/astro-ph/9506072}{{\ttfamily arXiv:astro-ph/9506072
  [astro-ph]}}.

\bibitem{Scoccimarro1996}
R.~Scoccimarro and J.~Frieman, ``{Loop corrections in nonlinear cosmological
  perturbation theory},'' \href{http://dx.doi.org/10.1086/192306}{{\em
  Astrophys.J.Suppl.} {\bfseries 105} (1996)  37},
\href{http://arxiv.org/abs/astro-ph/9509047}{{\ttfamily arXiv:astro-ph/9509047
  [astro-ph]}}.

\bibitem{Jain1996}
B.~Jain and E.~Bertschinger, ``Self-Similar Evolution of Cosmological Density
  Fluctuations,'' {\em Astrophys.J.} {\bfseries 456} (1996)  43,
  \href{http://arxiv.org/abs/astro-ph/9503025}{{\ttfamily astro-ph/9503025}}.

\bibitem{Zeldovich1965}
Y.~B. Zel'dovich {\em Adv.Astron.} {\bfseries 3} (1965)  241.

\bibitem{Bertschinger1995}
E.~Bertschinger, ``{COSMICS: cosmological initial conditions and microwave
  anisotropy codes},''
\href{http://arxiv.org/abs/astro-ph/9506070}{{\ttfamily arXiv:astro-ph/9506070
  [astro-ph]}}.

\bibitem{Dvorkin:2014lea}
C.~Dvorkin, M.~Wyman, D.~H. Rudd, and W.~Hu, ``{Neutrinos help reconcile Planck
  measurements with both Early and Local Universe},''
  \href{http://dx.doi.org/10.1103/PhysRevD.90.083503}{{\em Phys.Rev.}
  {\bfseries D90} (2014)  083503},
\href{http://arxiv.org/abs/1403.8049}{{\ttfamily arXiv:1403.8049
  [astro-ph.CO]}}.

\bibitem{Hamann:2013iba}
J.~Hamann and J.~Hasenkamp, ``{A new life for sterile neutrinos: resolving
  inconsistencies using hot dark matter},''
  \href{http://dx.doi.org/10.1088/1475-7516/2013/10/044}{{\em JCAP} {\bfseries
  1310} (2013)  044},
\href{http://arxiv.org/abs/1308.3255}{{\ttfamily arXiv:1308.3255
  [astro-ph.CO]}}.

\bibitem{Battye:2013xqa}
R.~A. Battye and A.~Moss, ``{Evidence for Massive Neutrinos from Cosmic
  Microwave Background and Lensing Observations},''
  \href{http://dx.doi.org/10.1103/PhysRevLett.112.051303}{{\em Phys.Rev.Lett.}
  {\bfseries 112} (2014) no.~5, 051303},
\href{http://arxiv.org/abs/1308.5870}{{\ttfamily arXiv:1308.5870
  [astro-ph.CO]}}.

\bibitem{Zhu:2013tma}
H.-M. Zhu, U.-L. Pen, X.~Chen, D.~Inman, and Y.~Yu, ``{Measurement of Neutrino
  Masses from Relative Velocities},''
  \href{http://dx.doi.org/10.1103/PhysRevLett.113.131301}{{\em Phys.Rev.Lett.}
  {\bfseries 113} (2014)  131301},
\href{http://arxiv.org/abs/1311.3422}{{\ttfamily arXiv:1311.3422
  [astro-ph.CO]}}.

\bibitem{Zhu:2014qma}
H.-M. Zhu, U.-L. Pen, X.~Chen, and D.~Inman, ``{Probing Neutrino Hierarchy and
  Chirality via Wakes},''
\href{http://arxiv.org/abs/1412.1660}{{\ttfamily arXiv:1412.1660
  [astro-ph.CO]}}.

\bibitem{Delves1992}
L.~M. Delves and J.~L. Mohamed, ``{Computional methods for integral
  equation},'' \href{http://dx.doi.org/10.1017/CBO9780511569609}{{\em Cambridge
  University Press} (1985)  }.

\end{thebibliography}\endgroup

\end{document}